\documentclass[prc,reprint,superscriptaddress,nofootinbib,longbibliography]{revtex4-1}

\usepackage[english]{babel}
\usepackage{graphicx}
\usepackage{color}
\usepackage{units}
\usepackage{xcolor,colortbl}
\definecolor{prediction}{rgb}{0.9,0.9,0.9}
\usepackage{amsmath,amssymb,bm}
\usepackage{amsfonts}
\usepackage[latin1]{inputenc}
\usepackage[caption=false]{subfig}
\usepackage[colorlinks,urlcolor={blue}]{hyperref}
\usepackage{soul}

\usepackage{todonotes}

\newcommand{\co}{}

  \def\nuc#1#2{\relax\ifmmode{}^{#1}{\protect\text{#2}}\else${}^{#1}$#2\fi}
  \def\itnuc#1#2{\setbox\@tempboxa=\hbox{\scriptsize\it #1}
    \def\@tempa{{}^{\box\@tempboxa}\!\protect\text{\it #2}}\relax
    \ifmmode \@tempa \else $\@tempa$\fi}
\newcommand{\rd}{\ensuremath{\mathrm{d}}}
\newcommand{\id}{\ensuremath{\,\rd}}

\newcommand{\bra}[1]{\ensuremath{\left\langle #1\right |}}
\newcommand{\ket}[1]{\ensuremath{\left | #1\right \rangle}}

\newcommand{\braopket}[3]{\ensuremath{\left\langle #1 \middle|#2\middle| #3\right \rangle}}
\newcommand{\LOsep}{LOsep}
\newcommand{\LOsim}{LOsim}
\newcommand{\NLOsep}{NLOsep}
\newcommand{\NLOsim}{NLOsim}
\newcommand{\NNLOsep}{NNLOsep}
\newcommand{\NNLOsim}{NNLOsim}
\newcommand{\xEFT}{\ensuremath{\chi}EFT}

\newcommand{\NN}{\ensuremath{N\!N}}
\newcommand{\piN}{\ensuremath{\pi N}}
\newcommand{\NNN}{\ensuremath{N\!N\!N}}

\newcommand{\ODd}{\ensuremath{D({}^2\text{H})}}
\newcommand{\OQd}{\ensuremath{Q({}^2\text{H})}}
\newcommand{\OHL}{\ensuremath{E^1_A({}^3\text{H})}}
\newcommand{\OcHL}{\ensuremath{fT_{1/2}({}^3\text{H})}}

\newcommand{\OEd}{\ensuremath{E({}^2\text{H})}}
\newcommand{\OEt}[1][{}]{\ensuremath{E^{#1}({}^3\text{H})}}
\newcommand{\OEh}{\ensuremath{E({}^3\text{He})}}
\newcommand{\OEa}{\ensuremath{E({}^4\text{He})}}
\newcommand{\OEO}{\ensuremath{E({}^{16}\text{O})}}

\newcommand{\Orp}[1][{}]{\ensuremath{r_p^{#1}}}
\newcommand{\Orn}[1][{}]{\ensuremath{r_n^{#1}}}

\newcommand{\Orpd}[1][{}]{\ensuremath{r_{\text{pt-p}}^{#1}({}^2\text{H})}}
\newcommand{\Orpt}[1][{}]{\ensuremath{r_{\text{pt-p}}^{#1}({}^3\text{H})}}
\newcommand{\Orph}[1][{}]{\ensuremath{r_{\text{pt-p}}^{#1}({}^3\text{He})}}
\newcommand{\Orpa}[1][{}]{\ensuremath{r_{\text{pt-p}}^{#1}({}^4\text{He})}}

\newcommand{\Orcd}[1][{}]{\ensuremath{r_{\text{ch}}^{#1}({}^2\text{H})}}

\newcommand{\Oann}{\ensuremath{a_{nn}^{\text{N}}}}
\newcommand{\Oanp}{\ensuremath{a_{np}^{\text{N}}}}

\newcommand{\OappC}{\ensuremath{a_{pp}^C}}

\newcommand{\Ornn}{\ensuremath{r_{nn}^{\text{N}}}}
\newcommand{\Ornp}{\ensuremath{r_{np}^{\text{N}}}}

\newcommand{\OrppC}{\ensuremath{r_{pp}^C}}

\newcommand{\fpi}{\ensuremath{F_{\pi}}}

\newcommand{\LECvec}{\ensuremath{\bm{\alpha}}}
\newcommand{\LECidx}{\LECvec}
\newcommand{\LEC}{\ensuremath{\alpha}}
\DeclareMathOperator*{\argmin}{arg\,min}
\DeclareMathOperator{\Cov}{Cov}
\DeclareMathOperator{\diag}{diag}

\graphicspath{{fig-article/}}

\begin{document}

\title{Uncertainty analysis and order-by-order optimization of chiral
  nuclear interactions} 

\author{B.~D.~Carlsson} \email{borisc@chalmers.se}
\affiliation{Department of Fundamental Physics,
  Chalmers University of Technology, SE-412 96 G\"oteborg, Sweden}

\author{A.~Ekstr\"om} \email{ekstrom@utk.edu}
\affiliation{Department of Physics and Astronomy, University of
  Tennessee, Knoxville, TN 37996, USA}
\affiliation{Physics Division, Oak Ridge National Laboratory, Oak Ridge,
  TN 37831, USA} 

\author{C.~Forss\'en} \email{christian.forssen@chalmers.se}
\affiliation{Department of Fundamental Physics,
  Chalmers University of Technology, SE-412 96 G\"oteborg, Sweden}
\affiliation{Department of Physics and Astronomy, University of
  Tennessee, Knoxville, TN 37996, USA}
\affiliation{Physics Division, Oak Ridge National Laboratory, Oak Ridge,
  TN 37831, USA} 

\author{D.~Fahlin~Str\"omberg}
\affiliation{Department of Fundamental Physics,
  Chalmers University of Technology, SE-412 96 G\"oteborg, Sweden}

\author{G. R. Jansen}
\affiliation{Physics Division, Oak Ridge National Laboratory, Oak
  Ridge, TN 37831, USA}
\affiliation{National Center for Computational
  Sciences, Oak Ridge National Laboratory, Oak Ridge, TN 37831, USA}

\author{O.~Lilja}
\affiliation{Department of Fundamental Physics,
  Chalmers University of Technology, SE-412 96 G\"oteborg, Sweden}

\author{M.~Lindby}
\affiliation{Department of Fundamental Physics,
  Chalmers University of Technology, SE-412 96 G\"oteborg, Sweden}

\author{B.~A.~Mattsson}
\affiliation{Department of Fundamental Physics,
  Chalmers University of Technology, SE-412 96 G\"oteborg, Sweden}

\author{K.~A.~Wendt}
\affiliation{Department of Physics and Astronomy, University of
  Tennessee, Knoxville, TN 37996, USA}
\affiliation{Physics Division, Oak Ridge National Laboratory, Oak Ridge,
  TN 37831, USA} 

\date{\today}

\begin{abstract}
Chiral effective field theory ($\chi$EFT) provides a systematic
approach to describe low-energy nuclear forces. Moreover, $\chi$EFT is
able to provide well-founded estimates of statistical and systematic
uncertainties --- although this unique advantage has not yet been fully
exploited.
We fill this gap by performing an optimization and statistical
analysis of all the low-energy constants (LECs) up to
next-to-next-to-leading order. Our optimization protocol corresponds
to a simultaneous fit to scattering and bound-state observables in the
pion-nucleon, nucleon-nucleon, and few-nucleon sectors, thereby
utilizing the full model capabilities of $\chi$EFT. Finally, we study
the effect on other observables by demonstrating
forward-error-propagation methods that can easily be adopted by future
works.
We employ mathematical optimization and implement automatic
differentiation to attain efficient and machine-precise first- and
second-order derivatives of the objective function with respect to the
LECs. This is also vital for the regression analysis.  We use
power-counting arguments to estimate the systematic uncertainty that
is inherent to $\chi$EFT and we construct chiral
interactions at different orders with quantified
uncertainties. Statistical error propagation is compared with Monte
Carlo sampling showing that statistical errors are in general small
compared to systematic ones.
In conclusion, we find that a simultaneous fit to different sets of
data is critical to (i) identify the optimal set of LECs, (ii) capture
all relevant correlations, (iii) reduce the statistical uncertainty,
and (iv) attain order-by-order convergence in $\chi$EFT. Furthermore,
certain systematic uncertainties in the few-nucleon sector are shown
to get substantially magnified in the many-body sector; in particlar
when varying the cutoff in the chiral potentials. The methodology and
results presented in this Paper open a new frontier for uncertainty
quantification in \textit{ab initio} nuclear theory.
\end{abstract}

\maketitle

%=======================================
\section{Introduction
\label{sec:introduction}} 
%-------------------
%
Uncertainty quantification is essential for generating new knowledge
in scientific studies.
This insight is resonating also in theoretical disciplines, and
forward error propagation is gaining well-deserved recognition.
For instance, theoretical error bars have been estimated in various
fields such as neurodynamics~\cite{valderrama2015b}, global climate
models~\cite{murphy2004}, molecular
dynamics~\cite{angelikopoulos2012}, density functional
theory~\cite{erler2012}, and high-energy physics~\cite{cacciari2011}.

In this paper, we present a systematic and practical approach for
uncertainty quantification in microscopic nuclear theory.  
For the first time, we provide a common statistical regression
analysis of two key frameworks in theoretical nuclear physics:
\emph{ab initio} many-body methods and chiral effective field theory
({\xEFT}). We supply a set of mathematically optimized interaction
models with known statistical properties so that our results can be readily
applied by others to explore uncertainties in related efforts.

The \emph{ab initio} methods for solving the many-nucleon
Schr\"odinger equation, such as the no-core shell model
(NCSM)~\cite{barrett2013} and the coupled cluster (CC)
approach~\cite{hagen2013}, are characterized by the use of controlled
approximations. This provides a handle on the error that is
associated with the solution method itself.
Over the past decade there has been significant progress in
first-principles calculations of bound, resonant, and scattering
states in light
nuclei~\cite{pieper2001,lee2009,barrett2013,leidemann2013,romeroredondo2014,deltuva2015}
and medium-mass
nuclei~\cite{mihaila2000,hagen2013,lahde2014,soma2013,hergert2013}.
The appearance of independently confirmed and numerically exact
solutions to the nuclear many-body problem has brought forward the
need for an optimized nuclear interaction model with high accuracy,
quantified uncertainties, and predictive capabilities.

\xEFT{} is a powerful and viable approach for describing the
low-energy interactions between constituent
nucleons~\cite{epelbaum2009,machleidt2011} --- a cornerstone for
the microscopically grounded description of the atomic nucleus and its
properties.
Most importantly, the inherent uncertainty of the \xEFT{} model can be
estimated from the remainder term of the underlying momentum-expansion
of the effective Lagrangian. We refer to this error as a
\emph{systematic} model uncertainty.

We use the common term low-energy constants (LECs) to denote the
effective parameters of a nuclear interaction model. Indeed, for the
description of atomic nuclei, the numerical values of the LECs play a
decisive role. In the \xEFT{} approach, the LECs can in principle be
connected to predictions from the underlying theory of quantum
chromodynamics (QCD), see e.g.\ Ref.~\cite{barnea2015}. However, the
currently viable approach to accurately describe atomic nuclei in
\xEFT{} requires that the LECs are constrained from experimental
low-energy data. The bulk of this fit data traditionally consists of
cross sections measured in nucleon-nucleon (\NN{}) scattering
experiments. Most often, this data is parameterized in terms of phase
shifts~\cite{stoks1993,arndt1999}. However, experimental data comes
with error bars, which implies that a thorough statistical error
analysis of the constructed nuclear Hamiltonian can only be performed
when fitting directly to nuclear scattering observables. This
optimization procedure gives rise to \emph{statistical} uncertainties
on the LECs.

In general, the determination of the LECs constitutes an extensive
nonlinear optimization problem. That is, the relatively large number
of parameters makes it challenging to find optimal values such that
the experimental fit data is best reproduced. Various methods and
objective functions have been used to solve this problem for a wide
array of available nuclear-interaction
models~\cite{stoks1994,wiringa1995,machleidt2001,pieper2001b,entem2003,epelbaum2005,epelbaum2014}. More
often than not, the parameters of the models were fitted by hand.
Mathematical optimization algorithms were only recently introduced in
this venture by~\citet{ekstrom2013} and by~\citet{navarro2013}. First
attempts to investigate the statistical constraints on the LECs of
mathematically optimized interactions have recently been performed in
the \NN{} sector with coarse-grained $\delta$-shell
interactions~\cite{navarro2013b,navarro2014,navarro2014b,navarro2015b}
and with \xEFT{} \NN{} interactions~\cite{ekstrom2015,navarro2015}.

The so called power-counting scheme of the \xEFT{} approach offers a
systematically improvable description of \NN{},
three-nucleon (\NNN{}), and pion-nucleon (\piN{}) interactions.
It provides a consistent framework in which LECs from
the effective \piN{} Lagrangian also govern the
strength of pion-exchanges in the \NN{} potential and of long- and
intermediate-range \NNN{} forces. This implies that \piN{} scattering
data can be used to constrain some LECs that enter the chiral nuclear
Hamiltonian.

Furthermore, \xEFT{} offers an explanation for the
appearance of many-nucleon interactions, such as \NNN{}-diagrams, and
the fact that they provide higher-order corrections in the hierarchy
of nuclear forces. Still, effective \NNN{} forces are known to
play a prominent role in nuclear
physics~\cite{pieper2001,navratil2007,hammer2013}. 
Most often, the LECs that are associated with the \NNN{} terms have
been determined relative to existing \NN{} Hamiltonians. These LECs
are optimized against a few select binding energies, excitation
energies, or other properties of light nuclei.

The extended approach that is presented here is conceptually
consistent with {\xEFT} in the sense that the $\NN+\NNN$ Hamiltonian
is constrained from a simultaneous mathematical optimization to \NN{}
and \piN{} scattering data, plus observables from \NNN{} bound states
including the electroweak process responsible for the $\beta$-decay of
\nuc{3}{H}. Furthermore, we include the truncation error of the chiral
expansion to take systematic theoretical errors into account. If
correctly implemented, the truncation error of an observable
calculated in this scheme should decrease systematically with
increasing order in the \xEFT{} expansion. Indeed, we will show that
the resulting propagated uncertainties of a simultaneous fit
are smaller and exhibit a more obvious convergence pattern compared to
the traditional separate or sequential approaches that have been
published so far.

Below, we summarize the work presented in this Paper
by listing three specific objectives:
\begin{itemize}
\item{Establish a systematic framework for performing mathematical
    optimization and uncertainty quantification of nuclear forces in
    the scheme of \xEFT{}. Our approach relies on the simultaneous
    optimization of the effective nuclear Hamiltonian to low-energy
    \piN{}, \NN{}, and \NNN{} data with the inclusion of experimental
    as well as theoretical error bars.}
\item{Demonstrate methods to propagate the statistical errors in the
    order-by-order optimized nuclear Hamiltonian to various nuclear
    observables and investigate the convergence of the chiral
    expansion.}
\item{Deliver optimized chiral interactions with well-defined
  uncertainties and thoroughly introduce the accompanying
  methodological development such that our results can be easily
  applied in other calculations.}
\end{itemize}

Our paper is organized as follows: In Section~\ref{sec:method} we
introduce the methodology. We start with the construction of the
nuclear potential from \xEFT{} and proceed to the calculation of
observables and the optimization of parameters. In particular, we
introduce automatic differentiation for numerically exact computation
of derivatives, and we discuss the error budget and error
propagation. The results of our analysis, for potentials at different
orders in the chiral expansion and using different optimization
strategies, is presented in Section~\ref{sec:results}. We study the
order-by-order convergence, the correlation between parameters, and we
present first results for few-nucleon observables with well-quantified
statistical errors propagated via chiral interactions. The
consequences of our findings in the few- and many-body sectors are
discussed in Section~\ref{sec:systematic}, and in
Section~\ref{sec:outlook} we present an outlook for further work.

%=======================================
\section{Method}\label{sec:method}
%-------------------
%
In this section we give an overview of the nuclear {\xEFT} that we
employ to construct a nuclear potential (Sec.~\ref{sec:chiralEFT}).
The optimal values for LECs are not provided by {\xEFT} itself;
they need to be constrained from a fit to data. For completeness
we will summarize the well-known methods to calculate the relevant
experimental observables: \NN{} scattering cross sections
(Sec.~\ref{sec:NNscatt}), \NN{} ${}^1S_0$ effective range
parameters (Sec.~\ref{sec:effr}) and bound state properties for $A\leq 4$
nuclei using the Jacobi-coordinate no-core shell model
(Sec.~\ref{sec:fewnucleon}). We also present the objective function
(Sec.~\ref{sec:objectivefunction}), the optimization algorithm
(Sec.~\ref{sec:optimizationalgorithm}), and the formalism for the
statistical regression (Sec.~\ref{sec:stat_err}).
%
%------------------------------
\subsection{\label{sec:chiralEFT}%
The nuclear potential from {\xEFT}}
%------------------------------
%
The long-range part of the nuclear interaction in \xEFT{} is governed
by the spontaneously broken chiral symmetry of QCD and mediated by the
corresponding Goldstone-boson; the pion ($\pi$). This groundbreaking
insight~\cite{weinberg1979} enables a perturbative
approach to the description of phenomena in low-energy nuclear
physics~\cite{weinberg1990}. High-energy physics that is not
explicitly important is accounted for through a process of
renormalization and regularization with an accompanying power counting
scheme. The expansion parameter is defined as $Q/\Lambda_{\chi}$,
where $Q$ is associated with the external momenta (soft scale) and
$\Lambda_{\chi} \approx M_{\rho}$ (hard scale), with
$M_{\rho}\approx\unit[800]{MeV}$ the mass of the rho meson. The
benefit of a small-parameter expansion is that higher orders
contribute less than lower orders. If the series is converging, an
estimate of the magnitude of the truncation error is given by the size
of the remainder.

\begin{figure}
    \centering
    \includegraphics[width=1.0\columnwidth]{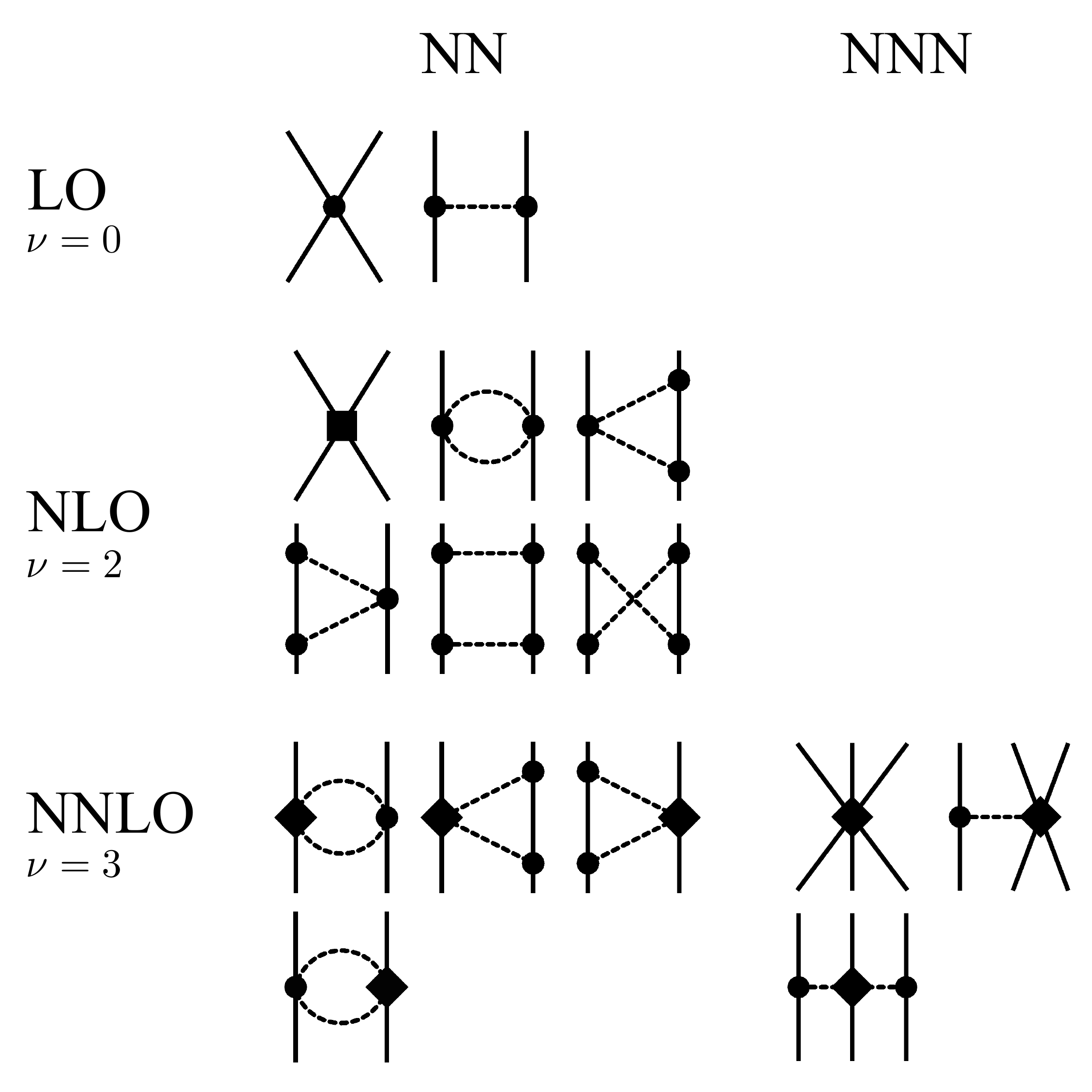}
    \caption{\label{fig:feynman_NNLO}Schematic overview of the Feynman
      diagrams present at leading order (LO), next-to-leading order
      (NLO), and next-to-next-to-leading order (NNLO).  Nucleons
      (pions) are represented by solid (dashed) lines. The
      three-nucleon (\NNN) interaction enters at NNLO\@. A circle,
      diamond and square represents a vertex of order $\Delta = 0$,
      $1$ and $2$ respectively. }
\end{figure}

The chiral order of a Feynman diagram is governed by the adopted
power-counting scheme. Given this, any chiral order $\nu \geq 0$ in
the expansion will be identified with a finite set of terms
proportional to $(Q/\Lambda_\chi)^{\nu}$. In this work we have
adopted the standard Weinberg power-counting (WPC) which is obtained
from the assumptions of naive dimensional analysis. For the scattering
of two or more nucleons without spectator particles, $\nu$ is determined by (see
e.g.\ Ref.~\cite{machleidt2011})
\begin{align}
  \label{eq:wpc}
    \nu = 2A - 4 + 2L + \sum_i \Delta_i
\end{align}
where $A$ is the number of nucleons and $L$ is the number of pion
loops involved. The sum runs over all vertices $i$ of the considered
diagrams and $\Delta_i$ is proportional to the number of nucleon
fields and pion-mass derivatives of vertex $i$. $\Delta_i\geq 0$ for
all diagrams allowed by chiral symmetry. In
Fig.~\ref{fig:feynman_NNLO} we show the different interaction diagrams
that enter at various orders.
For the \NN{} system, contributions at $\nu = 1$ vanish due to parity
and time-reversal invariance. Also, we consider nucleons and pions as
the only effective degrees of freedom and ignore possible nucleon
excitations, i.e., we use the so-called delta-less version of {\xEFT}.

The interaction due to short-range physics is parameterized by
contact terms, which also serve to renormalize the
infinities of the pion loop integrals. The order-by-order expansion of
this zero-range contribution is also organized in terms of increasing
powers of $Q/\Lambda_{\chi}$. Due to parity, only even powers of $\nu$
are non-zero. Furthermore, the contact terms of order $\nu=0$
contribute only to partial waves with angular momentum $L=0$,
i.e.\ $S-$waves, whereas $\nu=2$ contact terms contribute up to
$P$-waves. In general, the contact interaction at order $\nu$ acts in
partial waves with $L \leq \nu/2$. Following Eq.~\eqref{eq:wpc},
the terms in the {\xEFT} expansion, up to third order, are given by a
sum of contact interactions $V_{\rm ct}$ and one- plus two-pion
exchanges, denoted by $V_{1\pi}$ and $V_{2\pi}$, respectively:
\begin{align}
  \begin{split}
    \label{eq:nn_potential}
    V_{\rm LO} {}&= V_{\rm ct}^{(0)} + V_{1\pi}^{(0)}\\
    V_{\rm NLO} {}&= V_{\rm LO} + V_{\rm ct}^{(2)} + V_{1\pi}^{(2)} + V_{2\pi}^{(2)}\\
    V_{\rm NNLO} {}&= V_{\rm NLO} + V_{1\pi}^{(3)} + V_{2\pi}^{(3)} + V_{\NNN}.\\
  \end{split}
\end{align}
The superscript indicates the separate chiral orders $\nu=0,2,3$,
referred to as leading-order (LO), next-to-leading order (NLO), and
next-to-next-to-leading order (NNLO). For detailed expressions, see
e.g.\ Ref.~\cite{epelbaum2005}. The three-nucleon interaction,
$V_{\NNN}$, contains three different diagrams as shown in
Fig.~\ref{fig:feynman_NNLO}. These correspond to two-pion exchange,
one-pion exchange plus contact, and a pure \NNN{} contact term.
Insofar, the analytical expressions for the \NN{} potential have been
derived up to fifth order (N4LO)~\cite{machleidt2015,epelbaum2015}. The
partial-wave decomposition for the \NNN{} interaction at NNLO is well
known~\cite{epelbaum2002}, while the N3LO contribution was published
very recently~\cite{hebeler2015}. Note that the connected four-nucleon
diagrams also appear at this higher order. In the present work we
limit ourselves to NNLO for completeness.

The strengths of the terms in the {\xEFT} interaction are governed by
a set of LECs. These parameters play a central role in this work, and
we discuss in detail how they are constrained from measured data. In
general, for each chiral order there will appear a new set of
LECs. For the nuclear interactions used in this work, see
Eq.~\eqref{eq:nn_potential}, the corresponding LECs are denoted
\begin{align}
    \begin{split}\label{eq:LECs}
        V_{\text{ct}}^{(0)} &\sim \{\tilde{C}_{{}^1S_0}, \tilde{C}_{{}^3S_1}\} \\
        V_{\text{ct}}^{(2)} &\sim \{C_{{}^1S_0}, C_{{}^3S_1}, C_{E_1},
        C_{{}^3P_0}, C_{{}^1P_1}, C_{{}^3P_1}, C_{{}^3P_2}\}, \\ 
        V_{2\pi}^{(3)} &\sim \{c_1, c_3, c_4\},\\
        V_{\NNN} &\sim \{ c_1,c_3,c_4,c_D,c_E\}.
    \end{split}
\end{align}

Furthermore, there are additional constants that must be determined
before making quantitative predictions in {\xEFT}. Here, we set the
axial-vector coupling constant to the experimentally determined value
of $g_A=1.276$~\cite{Liu2010} for LO, whereas for the higher orders we
use the renormalized value of $g_A=1.29$ to account for the
Goldberger-Treiman discrepancy~\cite{epelbaum2005}.  At all orders we
use $\fpi = \unit[92.4]{MeV}$~\cite{epelbaum2005}.  All other physical
constants, such as nucleon masses and the electric charge, are taken
from CODATA 2010~\cite{mohr2012}, except the pion masses for which we
have used the values from the Particle Data Group~\cite{beringer2012}.

Note that LECs that determine the sub-leading \piN{} interaction
vertices occur in both the \NN{} interaction and the two-pion-exchange
part of the \NNN{}, see
Refs.~\cite{epelbaum2005,epelbaum2002}. Besides offering this
pion-vertex link between the \NN{} and the \NNN{} interaction,
the \piN{} interaction model of {\xEFT} allows to describe \piN{}
scattering processes. Consequently, experimental \piN{}
scattering data can be used to constrain the long-range part of the nuclear
interaction. The lowest order terms of the
effective \piN{} Lagrangian have $\nu = 1$ and are free from LECs,
besides $g_A$ and $\fpi$. At order $\nu = 2$ the LECs
$c_1$, $c_2$, $c_3$, and $c_4$ enter. Higher-order \piN{} LECs, such as $d_1+d_2$,
$d_3$, $d_5$ and $d_{14}-d_{15}$, enter at $\nu = 3$ while $e_{14}$ to
$e_{18}$ appear at $\nu = 4$. In total, there are $13$ LECs in the \piN{}
Lagrangian up to fourth order.

The different masses and charges of the up and down quarks give rise
to isospin-violating effects~\cite{epelbaum2005,machleidt2011}. There
are both short- and long-range isospin-violating effects. The
long-range effects are of electromagnetic (EM) origin and for this
contribution we use the well-known set of potentials
\begin{align}
    \begin{split}
        V_{\text{EM}}^{(pp)} =& V_{\text{C1}} + V_{\text{C2}} + V_{\text{VP}} + V_{\text{MM}}^{(pp)},\\
        V_{\text{EM}}^{({np})} =& V_{\text{MM}}^{({np})},
    \end{split}
\end{align}
where C1 is the static Coulomb potential, C2 the relativistic
correction to the Coulomb potential~\cite{austen1983}, VP is the vacuum
polarization potential~\cite{durand1957}, and MM the magnetic-moment
interaction~\cite{stoks1990}. The long-range effects become
increasingly important as the scattering energy approaches zero;
consequently we include all the above long-range effects at all orders
in the chiral expansion.
We also consider short-range isospin-breaking mechanisms.  At NLO, the
$\tilde{C}_{{}^1S_0}$ contact is split into three charge-dependent
terms: $\tilde{C}_{{}^1S_0}^{(pp)}$, $\tilde{C}_{{}^1S_0}^{({np})}$ and
$\tilde{C}_{{}^1S_0}^{({nn})}$. At this order, and above, we also take
the pion-mass splitting into account in one-pion exchange
terms~\cite{machleidt2011}.

An effective field theory often has to handle more than one expansion
parameter. In our case, the nucleon mass, $M_N \equiv
2M_pM_n/(M_p+M_n)$ where $M_p$ ($M_n$) is the proton (neutron) mass,
provides such an extra scale and the use of the heavy-baryon chiral
perturbation theory introduces relativistic corrections with factors
of $1/M_N$. We count these corrections as $Q/M_N \approx
(Q/\Lambda_\chi)^2$~\cite{weinberg1990,weinberg1991}. This choice
implies that no relativistic corrections appear in the \NN{} sector up
to the order considered in this paper.

To regularize the loop integrals that are present in the
two-pion exchange diagrams we employ spectral function regularization
(SFR)~\cite{epelbaum2006} with an energy cutoff $\tilde\Lambda =
\unit[700]{MeV}$. The nuclear interaction is calculated perturbatively
in {\xEFT}. A nuclear potential that can be used for bound and
scattering states is obtained by iterating the terms of the chiral
expansion in the Lippmann-Schwinger or Schr\"{o}dinger
equation~\cite{ordonez1994}. We employ the minimal-relativity
prescription from Ref.~\cite{brown1969} to obtain
relativistically-invariant potential amplitudes. The ultraviolet
divergent Lippmann-Schwinger equation also require regularization. We
remove high-momentum contributions beyond a cutoff energy $\Lambda$ by
multiplying the \NN{} and \NNN{} interaction terms with standard
(non-local) regulator functions $f_{\NN}(p)$ and $f_{\NNN}(p,q)$,
respectively,
\begin{align}
    f_{\NN}(p) = \exp\left[- \left(\frac{p}{\Lambda}\right)^{2n}\right]
\end{align}
and 
\begin{align}
  f_{\NNN}(p, q) &=
  \exp\left[-\left(\frac{4p^2+3q^2}{4\Lambda^2}\right)^n\right], 
\end{align}
where $p$ and $q$ are the Jacobi momenta of the interacting
nucleons. In this work, we mainly use $\Lambda = \unit[500]{MeV}$ and
$n=3$. However, we also explore the consequences of varying $\Lambda$
in steps of 25 MeV between $\unit[450-600]{MeV}$.
The canonical power-counting, i.e.\ WPC, and the non-perturbative
renormalization of nuclear {\xEFT} in its current inception is
currently under some debate~\cite{nogga2005,valderrama2015a}. In
relation to this it should be stressed that our implementation of
statistical regression methods and gradient-based optimization methods
furnishes an independent framework to extract well-founded estimates
of the uncertainties in theoretical few-nucleon physics and a tool to
assess the convergence properties of {\xEFT}.
%
%------------------------------
\subsection{\label{sec:NNscatt}%
Nuclear scattering}
%------------------------------
%
The \NN{} scattering observables are calculated from the
spin-scattering matrix $M$~\cite{bystricky1978,lafrance1980}. This is
a $4\times4$ matrix in spin-space that operates on the initial state
to give the scattered part of the final state. Thus, $M$ is related to
the conventional scattering matrix $S$ by $M = \frac{2\pi}{ip} (S-1)$,
where $p$ is the relative momentum between the nucleons. The
decomposition of $M$ into partial waves is given by (see
e.g.~\cite{stoks1993})
%
% widetext version of M-matrix PWD
\begin{widetext}
\begin{equation}
  \label{eq:scatt_matrix_pwd}
    M_{m'm}^{s's}(\theta,\phi) =
    \frac{\sqrt{4\pi}}{2ip}\sum_{J,L,L'}^\infty(-1)^{s-s'}i^{L-L'}\hat{J}^2 \hat{L} 
    Y_{m-m'}^{L'}(\theta,\phi)
        \begin{pmatrix}
            L' & s' & J \\
            m-m' & m' & -m
        \end{pmatrix}
        \begin{pmatrix}
            L & s & J \\
            0 & m & -m
        \end{pmatrix}
        \bra{L',s'}S^J-1\ket{L,s},
\end{equation}
where the big parentheses are Wigner $3j$-symbols, $s$ ($s'$) and $m$
($m'$) are initial (final) total spin and spin projection,
respectively, $\mathbf{J}=\mathbf{L}+\mathbf{s}$ is the total relative angular momentum
and $\hat{L}$ ($\hat{J}$) is $2L+1$ ($2J+1$). The
quantization axis is taken along the direction of the incoming
nucleon and $\theta$ gives the center-of-mass scattering angle. The $S$-matrix for
the scattering channel with angular momentum $J$ can be parameterized
by the Stapp phase shifts~\cite{stapp1957},
\begin{equation}
    S_{L=J\pm 1}^J =
        \begin{pmatrix}
            e^{2i\delta_{J-1,J}}\cos 2\epsilon_J & ie^{i(\delta_{J-1,J}+\delta_{J+1,J})}\sin 2\epsilon_J \\
            ie^{i(\delta_{J-1,J}+\delta_{J+1,J})}\sin 2\epsilon_J & e^{2i\delta_{J+1,J}}\cos 2\epsilon_J
        \end{pmatrix}
\end{equation}
for the coupled triplet channel, and
\begin{equation}
    S_{L=J}^J =
        \begin{pmatrix}
            e^{2i\delta_{J}}\cos 2\gamma_J & ie^{i(\delta_{J}+\delta_{J,J})}\sin 2\gamma_J \\
            ie^{i(\delta_{J}+\delta_{J,J})}\sin 2\gamma_J & e^{2i\delta_{J,J}}\cos 2\gamma_J
        \end{pmatrix}
\end{equation}
for the (coupled) singlet-triplet channel with $L=J$. 
\end{widetext}
The spin-singlet ($S=0$) phase shift is denoted by $\delta_{L=J}$, the
spin-triplet ($S=1$) phase shift by $\delta_{L,J}$, while $\epsilon_J$
represents the triplet-channel mixing angle and $\gamma_J$ is the
spin-flip mixing angle~\cite{gersten1977} ($\gamma_J = 0$ for $pp$
scattering).

In practice, the infinite sums in Eq.~\eqref{eq:scatt_matrix_pwd} are
truncated at $L,L' \leq L_{\max}$.  Calculations that involve
long-ranged EM effects require $L_{\max} \geq 1000$ in order to reach
convergence, while $L_{\max} =30$ is sufficient for the part coming
from the short-ranged nuclear interaction. This leads to a natural
separation of the terms in Eq.~\eqref{eq:scatt_matrix_pwd}, see
e.g.\ Ref.~\cite{stoks1990}. In brief, all EM amplitudes are
calculated independently in Coulomb Distorted-Wave Born Approximation
(CDWBA) using Vincent-Phatak matching~\cite{vincent1974} to handle
the difficulties of the Coulomb interaction in momentum space. For the
${}^1S_0$ channel, the C2 and VP interactions are strong enough that a
small correction to the bare phase shifts is needed, resulting in
\begin{align}
    \delta_{\text{total}} = \delta_{\text{C1}+\NN}^{(\text{CDWBA})} +
    \tilde{\Delta}_0 - \rho_0 - \tau_0
\end{align}
where $\delta_{\text{C1}+\NN}^{(\text{CDWBA})}$ is the phase shift of
the Coulomb and the chiral \NN{} interactions computed in CDWBA,
$\rho_0$ ($\tau_0$) is the C2 (VP) phase shifts in CDWBA, and
$\tilde{\Delta}_0$ is a correction calculated by interpolating between
the values tabulated by~\citet{bergervoet1988}. In
principle, $\tilde{\Delta}_0$ is dependent on the interaction model
for the strong force; this effect has been shown to be very
small~\cite{bergervoet1988} and was not considered here.

We compute the VP phase shifts, $\tau_L$, in CDWBA using the
variable-phase method~\cite{calogero1967}. The values we obtain agree
with the ones that are tabulated by~\citet{bergervoet1988}. The VP
amplitude is calculated in the first-order approximation derived by~\citet{durand1957}
using the expansion parameter $X \equiv
4m_e^2 / (T_{\text{lab}}M_p(1 - \cos(\theta)))$, where
$m_e$ is the electron mass. We find that $X \lesssim 0.031$ for all scattering
data that is employed in this work. The MM amplitude for $np$ and $pp$
scattering is given by Stoks\footnote{Note that Eq.~(24) in
  Ref.~\cite{stoks1990} has the wrong sign. Furthermore, Eq.~(25)
  should have $|\sin(\theta)|$.}~\cite{stoks1990}.

The Stapp phase shifts are calculated from the real-valued free
reaction matrix $R$~\cite{erkelenz1971}, which is defined through a
Lippman-Schwinger type equation~\cite{erkelenz1971}
\begin{align}\begin{split}
  \label{eq:LS_Rmatrix}
    &R_{L'L}^{S,J}(p',p) = V_{L'L}^{S,J}(p',p) 
    - 2\mu \\
    &\quad\times \sum_{L''}\mathcal{P}\int_0^\infty {p''}^2\id
    p''\frac{V_{L'L''}^{S,J}(p',p'')R_{L''L}^{S,J}(p'',p)}{{p''}^2-p^2}, 
\end{split}\end{align}
where $V$ is the potential, $\mu$ the reduced mass, and $\mathcal{P}$
denotes the Cauchy principal value.

Due to parity and time-reversal invariance, the scattering matrix $M$
has six linearly independent elements. We employ the Saclay
parameterization~\cite{bystricky1978}, with complex
amplitudes $a$ to $f$, to express
\begin{align}\begin{split}
    &M(\mathbf{q}, \mathbf{k}) = \frac{1}{2}\big\{(a+b) +
        (a-b)\bm{\sigma}_1\cdot \hat{\mathbf{r}}\bm{\sigma}_2\cdot\hat{\mathbf{r}} \\
        &\,\,\,+ (c+d)(\bm{\sigma}_1\cdot\hat{\mathbf{q}})(\bm{\sigma}_2\cdot\hat{\mathbf{q}}) 
        + (c-d)(\bm{\sigma}_1\cdot\hat{\mathbf{
          k}})(\bm{\sigma}_2\cdot\hat{\mathbf{k}}) \\
        &\,\,\,-e(\bm{\sigma}_1+\bm{\sigma}_2)\cdot \hat{\mathbf{r}} -
        f(\bm{\sigma}_1-\bm{\sigma}_2)\cdot \hat{\mathbf{r}}\big\},
\end{split}\end{align}
where $\mathbf q = \mathbf{p}' - \mathbf{p}$ is the momentum transfer, $\mathbf k = (\mathbf{p}' + \mathbf{p})/2$
and $\mathbf{r} = \mathbf{q}\times\mathbf{k}$.
For identical particles, $f$ will be zero. Expressions for
the scattering observables in terms of the Saclay parameters can be
found in Ref.~\cite{bystricky1978} for identical particles
and in Ref.~\cite{lafrance1980} for the more general case
of non-identical particles. 

For the theoretical description of the \piN{} scattering observables
we use the fourth order {\xEFT} expressions according to
Refs.~\cite{wendt2014arxiv,krebs2012}. A detailed description
of the EM amplitudes that we employ are given in
Refs.~\cite{tromborg1978,tromborg1974,tromborg1977,bugg1973}
%
%------------------------------
\subsection{\label{sec:effr}%
Effective range parameters}
%------------------------------
%
The effective-range expansion (ERE) of low-energy phase
shifts~\cite{bethe1949} provides parameters that can be
directly compared to experimentally inferred values.  The ERE
can be expressed in the general form
\begin{align}
    A(p) + B(p)p\cot(\delta_{\text{LR}+\NN}^{\rm LR}) = -\frac{1}{a} +
    \frac{1}{2}r^2p^2 + O(p^4). 
\end{align}
The functions $A(p)$ and $B(p)$ depend on the choice of included
long-range EM effects and $\delta_{\text{LR}+\NN}^{\rm LR}$ is the
phase shift of the total nuclear potential (long-range plus strong
\NN{}) relative to the phase shift of only the long-range part.

For $nn$ and $np$ scattering we have $A(p) = 0$ and
$B(p) = 1$~\cite{bethe1949} since
there are no EM effects. The corresponding ERE
parameters are denoted \Oann, \Ornn, \Oanp{} and \Ornp.

For $pp$ scattering we calculate ERE parameters for the nuclear plus
Coulomb potential, i.e., using the phase shifts
$\delta_{\text{C1}+\NN}^{(\text{CDWBA})}$.
The expressions for $A_C(p)$ and $B_C(p)$ can be found in
Refs.~\cite{bethe1949,bergervoet1988}.
 The corresponding ERE parameters are denoted
\OappC{} and \OrppC.

In practice, the ERE parameters are determined using a linear
least-squares fit to $20$ equally-spaced phase shifts in the
$T_{\rm lab} = \unit[10-100]{keV}$ range.
%------------------------------
\subsection{\label{sec:fewnucleon}%
Few-nucleon observables}
%------------------------------
%
We employ the Jacobi-coordinate version of the
NCSM~\cite{navratil2000} to compute bound-state observables for
\nuc{2,3}{H} and \nuc{3,4}{He}. Apart from binding
energies and radii we also compute the deuteron quadrupole moment,
\OQd{}, and the comparative half-life for the triton \OcHL{}.

In the NCSM, observables and wave functions are obtained from the
exact solution of the eigenvalue problem $\mathcal{H}\ket{\psi} =
E\ket{\psi}$. In this work, the nuclear Hamiltonian $\mathcal{H}$ is
given by
\begin{equation}\label{eq:ncsm}
\mathcal{H}=\sum_{i<j=1}^A T_{ij} + \sum_{i<j=1}^A V_{ij} +
\sum_{i<j<k=1}^{A} V_{ijk}, 
\end{equation} 
where $T_{ij}$ are relative kinetic energies while $V_{ij}$ and
$V_{ijk}$ are the \NN{} and \NNN{} interactions, respectively. 
In our calculations we use the isoscalar approximation
as presented in Ref.~\cite{kamuntavicius1999}.
The model-space dimension is determined from the maximal number of
allowed harmonic-oscillator (HO) excitations $N_{\rm max}$. We obtain
essentially converged results in a HO basis with oscillator energy
$\hbar \omega = \unit[36]{MeV}$ and model-space dimension
$N_{\rm max}=40(20)$ for $A=3(4)$.

The experimentally measured electric-charge radius can be
related to the theoretically calculated point-proton 
radius through the relation~\cite{friar1976}
\begin{align}
    r_{\text{pt-p}}^2 = r_{\text{ch}}^2 - \Orp[2] - \frac{N}{Z}\Orn[2] -
    r_{\text{DF}}^2 - \Delta r^2,
\end{align}
where \Orp[2] (\Orn[2]) is the proton (neutron) charge mean-squared
radius and $Z$ ($N$) is proton (neutron) number.  Furthermore,
$r_{\text{DF}}^2 \equiv \frac{3}{4M_N^2}$ is the Darwin-Foldy
correction~\cite{jentschura2011} and $\Delta r^2$ includes effects of
two-body currents and further relativistic corrections.  We use $r_p =
\unit[0.8783(86)]{fm}$ and $r_n^2 =
\unit[-0.1149(27)]{fm^2}$~\cite{angeli2013}. For all nuclei, we use
$\Delta r^2 = 0$.

Precise results for electroweak observables depend on two-body
nuclear currents and relativistic effects.
{\xEFT} provides a consistent framework for including such corrections
and for deriving quantum-mechanical currents, such as the electroweak
one, from the same Lagrangian as the nuclear force. We 
follow the approach by \citet{gazit2009} and compute the triton
half-life from the reduced matrix element for $E_{1}^{A}$, the $J = 1$ electric
multipole of the axial-vector current 
\begin{align}
    \left\langle E_1^A\right\rangle \equiv \big|\!\left\langle
    {}^3\text{He}\middle\|E_1^A\middle\| {}^3\text{H}\right\rangle\!\big|.
\end{align}
This matrix element is proportional to $c_D$, the LEC that also
determines the strength of the $\NN-\pi N$ diagram of the \NNN{}
interaction. As a consequence, the triton half-life provides a further
constraint of the nuclear force. The experimentally determined
comparative half-life, $fT_{1/2} = \unit[1129.6 \pm
3]{s}$~\cite{akulov2005}, leads to an empirical value for
$\left\langle E_1^A\right\rangle = 0.6848\pm 0.0011$~\cite{gazit2009}.

For the deuteron quadrupole moment we choose, instead, to fit to the
theoretical value obtained from the high-precision meson-exchange
\NN{} model CD-Bonn, $Q_{\text{d}} = 0.27$~\cite{machleidt2001}, with a
$4\%$ error bar that more than well covers the spread in values using
other \NN{} potential models~\cite{machleidt2011}.
%
%------------------------------
\subsection{\label{sec:objectivefunction}%
Objective function}
%------------------------------
%
Using the methods to compute observables outlined above, the vector
\LECvec{} of numerical values for the LECs at a given order in
{\xEFT} is constrained using experimental data. This is accomplished
by minimizing an objective function defined as
\begin{align}
  \label{eq:objective_function}
    \chi^2(\LECvec) \equiv \sum_{i\in \mathbb{M}}
    \left(\frac{\mathcal{O}_i^{\text{theo}}(\LECvec) - \mathcal{O}_i^{\text{exp}}}{\sigma_i}\right)^2 \equiv
    \sum_{i\in \mathbb{M}} r_i^2(\LECvec),
\end{align}
where $\mathcal{O}_i^{\text{theo}}$ and $\mathcal{O}_i^{\text{exp}}$
denote the theoretical and experimental values of observable
$\mathcal{O}_i$ in the pool of fit data $\mathbb{M}$, and the total
uncertainty $\sigma_i$ determines the weight of the residual, $r_i$. The
optimal set of LECs $\LECvec_{\star}$ is defined from
\begin{equation}
\LECvec_{\star} = \argmin_{\LECvec} \chi^2(\LECvec)
\end{equation}

We wish to explore the physics capabilities and limitations of nuclear
{\xEFT} by forming different objective functions and subsequently
probing the precision and accuracy of each one in a statistical
regression analysis~\cite{dobaczewski2014}. At each chiral order (LO,
NLO, or NNLO) we compare two different strategies of minimization:
\emph{simultaneous} (sim) and \emph{separate} (sep). In the
``separate'' approach we first optimize the sub-leading \piN{} LECs
($c_i,d_i,e_i)$ using \piN{} data. Subsequently, we optimize the \NN{}
contact potential of the nuclear interaction using \NN{} scattering
data, and finally (at NNLO) the \NNN{} interaction is determined by
fitting $c_D$ and $c_E$ to the known binding energies and radii of
\nuc{3}{H} and \nuc{3}{He}, and the comparative $\beta$-decay half
life of \nuc{3}{H}. Besides the first-ever application of novel
derivative-based optimization techniques to this problem, the
``separate'' approach is very similar to the conventional procedure to
constrain the description of the nuclear interaction. In contrast, with
the ``simultaneous'' approach we optimize all the LECs up to a
specific-order in {\xEFT} at the same time with respect to \NN{} and
\piN{} scattering data as well as experimentally determined
bound-state observables in the two- and three-nucleon systems:
\nuc{2,3}{H} and \nuc{3}{He}. At LO and NLO, the \NN{} interaction does
not involve any sub-leading \piN{} amplitudes, nor are there any \NNN{}
force terms. Therefore, at these orders the sim-potentials are
optimized using only \NN{} scattering data and the binding energy,
radius, and quadrupole moment of the deuteron. A summary of the data
types that were included in the objective function for each potential
is given in Table~\ref{tab:optimmethod:included_data}.

\begin{table}[htb]
  \caption{\label{tab:optimmethod:included_data} Objective functions
    for the various nuclear interactions in this work. Included data
    types are marked with 'X'. For sequential optimization, the
    subscript '$i$' indicates at what stage the model is optimized to
    that data. Excluded
    data-types are indicated with '--'.}
    \begin{ruledtabular}
        \begin{tabular}{cccccc}
          \mbox{} & \multicolumn{2}{c}{Scattering data} & $nn$ ERE &
          \multicolumn{2}{c}{bound-state data} \\
          Potential & \NN{} & \piN{} & parameters & \nuc{2}{H} &
          \nuc{3}{H}, \nuc{3}{He} \\ \hline
            \LOsep   & X & -- & -- & -- & -- \\
            \LOsim   & X & -- & -- & X  & -- \\
            \NLOsep  & X$_1$ & -- & X$_2$  & -- & -- \\
            \NLOsim  & X & -- & X  & X  & -- \\
            \NNLOsep & X$_2$ & X$_1$  & -- & -- & X$_3$  \\
            \NNLOsim & X & X  & -- & X  & X 
        \end{tabular}
    \end{ruledtabular}
\end{table}

The bulk of the experimental data consists of \NN{} and \piN{}
scattering cross sections. For the \NN{} data we take the SM99
database~\cite{arndt1999} entries with laboratory scattering energies
$T_{\rm Lab}^{\rm max}\leq\unit[290]{MeV}$, i.e. the pion-production
threshold, which constitutes a natural limit of applicability for
\xEFT.  This results in $N_{\text{data}}^{(pp)} = 2045$ and
$N_{\text{data}}^{({np})} = 2400$ data points, including normalization
data. The number of normalization constants are
$N_{\text{norm}}^{(pp)} = 124$ and $N_{\text{norm}}^{({np})} = 148$.
However, we also explore the consequences of varying $T_{\rm Lab}^{\rm
  max}$ between \unit[125-290]{MeV}. Unless otherwise stated, our
canonical choice is $T_{\rm Lab}^{\rm max} = \unit[290]{MeV}$ and
$\Lambda =\unit[500]{MeV}$. As there is no neutron-neutron scattering
data, we use the neutron-neutron ${}^1S_0$ scattering length $\Oann =
\unit[-18.95(40)]{fm}$~\cite{machleidt2011} and effective range $\Ornn
= \unit[2.75(11)]{fm}$~\cite{miller1990} to constrain the parameter
$\tilde{C}_{{}^1S_0}^{({nn})}$ at order NLO.  For the \piN{} scattering
observables we employ the database from the Washington Institute
group~\cite{workman2012}, here referred to as the WI08 database.
The \piN{} data consists mainly of differential cross sections and
some singly-polarized differential cross sections for the processes
$\pi^\pm + p \rightarrow \pi^\pm + p$ and $\pi^- + p \rightarrow \pi^0
+ n$.  
Unfortunately, the WI08 database contains very little data at low
scattering energies, which would have been preferred to constrain the
low-energy theory of \xEFT{}. In fact, there is no scattering data
below $T_{\text{lab}} = \unit[10.6]{MeV}$. For this reason, we include
all data up to lab energy $T_{\text{lab}} = \unit[70]{MeV}$ and keep
all terms up to, and including, $\nu = 4$ when calculating \piN{}
observables. A lower chiral order does not give a reasonable
description of the data. This results in
$N_{\text{data}}^{(\pi\text{N})} = 1347$ data points including
$N_{\text{norm}}^{(\pi\text{N})} = 110$ normalization data.  At the
optimum, it is usually assumed that the residuals are normally
distributed, and that they are all independent of each other. If so,
then $\chi^2(\LECvec_{\star})$ will comply with a chi-squared
distribution with $N_{\mathbb{M}} - N_{\text{norm}} - N_{\LECidx}
\equiv N_{\text{edf}} - N_{\LECidx} \equiv N_{\text{dof}}$ degrees of
freedom, where $N_{\LECidx}$ denotes the number of LECs (i.e., the
number of model parameters). In turn, this allows for a standard
regression analysis. These rather strong assumptions of both the model
and the data are only approximately fulfilled, mainly due to the
inherent systematic error in \xEFT.

\begin{figure}
    \centering
    \includegraphics[width=\columnwidth]{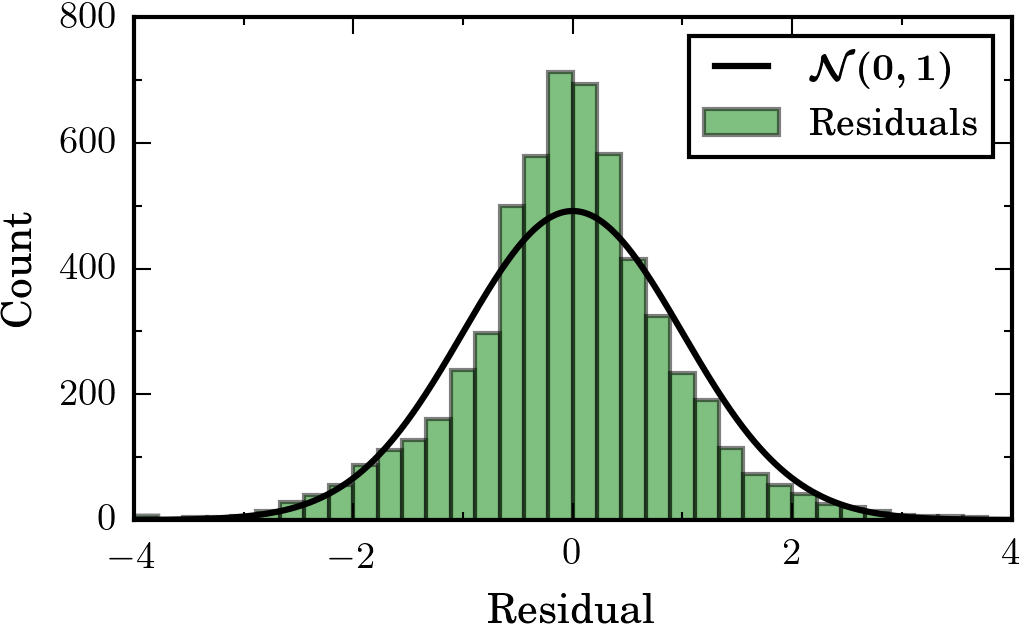}
    \caption{\label{fig:res_dist_NNLOsim}\co Residual distribution for
      the \NNLOsim{} potential, with a sample mean and standard
      deviation of $-0.04(1)$ and $0.977(9)$, respectively. The
      deviations from normality, as discussed in the text, are mainly
      due to the model error of \xEFT{}.}
\end{figure}

The distribution of residuals, $r_i$, for the \NNLOsim{} potential,
which will be thoroughly introduced in Sec.~\ref{sec:optimresults}, is
shown in Fig.~\ref{fig:res_dist_NNLOsim}. It is clear that the
residuals are not entirely normally distributed, with a skewness of
$-0.38(3)$ and excess kurtosis of $5.39(6)$. The main reason for this
deviation can be traced to the inclusion of a systematic error in the
fit. This can produce a consistent over- or underestimate of
observables, resulting in a non-zero skewness. A non-zero excess
kurtosis indicates that the model error sometimes overestimates the
uncertainty and in other cases underestimates it, causing a too sharp
peak near zero in the histogram in Fig.~\ref{fig:res_dist_NNLOsim}. We
stress that the deviations from normality does not invalidate the use
of $\chi^2(\LECvec)$ as an objective function to fit the parameters;
it just indicates that the minimizer $\LECvec_{\star}$ will not be a
maximum-likelihood estimator. In fact, we find that when optimizing
\NNLOsim{} using \NN{} scattering data up to $\unit[125]{MeV}$ only,
to avoid large model errors, the skewness and excess kurtosis of the
\NN{} scattering residuals are significantly reduced; $-0.01(6)$ and
$0.6(1)$, respectively.
Still, the propagated uncertainties are very similar in these two
cases Thus, the minimization and subsequent regression analysis of the
$\chi^2(\LECvec)$ function will provide valuable insights into both
the model and the data~\cite{dobaczewski2014}.

\subsubsection{Total error budget}
For each residual, the total uncertainty $\sigma^2$ is divided into an
experimental part and a theoretical part
\begin{equation}
\begin{aligned}
    \sigma^2 &= \sigma_{\text{exp}}^2 +
    \sigma_{\text{theo}}^2 \\
    & = \sigma_{\text{exp}}^2 + \sigma_{\text{numerical}}^2 +
        \sigma_{\text{method}}^2 + \sigma_{\text{model}}^2
\end{aligned}
\end{equation}
The experimental uncertainty (statistical or systematic) is provided
by the experimenter. Here, we focus on estimating the theoretical
uncertainty. As a first step, we identify three different components:
(1) the \emph{numerical} error originating in finite computational
precision, (2) the \emph{method} error due to mathematical
approximations in the solution of the bound-state or scattering
problem, (3) the \emph{model} error that is inherent to the
truncation of the momentum expansion in {\xEFT}. 

The numerical error is the smallest one and several new technical
developments, such as automatic differentiation for computing
derivatives, allow us to generally ignore
$\sigma_{\text{numerical}}^2$. However, some elements of the
statistical analysis can potentially become numerically unstable if
the relative errors are too small. In particular, this concerns the
computation of the covariance matrix through the inversion of the
Hessian~\eqref{eq:stat_err:C}. For this reason we impose a minimum
relative uncertainty of 0.01\%. In practice this requirement only
affects the error of the deuteron binding energy.

Regarding the method error, the only significant contributions come
from truncating the NCSM model space and from the use of the isoscalar
approximation in calculations of bound-state
observables. Indeed, for all scattering cross sections we include
sufficiently many partial waves to construct an exact scattering
matrix. We estimate the method error of the NCSM calculations using a
simple exponential extrapolation, $E(N_{\max}) = E_\infty +
a\exp(-bN_{\max})$, for a range of different {\xEFT} potentials.
However, the uncertainties from the isoscalar
approximation dominate the truncation error
by an order of magnitude. We therefore use the uncertainties presented
in Ref.~\cite{kamuntavicius1999} as our method error.

In practice, we combine the method errors with the experimental ones
to obtain the resulting weight of each bound-state observable in the
optimization, see Table~\ref{tab:fewnucleon:values}. In certain cases,
the method error is comparative to, or larger than, the experimental
error. 

\begin{table}[htb]
  \caption{\label{tab:fewnucleon:values}Experimentally determined
    values and uncertainties for ground-state energies (in MeV) and
    radii (in fm) for \nuc{2,3}{H} and \nuc{3,4}{He}. The quadrupole
    moment \OQd{} of the deuteron is given in fm${}^2$ and $E_1^A$
    denotes the reduced transition matrix element related to the $\beta$-decay of
    \nuc{3}{H}. The last column is the combined experimental and method
    errors. For the ground-state energies the method error is much
    larger than the experimental
    one. Table~\ref{tab:optimmethod:included_data} indicates which
    observables are included in the optimization. Note that the
    \nuc{4}{He} properties are not included in the objective
    function.}
    \begin{ruledtabular}
        \begin{tabular}{c@{\phantom{\hspace{15pt}}}r@{.}l@{\phantom{\hspace{15pt}}}c@{\phantom{\hspace{15pt}}}r@{.}l}
            & \multicolumn{2}{c}{Exp.\ value} & Ref. & \multicolumn{2}{c}{$\sigma_{\text{exp+method}}$\phantom{\hspace{0pt}}} \\\hline\\[-1em]
            \OEd & $-2$&$22456627(46)$ & \cite{mohr2012} & $0$&$22\times 10^{-3}$ \\
            \OEt & $-8$&$4817987(25)$ & \cite{mohr2012} & $0$&$028$ \\
            \OEh & $-7$&$7179898(24)$ & \cite{mohr2012} & $0$&$019$ \\
            \OEa & $-28$&$2956099(11)$ & \cite{mohr2012} & $0$&$11$ \\
            \Orpd & $1$&$97559(78)^{\text{a}}$ & \cite{huber1998,angeli2013} & $0$&$79\times 10^{-3}$ \\
            \Orpt & $1$&$587(41)$ & \cite{angeli2013} & $0$&$041$ \\
            \Orph & $1$&$7659(54)$ & \cite{angeli2013} & $0$&$013$ \\
            \Orpa & $1$&$4552(62)$ & \cite{angeli2013} & $0$&$0071$ \\
            \OQd & $0$&$27(1)^{\text{b}}$ & & $0$&$01$ \\
            \OHL & $0$&$6848(11)$ & \cite{gazit2009} & $0$&$0011$
        \end{tabular}
    \end{ruledtabular}
    \begin{flushleft}
        ${}^{\text{a}}$ The experimental value is $\Orcd[2] - \Orp[2]$, we still use the value of \Orn[2] from Ref.~\cite{angeli2013}\\
        ${}^{\text{b}}$ This is not an empirical value, see the text for details.
    \end{flushleft}
\end{table}

The model errors can be labeled as systematic and are the most
difficult to assess. We follow the most naive {\xEFT} estimate and
associate a truncation error with the effect of excluded higher-order
Feynman diagrams. The {\xEFT} expansion up to a given chiral order
$\nu$ includes all diagrams that scale as $(Q/\Lambda_\chi)^\nu$ where
$Q \in \{p, m_\pi\}$. The remainder of the diagrams could \emph{a
  priori} be assumed proportional to $(Q/\Lambda_\chi)^{\nu+1}$.

For bound-state properties it is not straightforward to associate a
relevant and system-dependent momentum scale; therefore, we will not
include systematic theoretical errors for these
observables. Scattering observables, on the other hand, have a
well-defined center-of-mass momentum. As described in
section~\ref{sec:NNscatt}, $M$-matrix elements are the fundamental
quantities that are needed to calculate \NN{} scattering observables
and can be parameterized by the complex-valued Saclay amplitudes $a$
to $f$. Similarly, the \piN{} non-spin-flip and spin-flip amplitudes
$g^{\pm}$ and $h^{\pm}$ determine the \piN{} scattering
observables~\cite{krebs2012}. Therefore, from the above scaling
argument we introduce a model error in the scattering amplitudes of
the form
\begin{align}
\label{eq:NNscatt:erroramp}
\sigma_{\text{model,x}}^{(\text{amp})} = C_{\text{x}}\left(\frac{Q}{\Lambda_\chi}\right)^{\nu_{\text{x}}+1}
    \quad,\quad \text{x}\in\{\NN, \piN\},
\end{align}
where $C_{\NN}$ and $C_{\pi N}$ are two overall constants that need to
be determined.

\begin{figure}
    \centering
    \includegraphics[width=1.0\columnwidth]{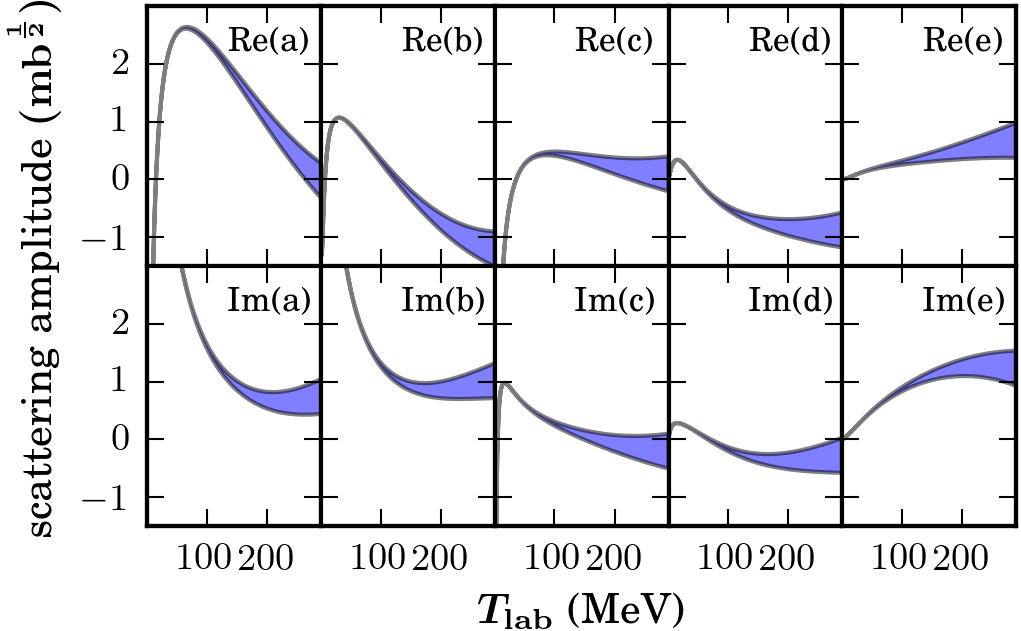}
    \caption{\label{fig:saclay_amplitudes}\co Saclay amplitudes $a$ to
      $e$ at $\theta_{\rm cm} = 45^\circ$ scattering angle for the
      potential \NNLOsim{}. The model error bands were extracted
      according to the discussion in the text.}
\end{figure}

We assume that
both the real and the imaginary parts of the Saclay amplitudes $a-e$
scale in this manner. The nuclear force does not contribute to the $f$
amplitude so we do not impose a model error in that amplitude. Since
the order of magnitude of each scattering amplitude is the same, we
assign the same constant of proportionality to all of them, see
e.g.\ Fig.~\ref{fig:saclay_amplitudes}. The same argument applies to
the \piN{} amplitudes.

We set $Q=p$ to capture the increasing uncertainty in the model as the
energy increases. The definition
$Q=\max\{p,m_\pi\}$~\cite{epelbaum2009} seems to have a comparatively
small impact on the theoretical predictions of the model as discussed
further in Sec.~\ref{sec:optimizationprotocol}.

To determine $C_{\NN}$ and $C_{\piN}$ we use the statistical
guiding principle that $\chi^2/N_{\text{dof}}$ for both \NN{} and
\piN{} scattering should be $1$ if the objective function $\chi^2$
follows a chi-squared distribution and all errors have been correctly
accounted for. This leads to an iterative process where first the
$C_{\rm x}$ constants are updated, then the LECs are optimized using
the previously determined $C_{\rm x}$, and so on until the values of
the constants have stabilized.  This usually requires no more than
three iterations.
%
%------------------------------
\subsection{\label{sec:optimizationalgorithm}%
Optimization algorithms}
%------------------------------
%
The minimization of $\chi^2(\LECvec)$,
Eq.~\eqref{eq:objective_function}, is a non-linear optimization
problem. In this work we have employed three different non-linear
least-squares minimization methods at different stages during the
optimization: POUNDerS~\cite{wild2014}, Levenberg-Marquardt (LM) and
Newton's method. POUNDerS is part of the TAO package~\cite{munson2012}
and is a so-called derivative-free method. As the label indicates, it
does not require the computation of any derivatives. This makes it
very attractive for use with applications where differentiation is a
formidable task; e.g.\ nuclear energy density
optimization~\cite{kortelainen2010} and previous optimizations of
chiral
interactions~\cite{ekstrom2013,ekstrom2015,ekstrom2015b}. However, in
this work, we have managed to make significant progress in the
optimization problem by implementing automatic differentiation, which
enables us to extract machine-precise derivatives of the objective
function. Consequently, the whole class of derivative-based
optimization algorithms becomes readily available.  The convergence
rate is increased considerably with the LM method that employs
first-order derivatives of the residuals with respect to the LECs. A
further improvement can be achieved with Newton's method that uses
also the second-order derivatives.
At LO, the presence of only two LECs to parameterize the potential
makes it a trivial task to minimize the corresponding objective
functions. However, already at the next order, NLO, the optimization
requires quite an effort. There are 11 LECs, and in order to provide a
reasonable start vector $\LECvec_0$ of numerical values for these we
make an initial fit to the \NN{} scattering phase-shifts published by
the Nijmegen group~\cite{stoks1993}. At NNLO there is a total of $26$
LECs, since we also need to include all the $13$ \piN{} LECs up to
order $\nu = 4$. Also at this order we carry out an initial fit to
\NN{} phase-shifts before proceeding with the optimization of the
complete objective function. The optimization with respect to
scattering observables in the \piN{} sector could proceed without any
fits to phase shifts.

There is always a risk of getting trapped in local minima and the
success of the minimization strongly depends on the starting point
$\LECvec_0$. Extensive searches were performed to search for a global
minimum, which is described in more detail in
Sec.~\ref{sec:optimresults}.
\subsubsection{Automatic differentiation}
\begin{figure}
    \centering
    \includegraphics[width=\columnwidth]{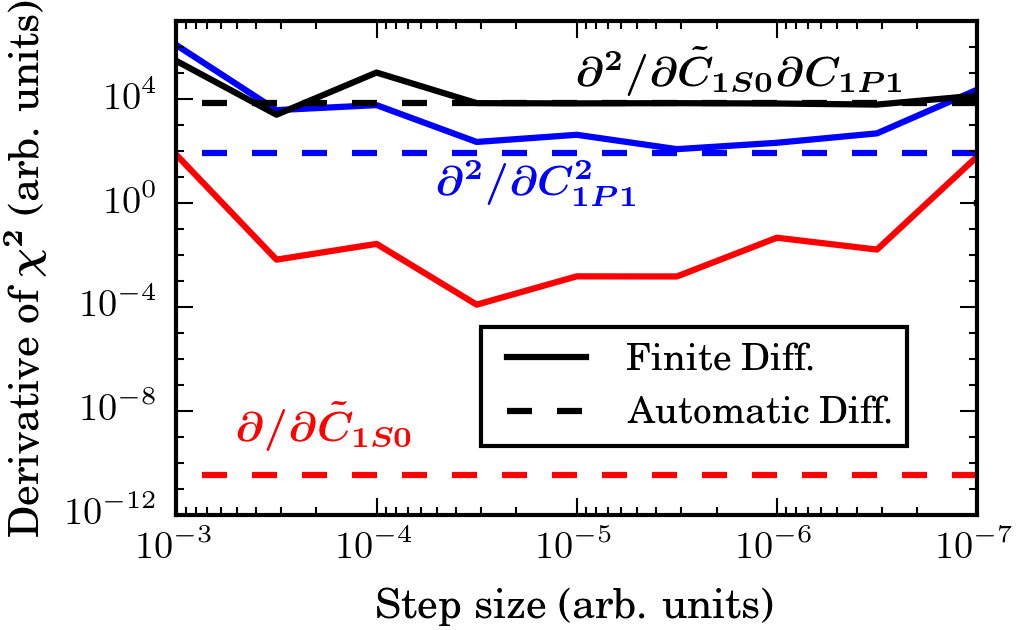}
    \caption{\label{fig:AD_num_deriv}\co Comparison between
      calculated first and second derivatives of an objective function
      using finite differences (third order) with different step sizes
      (filled lines) and automatic differentiation (dashed lines). The
      calculation is done at a minimum where the first derivatives
      should be approximately zero. Due to cancellation effects the
      finite-difference method cannot correctly reproduce the low
      values of the derivatives for any step size.}
\end{figure}

First- and second-order derivatives of $\chi^2(\LECvec)$ with respect
to the LECs are needed during the minimization process and the
subsequent statistical regression analysis, i.e., we need to compute
\begin{align}
  \begin{split}
    \label{eq:derivatives}
          {}&\frac{\partial \mathcal{O}_i^{(\text{theo})}(\LECvec)}{\partial \LEC_m}\quad,\quad\forall \,i, m\\
          {}&\frac{\partial^2 \mathcal{O}_i^{(\text{theo})}(\LECvec)}{\partial \LEC_m\partial \LEC_n}\quad,\quad\forall\, i,m,n.
  \end{split}
\end{align}
The straightforward numerical approach is to approximate the
$n$th-order derivatives with finite differences. The general idea is
to form appropriate linear combinations of $M$ function evaluations in
the vicinity of the point of interest. There are, however, a number of
issues with this method. First, it is prone to large numerical errors
since differences of large, almost equal, numbers are needed. Second,
the result can be very sensitive to the choice of step
size. Furthermore, it is also a computationally demanding method since
the number of required function evaluations grows quickly with the
number of dependent variables and order of the derivative. For
instance, a third-order, finite-difference calculation of first and
second derivatives with respect to all 26 LECs requires $M=3653$
function evaluations. For these reasons, we abandon finite-difference
methods and employ instead forward-mode automatic differentiation
(AD). The basic idea of AD is the following: A computer implementation
for calculating the observables, or any computational algorithm for
that matter, will consist of a chain of simple (or intrinsic)
mathematical operations; e.g.\ addition and multiplication, elementary
functions such as $\sin$ and $\exp$, and matrix operations. Therefore,
by repeatedly employing the \emph{chain rule}, derivatives with
respect to the LECs can be calculated alongside the usual function
evaluations.
Using AD, the derivatives of Eq.~\eqref{eq:derivatives} can
actually be computed to machine precision, which is far beyond the
precision of any reasonable finite-difference scheme. This accomplishment is 
illustrated in Fig.~\ref{fig:AD_num_deriv}, where also the dependence
on the step size for the finite difference method is shown for comparison.

We implement forward-mode AD using the Rapsodia computational
library~\cite{charpentier2014}. For the calculation of first and
second derivatives with respect to $N_{\LECidx}$ different LECs, Rapsodia
requires a total of
\begin{align}
    M = 2\binom{N_{\LECidx}+1}{2}
\end{align}
derivative calculations. For $N_{\LECidx}=26$, this results in $M = 702$, thus
considerably more efficient than the finite-difference
approach. Furthermore, all calculations that do not depend on the LECs
are performed only once, compared to the brute-force implementation of
the finite-difference scheme that requires a full calculation for
every function evaluation. Furthermore, since all LECs enter linearly
in the momentum-space formulation of the chiral potential it is very
easy to calculate the derivatives of the potential with respect to the
LECs. Thus, the only workhorses in our calculations are the $R$-matrix
evaluation (matrix inversion) of the scattering process and the
solution to the NCSM eigenvalue problem (matrix diagonalization) as we
will discuss next.

To solve for the two-nucleon $R$-matrix~\eqref{eq:LS_Rmatrix} at
a given on-shell scattering energy we use the well-known method of
Ref.~\cite{haftel1970}. It recasts the Lippmann-Schwinger equation
into a matrix equation
\begin{align}
    \label{eq:NNscatt:R_equation}(I + VZ)R &= V,
\end{align}
where $I$ is the identity matrix, $V$ is the two-nucleon potential,
and $Z$ is a simple diagonal matrix defined in
Ref.~\cite{haftel1970}. The $R$-matrix is easily obtained after
inverting
$(I+VZ)$ using e.g.\ LU factorization. First- and second-order
derivatives of the $R$-matrix with respect to LECs $\LEC_x$ and $\LEC_y$ are
easily obtained using the AD technology and the same LU factorization,
\begin{align}
    (I+VZ)\frac{\partial R}{\partial \LEC_x} &= \frac{\partial
      V}{\partial \LEC_x}(I - ZR)\\
    \begin{split}
        (I+VZ)\frac{\partial^2 R}{\partial \LEC_x\partial \LEC_y} &=
        \frac{\partial^2 V}{\partial \LEC_x\partial \LEC_y}(I - ZR) \\
        &\quad - \frac{\partial V}{\partial \LEC_x}Z\frac{\partial
          R}{\partial \LEC_y} 
        - \frac{\partial V}{\partial \LEC_y}Z\frac{\partial R}{\partial \LEC_x}.
    \end{split}
\end{align}

We also use the fact that many derivatives are exactly zero, for
example the \piN{} LECs $d_i$ and $e_i$ do not appear in the formalism
for \NN{} scattering at the present chiral orders. The computational
overhead of AD in terms of wall time is very small. On a single
computational node, the calculation of all first- and second-order
derivatives of the 4450 \NN{} scattering observables with respect to
the $26$ LECs at NNLO only takes twice as long as computing just the
central values.

It is straightforward, but slightly more costly, to apply the AD
technology to the NCSM diagonalization of the nuclear Hamiltonian
$\mathcal{H}$ for $A\leq 4$. If the eigenvalue spectrum is
non-degenerate, the first-order derivatives of the ground-state
energy $E_0$ and wave function $\ket{\psi_0}$ with respect to the LEC
$\LEC_x$ are given by~\cite{carlsson2015}
\begin{align}
    \frac{\partial E_0}{\partial \LEC_x} &=
    \braopket{\psi_0}{\frac{\partial\mathcal{H}}{\partial
        \LEC_x}}{\psi_0}, \\ 
    \frac{\partial}{\partial \LEC_x}\ket{\psi_0} &=
    \sum_{i\neq 0}
    \frac{\bra{\psi_i}\frac{\partial\mathcal{H}}{\partial
        \LEC_x}\ket{\psi_0}}{E_0-E_i}\ket{\psi_i}.
\end{align}
Higher-order derivatives are simply obtained by repeated differentiation.

For bound-state observables, the computational overhead in terms of
wall time is slightly larger than for two-body scattering since we
must compute all eigenvalues and eigenvectors of $\mathcal{H}$\@. The
calculation of all first and second derivatives for all $26$ LECs at
NNLO for the $A = 3$ observables is approximately $20$ times slower
than just calculating the central values.
%------------------------------
\subsection{Uncertainty Quantification}\label{sec:stat_err}
%------------------------------
%
We employ well-known methods from statistical regression analysis to
study the sensitivities and quantify the uncertainties at the optimum
$\chi^2(\LECvec_{\star})$, see e.g.~\citet{dobaczewski2014}.
The $N_{\LECidx} \times N_{\LECidx}$ covariance
matrix $\Cov(\LECvec_{\star})$ defines the permissible
variations $\Delta \LECvec$ in the LECs that maintain an objective
function value such that
\begin{align}
    \label{eq:optimmethod:chi2_surface} 
    \chi^2(\LECvec_{\star}+\Delta \LECvec) - \chi^2(\LECvec_{\star}) \leq T,
\end{align}
where $T$ is some chosen tolerance. We can assume rather small
variations $\Delta \LECvec$, and therefore truncate a Taylor expansion of
the objective function at the second order
\begin{equation}\label{eq:stat_err:chi2_quad_approx}
    \begin{aligned}
        \chi^2(\LECvec_{\star} + \Delta\LECvec) -
        \chi^2(\LECvec_{\star}) &\approx \frac{1}{2}(\Delta \LECvec
        )^T \mathbf H (\Delta\LECvec),\\ \mathrm{where~} H_{ij} &=
        \left. \frac{\partial^2\chi^2(\LECvec)}{\partial \LEC_i\partial
          \LEC_j}\right|_{\LECvec = \LECvec_{\star}},
    \end{aligned}
\end{equation}
are matrix elements of the Hessian $\mathbf H$. This should be positive
definite. It can be decomposed into $\mathbf H = \mathbf U \mathbf D
\mathbf U^T$, where the columns of $\mathbf U$ are the eigenvectors of
$\mathbf H$ and $\mathbf D$ is a diagonal matrix with the eigenvalues
of $\mathbf H$. Defining $\mathbf x \equiv \mathbf U^T(\Delta\LECvec
)$, Eq.~\eqref{eq:optimmethod:chi2_surface} becomes
\begin{align}
    \frac{1}{2}\mathbf x^T \mathbf D\mathbf x =
    \frac{1}{2}\sum_{i=1}^{N_{\LECidx}} x_i^2D_{ii} \leq T.
\end{align}
The $N_{\LECidx}$ parameters $\mathbf{x}$ can be viewed as ``rotated''
LECs. They are very convenient since they are independent of each
other, which simplifies the previous equation and gives
\begin{align}
    \frac{1}{2}x_i^2D_{ii} \leq T_1\quad\quad\forall i,
\end{align}
where $T_1$ is the limit to use when considering only variations in
one parameter and keeping the others fixed.  If $\chi^2(\LECvec)$
follows a chi-squared distribution, then $x_i^2D_{\chi^2,ii}/2$ will
also follow a chi-squared distribution with one degree of freedom,
meaning that the $1\sigma$ confidence level is given by $T_1 = 1$, and
$x_i\sim \mathcal{N}(0, 2/D_{\chi^2,ii})$. In practice,
$\chi^2(\LECvec_{\star})$ will only be an approximate chi-squared
distribution, which modifies $T_1$ slightly. Here we set $T_1 =
\chi^2(\LECvec_{\star})/N_{\text{dof}}$ which corresponds to a
rescaling of the $\chi^2(\LECvec_{\star})$-function~\cite{dobaczewski2014},
\begin{align}\label{eq:stat_err:chi2_scaled}
    \chi_{\text{scaled}}^2(\LECvec) \equiv \chi^2(\LECvec
    )\frac{N_{\text{dof}}}{\chi^2(\LECvec_{\star})}.
\end{align}
The covariance matrix is then given by
\begin{align}
    \label{eq:stat_err:C}\Cov(\LECvec_{\star}) =
    2\frac{\chi^2(\LECvec_{\star})}{N_{\text{dof}}}\mathbf{H}^{-1}
    \equiv \mathbf{U}\bm{\Sigma}\mathbf{U}^T,
\end{align}
where $\bm{\Sigma}$ is the diagonal matrix with the vector of
variances, $\bm{\sigma}^2$, of the rotated LECs, on the diagonal.
Since $T_1$ only affects $\Cov$ with a constant factor,
correlations remain invariant under changes in $T_1$.

\subsubsection{Error propagation}

Starting from the covariance matrix $\Cov(\LECvec_{\star})$ we can
propagate the statistical uncertainties in the LECs to any observable
$\mathcal{O}_A$, and compute the linear correlation coefficient between any
two observables $\mathcal{O}_A$ and $\mathcal{O}_B$\@. To this aim, it
is most convenient to use the rotated and independent LEC
representation $\mathbf x$ defined above. Each LEC $x_i$ is normally
distributed with zero mean. Next we use a quadratic approximation of
the observable $\mathcal{O}_A$,
\begin{align}\begin{split}\label{eq:stat_err:obs_quad_approx}
         \mathcal{O}_{A}(&\LECvec_{\star} + \Delta\LECvec) -
         \mathcal{O}_A(\LECvec_{\star}) \\
         &\approx (\Delta\LECvec
         )^T\mathbf{J}_A + \frac{1}{2}(\Delta\LECvec)^T
         \mathbf{H}_A(\Delta\LECvec) \\ &= \mathbf{x}^T\mathbf U^T
         \mathbf J_A + \frac{1}{2}\mathbf{x}^T \mathbf U^T\mathbf H_A
         \mathbf U \mathbf{x} \\&\equiv \mathbf{x}^T\tilde{\mathbf{J}}_A +
         \frac{1}{2}\mathbf{x}^T\tilde{\mathbf H}_A \mathbf{x},
\end{split}\end{align}
where $\mathbf{J}_A$ is the Jacobian vector of partial derivatives, $
J_{A,i} = \frac{\partial \mathcal{O}_{A}}{\partial \LEC_i}$, $\mathbf{H}_A$ is the
corresponding Hessian matrix, and the tilde
notation in the last line indicates the similarly rotated Jacobian and
Hessian. The corresponding statistical expectation value
$\mathbb{E}(\cdot)$ is given by
\begin{align}
    \begin{split}\label{eq:stat-expect}
        \mathbb{E}[\mathcal{O}_A(\LECvec)] &\approx \mathcal{O}_A(\LECvec_{\star}) +
        \frac{1}{2}\sum_{ij}^{N_{\LECidx}}\tilde{H}_{A,ij}\mathbb{E}[x_ix_j] \\ 
        &= \mathcal{O}_{A}(\LECvec_{\star}) +
		\frac{1}{2}(\bm{\sigma}^2)^T\diag(\tilde{\bm{H}}_A)
    \end{split}
\end{align}
Finally, we define the covariance of $\mathcal{O}_A$ and $\mathcal{O}_B$ by
\begin{align}\begin{split}
    \Cov(&A,B) \equiv \mathbb{E} \big[ \left( \mathcal{O}_A(\LECvec
        ) - \mathbb{E}[\mathcal{O}_A(\LECvec)] \right)\\
        &\quad\quad\quad\quad\times\left(
        \mathcal{O}_B(\LECvec) - \mathbb{E}[\mathcal{O}_B(\LECvec
        )] \right) \big] \\
    &\approx \sum_{ijkl}^{N_{\LECidx}}\mathbb{E} \big[ (\tilde{J}_{A,i}x_i +
    \frac{1}{2}\tilde{H}_{A,ij}x_ix_j 
    - \frac{1}{2}\tilde{H}_{A,ii}\sigma_i^2) \\
    &\quad\times (\tilde{J}_{B,k}x_k 
    + \frac{1}{2}\tilde{H}_{B,kl}x_kx_l -
    \frac{1}{2}\tilde{H}_{B,kk}\sigma_k^2) \big] \\
    &= \tilde{\mathbf{J}}_A^T\bm{\Sigma}\tilde{\mathbf{J}}_B +
    \frac{1}{2}(\bm{\sigma}^2)^T(\tilde{\mathbf{H}}_A\circ\tilde{\mathbf{H}}_B)\bm{\sigma}^2, 
\end{split}\end{align}
where $\circ$ denotes the Hadamard product.
The statistical uncertainty of an observable $\mathcal{O}_A$ is then given by
$\sigma_A\equiv\sqrt{\Cov(A,A)}$.
This approximation of the covariance is valid as long as the quadratic
approximations~\eqref{eq:stat_err:chi2_quad_approx}
and~\eqref{eq:stat_err:obs_quad_approx} are valid and the normalized
objective function can be assumed to follow a chi-squared
distribution.

Using a linear approximation, the probability distribution for an
observable $\mathcal{O}_A$ will follow the well-known Gaussian
form. However, for the quadratic approximation there is no such
analytic expression. Instead, it is easy to reconstruct the
probability distribution numerically by using
Eq.~\eqref{eq:stat_err:obs_quad_approx} with a large sample of
parameter sets.
%=======================================
\section{Results
\label{sec:results}}
%-------------------
In this section we discuss our results from the optimization of
\xEFT{} at LO, NLO, and NNLO (Sec.~\ref{sec:optimresults}), the
subsequent error propagation (Sec.~\ref{sec:errorpropagation}), as
well as an expanded discussion on the implications and advantages of a
simultaneous optimization protocol
(Sec.~\ref{sec:optimizationprotocol}.) In particular we discuss the
important consequences of correlations between the LECs in the case of
simultaneous versus separate optimization strategies.
%------------------------------
\subsection{\label{sec:optimresults}%
Optimization}
%------------------------------
%
With all the necessary tools in place we can perform the fits to
experimental data. For all cases we implicitly assume that the LECs
are of natural size~\cite{weinberg1979} by choosing starting points
in this region of the parameter space. We did not in any other way
force the LECs to be natural.
A possible problem in multi-parameter optimization is the existence of
several local minima.  At LO, with just two parameters, there is only
one minimum.
\begin{table}[htb]
  \caption{\label{tab:optimresults:good_bad_minima}Comparison of
    different minima at various chiral orders. \NN{}-LECs are
    optimized using only
    \NN{} scattering data (at NNLO, the \piN{} LECs are fixed). The minima are equally
    good for $A=2$ observables, but differ significantly in $A=3$
    bound-state properties, 
    calculated here without a three-body force. The last row
    corresponds to parameters and results (with \NN{} forces only) of
    the simultaneously optimized \NNLOsim{} interaction.
    The $\tilde{C}$ LECs are in units of $\unit[10^4]{GeV^{-2}}$. The
    scattering $\chi^2/N_{\text{dof}}$ shown are for data up to
    $\unit[125]{MeV}$ without 
    model errors included. $\OEt[(\text{exp})]\approx
    \unit[-8.48]{MeV}$. Energies are in MeV.} 
    \begin{ruledtabular}
        \begin{tabular}{c@{\phantom{\hspace{5pt}}}
  r@{.}l@{\phantom{\hspace{18pt}}}
  r@{.}l@{\phantom{\hspace{18pt}}}
  r@{}l@{\phantom{\hspace{18pt}}}
  r@{.}l@{\phantom{\hspace{18pt}}}
  r@{.}l}
            & \multicolumn{2}{c}{$\tilde{C}_{{}^1S_0}^{({np})}$} &
            \multicolumn{2}{c}{$\tilde{C}_{{}^3S_1}$} &
            \multicolumn{2}{c}{$\chi^2/N_{\text{dof}}$} &
            \multicolumn{2}{c}{\OEd} &
            \multicolumn{2}{c}{\OEt}\\\hline 
            \LOsep & $-0$&$11$ & $-0$&$072$ & $350$& & $-2$&$21$ & $-11$&$4$ \\\hline
            NLO-1 & $+0$&$81$ & $+0$&$69$ & $14$& & $-2$&$17$ & $-3$&$03$ \\
            NLO-2 & $+0$&$81$ & $-0$&$17$ & $14$& & $-2$&$16$ & $-3$&$30$ \\
            NLO-3 & $-0$&$15$ & $+0$&$68$ & $14$& & $-2$&$17$ & $-2$&$92$ \\
            NLO-4 & $-0$&$15$ & $-0$&$17$ & $14$& & $-2$&$16$ & $-8$&$22$ \\\hline
            NNLO-1 & $+0$&$49$ & $+0$&$53$ & $2$&$.4$ & $-2$&$19$ & $-3$&$64$ \\
            NNLO-2 & $+0$&$49$ & $-0$&$17$ & $2$&$.4$ & $-2$&$21$ & $-3$&$71$ \\
            NNLO-3 & $-0$&$15$ & $+0$&$53$ & $2$&$.4$ & $-2$&$19$ & $-3$&$23$ \\
            NNLO-4 & $-0$&$15$ & $-0$&$17$ & $2$&$.4$ & $-2$&$22$ & $-8$&$21$ \\\hline
            \NNLOsim & $-0$&$15$ & $-0$&$17$ & $1$&$.7$ & $-2$&$22$ & $-8$&$54$ \\
        \end{tabular}
    \end{ruledtabular}
\end{table}
However, at NLO we find four local minima. They correspond to
combinations of two optima in the ${}^1S_0$ channel and two optima in
the coupled ${}^3S_1-{}^3D_1$ channel. As shown in
Table~\ref{tab:optimresults:good_bad_minima}, all four combinations
describe scattering data and the deuteron properties equally well,
thus making them indistinguishable from this point of view.
Furthermore, a similar set of minima exists at NNLO when
fitting the \piN{} and \NN{} data separately.

A theoretical argument can provide partial guidance in the choice
between these parameter sets. The nuclear interaction will have an
approximate Wigner SU(4) symmetry~\cite{mehen1999} due to the large
scattering lengths in the S-waves, which implies $\tilde{C}_{{}^1S_0}
\approx \tilde{C}_{{}^3S_1}$.  This approximate constraint rules out
the second and third of the four candidate NLO and NNLO minima in
Table~\ref{tab:optimresults:good_bad_minima}. Furthermore, we might
argue that the fourth minimum (NLO-4 and NNLO-4, respectively) is the
physical one since its $\tilde{C}$ LECs most resemble the values
obtained at LO. This is not a strong justification since LECs are
allowed to vary between orders. In the end, it does turn out that both
NLO-4 and NNLO-4 are indeed close to the single minimum that exists in
the simultaneous NNLO optimization.

A much more interesting difference between the four minima occurs in
the few-nucleon sector. It turns out that minima 1--3 give significant
underbinding of the triton. 
Since the measured ground-state energy is \unit[-8.48]{MeV}, these results
imply that three-nucleon forces, which appear at NNLO, would have to
contribute \unit[5\text{--}6]{MeV} of the missing binding energy.
This difference is smaller for the NLO-4 and NNLO-4 minima and they
most likely represent the physical minima. This is also more in line
with the power-counting arguments that the three-nucleon force should
be weaker than the two-nucleon force, see e.g. Ref~\cite{friar1997b}.
Furthermore, with the subsequent addition of the \NNN{} terms
at NNLO (as it is done in the sequential optimization strategy) it
turns out that only the NNLO-4 minimum allows to reproduce all $A=3$
observables within one standard deviation.
For these reasons, NLO-4 and NNLO-4 define the \NN{}-only parts of the
\NLOsep{} and \NNLOsep{} potentials, respectively.

The values for the LECs of our optimized potentials at LO, NLO, and
NNLO are tabulated in the Supplemental Material~\cite{supplemental}
together with their estimated statistical uncertainties.
The statistical uncertainty of the $i$th LEC, i.e.\
$\sqrt{\Cov(\LECvec_*)_{ii}}$, is a measure of how much this
particular parameter can change while maintaining a good description
of the fitted data, as detailed in section~\ref{sec:stat_err}. That
is, the uncertainty for a given LEC represents its \emph{maximal
  variation} while assuming that all other LECs are
fixed at the $\chi^2$ minimum. Note, however, that the LECs really cannot be
varied independently of each other due to mutual correlations. A full
error analysis requires a complete covariance matrix as we demonstrate below.

The appearance of \NNN{} diagrams and sub-leading terms from the \piN{}
sector does not occur until NNLO in our chiral expansion. This implies
small differences between the separately and simultaneously optimized
interactions at lower orders. The deuteron properties are included in
the optimization of \LOsim{} and \NLOsim{}, but not in \LOsep{} and
\NLOsep{}.
We find that the statistical $\chi^2$ values (not including the model
errors) with respect to \NN{} scattering data are almost identical for
\LOsim{} and \LOsep{}, and so are the values of the LECs. The small
value of $\sigma_{\rm exp + \rm method}$ for the deuteron binding
energy constrains the statistical error for $\tilde{C}_{{}^3S_1}$ in
\LOsim{} correspondingly.
For the contact potential at NLO there are three LECs that operate in
the deuteron channel, more than in any other \NN{} partial wave. The
presence of mutual correlations cannot be neglected. This explains why
the individual statistical errors for the LECs in \NLOsim{} and
\NLOsep{} in this channel are similar and larger than at LO.  The
covariances will also impact the value for the forward error in the
deuteron binding energy, discussed further in Sec.~\ref{sec:errorpropagation}.

We find that the description of the $pp$ scattering data is not
influenced much by the inclusion of the deuteron in the optimization,
while the agreement with $np$ data is notably worse above
$\unit[35]{MeV}$. At this order it is mainly the $C_{{}^3S_1}$ and
$C_{{}^1P_1}$ LECs that have changed, see Ref.~\cite{supplemental},
which only affect $np$ scattering.

\begin{figure}
    \centering
    \includegraphics[width=\columnwidth]{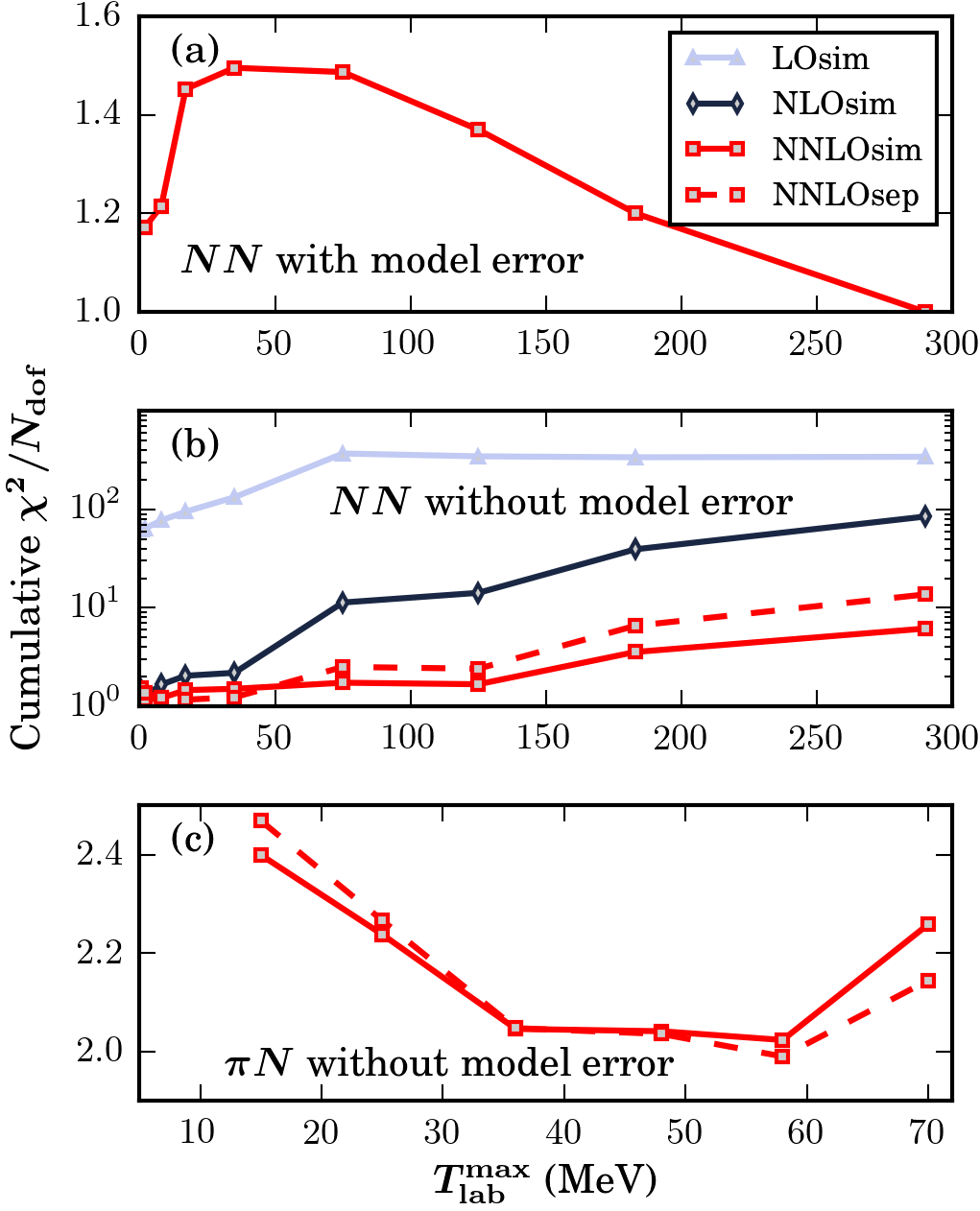}
    \caption{\label{fig:optimresults:scatt_chi2_tot}\co (a) Cumulative
      $\chi^2/N_{\text{dof}}$ for \NN{} scattering data including the
      model error (see Sec.~\ref{sec:objectivefunction}).  Note that
      the amplitude of the model error is chosen so that
      $\chi^2/N_{\text{dof}}=1$ when all data up to
      $T_\mathrm{lab}=290$~MeV is included. (b,c) Cumulative
      $\chi^2/N_{\text{dof}}$ without the model error for \NN{} and
      \piN{} scattering data, respectively.}
\end{figure}

As previously mentioned, the \xEFT{} interaction becomes significantly
more involved at NNLO as \NNN{} and sub-leading \piN{} terms enter at
that order. The simultaneous optimization of all data listed in
Table~\ref{tab:optimmethod:included_data} leads to the construction of
the \NNLOsim{} interaction.
The consequences of the simultaneous approach are dramatic.
First of all, we find a single optimum as this strategy eliminates all
but one of the local minima that were obtained in the sequential
optimization. Moreover, a possible concern turns out to be
unwarranted: an improved overall description of scattering data does
not detoriate the description of different subsets. In fact, the
result is quite the opposite. 
With the simultaneous-optimization strategy we find that the
description of the $pp$ scattering data is actually significantly
improved. For scattering energies $T_{\text{lab}}\leq\unit[290]{MeV}$
the statistical $\chi^2_{(pp)}/N_{\text{dof}} = 9.1$ for \NNLOsim{} compared to
$\chi^2_{(pp)}/N_{\text{dof}} = 26$ for \NNLOsep, not including the
model error.
At the same time, the $\chi^2$ for $np$ scattering and \piN{}
scattering are similar for the two potentials. Measured $np$
scattering cross sections are characterized by larger uncertainties
and it is therefore not surprising that this data remains well
described.
However, it is noteworthy that the \NNLOsim{} potential reaches a
better description of the \NN{} data while maintaining a description
of the \piN{} data that is comparable to the one of \NNLOsep{}. Keep
in mind that \NNLOsep{} is separately optimized to the \piN{}
scattering data. In the simultaneous optimization protocol we are
effectively introducing additional constraints on the $c_i$ LECs via
the \NN{} data set. One could be concerned that the short-range \NN{}
physics would impact and worsen the description of the long-range pion
physics. It is not unlikely that we would have seen such unphysical
effects if the \piN{} database would have been more comprehensive.
The existing \piN{} data does not constrain all directions in the
\piN{} LEC parameter space, which allows for large variations in the
parameter values and a better description of the $pp$ scattering data
with \NNLOsim{}.
The $\chi^2/N_{\text{dof}}$ for \NN{} and \piN{} scattering up to
different $T_{\text{lab}}^{\max}$ are presented in
Fig.~\ref{fig:optimresults:scatt_chi2_tot}.

\begin{figure*}[tbh]
  \includegraphics[width=\textwidth]{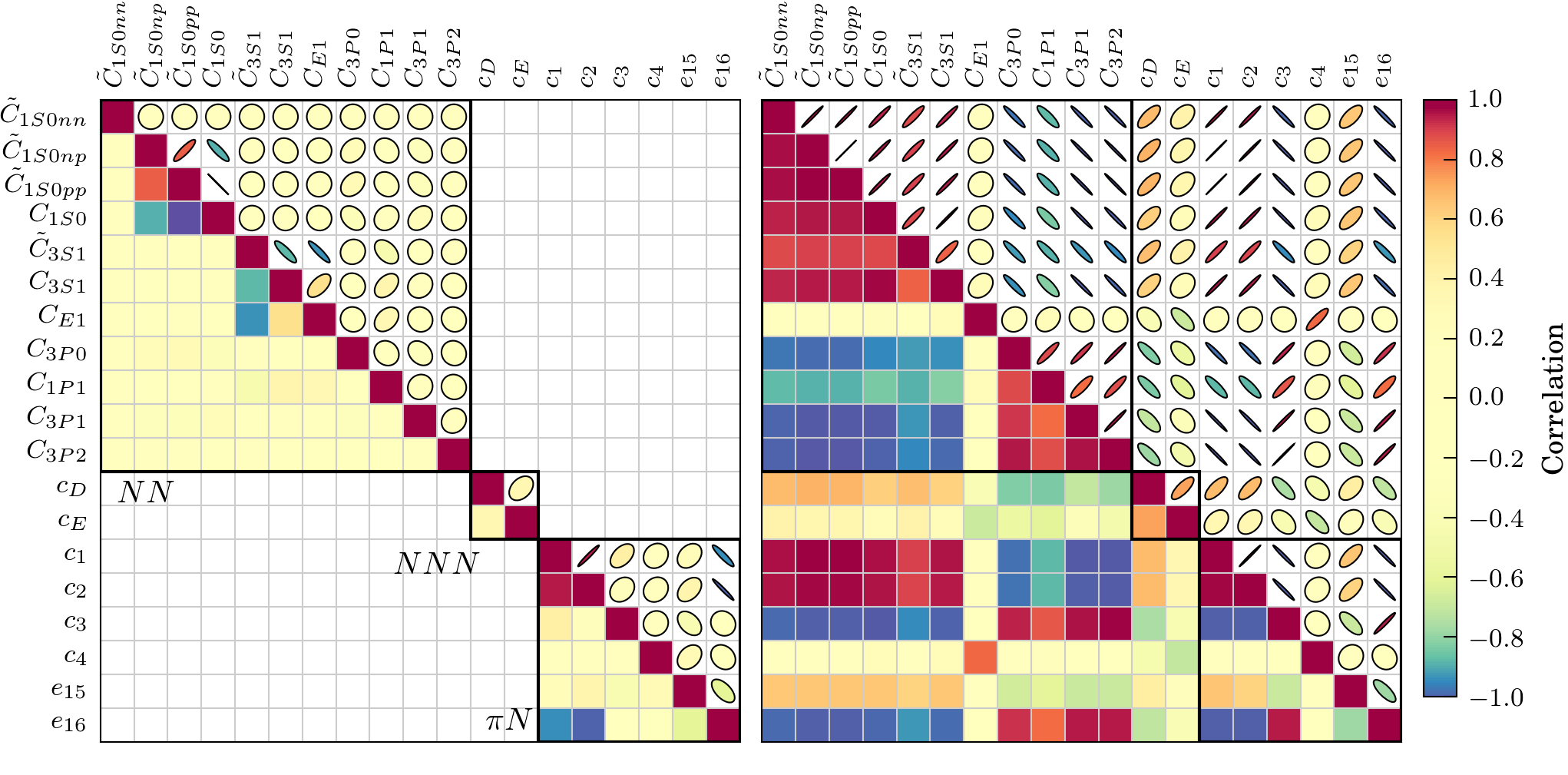} 
  \caption{\label{fig:corr_par}\co Graphical representation of the
    linear correlation matrix for \NNLOsep{} (left) and \NNLOsim{}
    (right) including selected LECs. The separately optimized \NNLOsep{}
    potential does not probe the statistical correlation between LECs
    entering different optimization stages. It is striking that there
    are almost no correlations for the \NNLOsep{} potential, while for
    the \NNLOsim{}-potential the situation is quite the opposite.}
\end{figure*}

The predominant advantage of the simultaneous optimization is the
correct treatment of correlations.
Although the uncertainties of the LECs presented in the Supplemental
Material~\cite{supplemental} are similar for \NNLOsep{} and
\NNLOsim{}, the propagated statistical errors of observables can be
several orders of magnitude larger for \NNLOsep{} due to missing
correlations, see Sec.~\ref{sec:errorpropagation}. To visualize the
correlations between all LECs, we plot the linear correlation matrix
in Fig.~\ref{fig:corr_par}.
The linear correlation between two LECs, or any observables $A$ and
$B$, indicates their linear relationship and is defined as the
normalized covariance, $\Cov(A, B) / (\sigma_A\sigma_B)$. This
quantity assumes values between $-1$ (fully anti-correlated) and $+1$
(fully correlated). A positive (negative) value for the correlation
indicates that a larger value for $A$ most likely requires a larger
(smaller) value for $B$.
The correlation coefficients between LECs that belong to different
objective functions are zero. For \NNLOsep{} this implies that the
correlation matrix is block-diagonal in terms of the \piN{}, \NN{},
and \NNN{} sectors. For \NNLOsim{}, however, such inter-block
correlations are revealed. In addition, we observe an increase of the
correlations within each group. This can be traced to the fact that
the \piN{} LECs, $c_1$, $c_3$ and $c_4$, occur in the description of
\NN{}-, \piN{}-, and \NNN{}-data.
The failure to capture these correlations within the sequential
optimization approach, such as with the \NNLOsep{} potential, will
induce very large propagated statistical errors. In conclusion,
simultaneous optimization is key for a realistic forward propagation
of parametric uncertainties.
%
%------------------------------
\subsection{\label{sec:errorpropagation}%
Error propagation}
%------------------------------
%
Statistical errors and covariances between computed observables are
calculated under the assumption that each observable depends
quadratically on the LECs in the vicinity of the minimum, see
Eq.~\eqref{eq:stat_err:obs_quad_approx}. Our estimate of the
statistical uncertainty, $\sigma_A$, of an observable,
$\mathcal{O}_A$, rests on this assumption, which also explains
why we have asymmetric error bars.
We have performed extensive Monte Carlo samplings to verify the
validity and necessity of using the second-order approximation. A
linear truncation is more common. In particular, we compare the
probability density function for various observables obtained from:
(i) Monte Carlo samplings of the multivariate Gaussian spanned by the
covariance matrix, (ii) the quadratic approximation, and (iii) the
linear approximation of Eq.~\eqref{eq:stat_err:obs_quad_approx}. The
Monte Carlo calculations use $10^5$ sets of normally distributed LEC
vectors. 

\begin{figure}[h]
  \centering
  \includegraphics[width=\columnwidth]{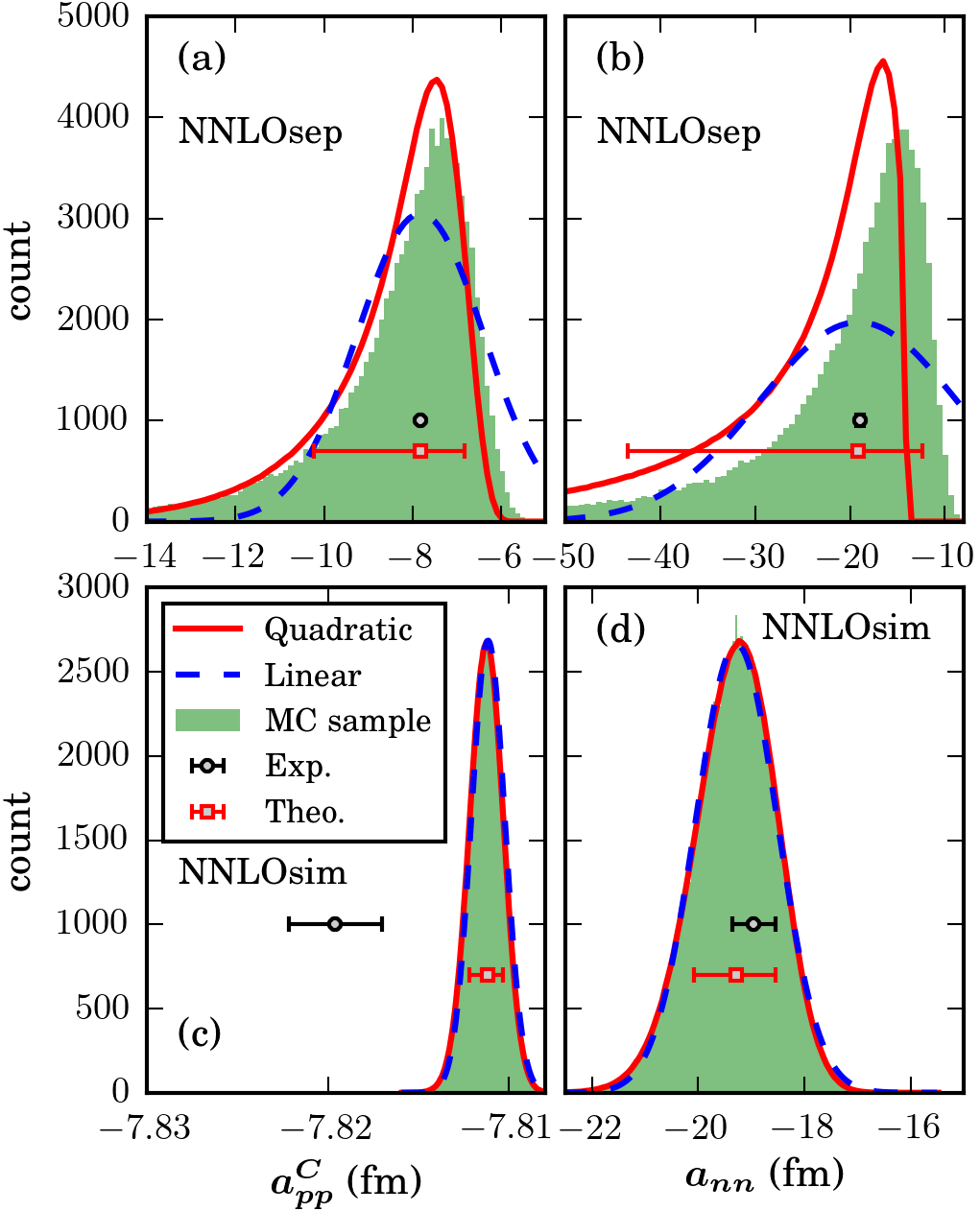}
  \caption{\label{fig:scatt_length_hist}\co Histograms (filled green
    area) for the sampled probability distribution of the $nn$ (b,d)
    and $pp$ (a,c) scattering lengths (including Coulomb) using the
    NNLO potentials: \NNLOsep{} (a,b) and \NNLOsim{} (c,d). The dashed
    (solid) lines show error estimates from the sample assuming that
    the scattering length depends linearly (quadratically) on the fitting
    parameters. The final theory result (red square) from
    Eq.~\eqref{eq:stat_err:obs_quad_approx} agrees well with the
    sampled distribution.}
\end{figure}

The probability distributions for the scattering lengths \OappC{} and
\Oann{} for the potentials \NNLOsep{} and \NNLOsim{} are shown in
Fig.~\ref{fig:scatt_length_hist}. Note that these results are
predictions since the scattering lengths are not included in the
objective function at NNLO.
The statistical errors for \OappC{} and \Oann{} obtained in the Monte
Carlo calculations with the \NNLOsim{} potential are small and well
reproduced already by the corresponding linear approximation, as
expected. With \NNLOsep{}, the errors are much larger and require at
least a quadratic approximation for the forward error. The
uncertainties of the ERE parameters differ quite a lot between these
two potentials. It is important to remember that for the \NNLOsim{}
potential, all LECs are constrained by \piN{}, \NN{} as well as \NNN{}
data. Hence, in the error analysis, the LECs that fulfill
$\chi^2_{\text{scaled}}(\vec p)\approx N_{\text{dof}}$ will provide a
reasonable description of most scattering data. The \piN{} LECs for
\NNLOsep{} on the other hand, are constrained only by the \piN{}-data
and the missing statistical correlations allow for wide permissible
ranges for the \NN{} scattering lengths.

\begin{figure}
    \centering
    \includegraphics[width=\columnwidth]{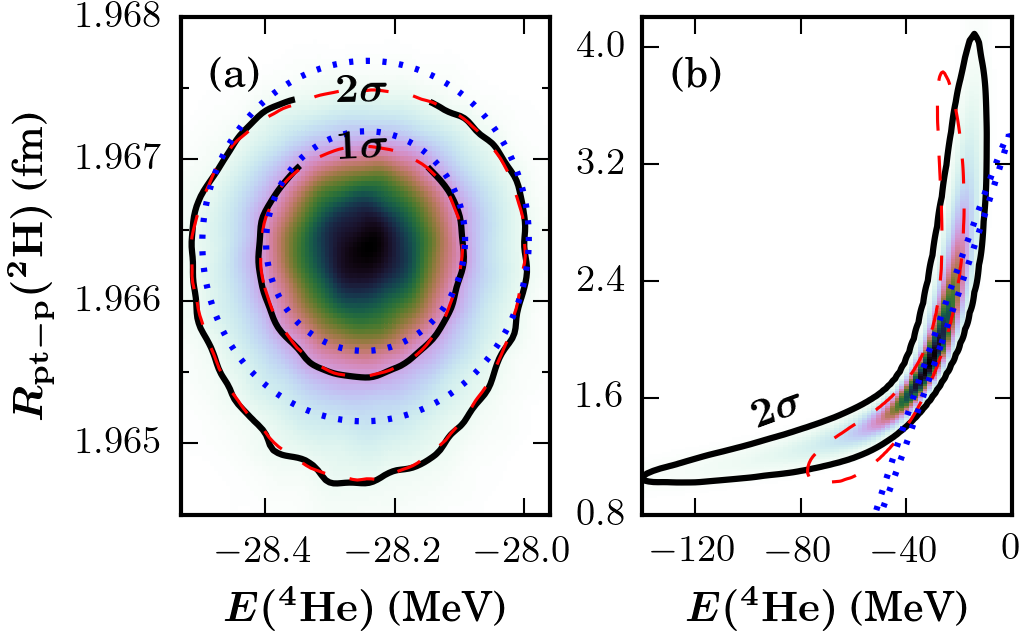}
    \caption{\label{fig:err_prop_hist}\co Joint statistical
      probability distribution for $\OEa$ and $\Orpd$ for (a)
      \NNLOsim{} and (b) \NNLOsep{} obtained in a Monte Carlo sampling
      ($N_\mathrm{sample} = 10^5$) as described in the text. Contour
      lines for this distribution are shown as black, solid lines,
      while blue dotted (red dashed) contours are obtained assuming a
      linear (quadratic) dependence on the LECs for the observables.}
\end{figure}

It is possible to explore correlations between any pair of observables
by looking at joint probability distributions. As an example, we plot
the statistical distribution of binding energies of \nuc{4}{He} and
corresponding radii of the deuteron for the NNLO potentials in
Fig.~\ref{fig:err_prop_hist}.
The contour lines indicate the regions that encompass 68\% ($1\sigma$)
and 95\% ($2\sigma$) of the probability density. It is remarkable that
the quadratic approximation (dashed lines) reproduces even the fine
details of the full calculation (solid lines) for the \NNLOsim{}
interaction. Again, the magnitude of variations is strikingly large
for \NNLOsep{}, but the quadratic approximations does rather well in
reproducing them. In particular, we see a large improvement when going
from a linear (dotted lines) to a quadratic dependence on the
LECs. This even captures the departure from the standard first-order
ellipse.

We present final results for bound-state observables in few-body
systems ($A=2-4$) as well as ERE parameters in
Table~\ref{tab:nonscatt_values} for the LO, NLO and NNLO
potentials. Observables that were part of the respective objective
function are indicated by a white background, while entries with grey
background are predictions. Note that the errors that are given in
this table do not include a model error from the \xEFT{} truncation,
only the propagated statistical uncertainties as described in
Sec.~\ref{sec:stat_err}. It is therefore difficult to make strong
conclusions regarding the order-by-order convergence but we certainly
observe improved predictions when going to higher orders. Note that LO
results in general are characterized by small statistical
uncertainties since very little freedom is allowed with just two
parameters.
At NNLO we observe large statistical errors in the predictions
following the sequential approach, e.g., with \NNLOsep{} the
statistical error for \OEa{} is more than \unit[10]{MeV}.

Energies and radii of few-nucleon systems are well reproduced by
\NNLOsim{} as shown in Table~\ref{tab:nonscatt_values}, with the
deuteron radius being the possible exception. This can be traced back
to omitted relativistic effects. For the deuteron, $\Delta r^2$ has
been estimated to be of the size
$\unit[0.013]{fm^2}$~\cite{klarsfeld1986} and
$\unit[0.016]{fm^2}$~\cite{friar1997a}.

We have also extracted correlations between other observables in
the few-nucleon sector. As expected, for both \NNLOsep{} and
\NNLOsim{} there exist a significant correlation between the D-state
probability and the quadrupole moment of the deuteron. More
interestingly, at the present optima the triton $\beta$-decay
half-life does not correlate strongly with any other bound-state observable in
Table~\ref{tab:nonscatt_values}. This corroborates the importance of
using this observable to constrain nuclear forces, as was done already
in Ref.~\cite{gazit2009}.
\begin{table*}[tbh]
  \caption{\label{tab:nonscatt_values}Statistical uncertainties
    propagated from
    the \NN{}, \NNN{}, and \piN{} LECs to the ground-state energies (in MeV) and
    radii (in fm) for $A\leq 4$ nuclei, the deuteron D-state
    probability \ODd{} (in percent) and quadrupole moment \OQd{} (in
    fm${}^2$) and effective range observables for the ${}^1S_0$
    channel (in fm).  Gray background indicates that the corresponding result is a
    prediction. Asymmetrical errors are due to the quadratic
    dependence of the observables on the LECs.  The
    error bars on the experimental values for bound-state observables include both
    experimental and method uncertainties as detailed in
    Table~\ref{tab:fewnucleon:values}.}
    \begin{ruledtabular}
        \begin{tabular}{cc@{\phantom{\hspace{0pt}}}
  r@{}lc@{\phantom{\hspace{6pt}}}
  r@{}lc@{\phantom{\hspace{6pt}}}
  r@{}lc@{\phantom{\hspace{6pt}}}
  r@{}lc@{\phantom{\hspace{6pt}}}
  r@{}lc@{\phantom{\hspace{6pt}}}
  r@{}lc@{\phantom{\hspace{6pt}}}
  r@{}lc@{\phantom{\hspace{0pt}}}
  c}
 && \multicolumn{2}{c}{\LOsep} && \multicolumn{2}{c}{\NLOsep} && \multicolumn{2}{c}{\NNLOsep} && \multicolumn{2}{c}{\LOsim} && \multicolumn{2}{c}{\NLOsim} && \multicolumn{2}{c}{\NNLOsim} && \multicolumn{2}{c}{Exp.} && Ref.\\\hline
\OEd && \cellcolor{prediction}$-2$&\cellcolor{prediction}$.211(15)$ && \cellcolor{prediction}$-2$&\cellcolor{prediction}$.163_{(-16)}^{(+9)}$ && \cellcolor{prediction}$-2$&\cellcolor{prediction}$.2_{(-25)}^{(+12)}$ && $-2$&$.223$ && $-2$&$.224_{(-6)}^{(+1)}$ && $-2$&$.224_{(-1)}^{(+0)}$ && $-2$&$.225$ && Table~\ref{tab:fewnucleon:values}\\
\OEt && \cellcolor{prediction}$-11$&\cellcolor{prediction}$.40(4)$ && \cellcolor{prediction}$-8$&\cellcolor{prediction}$.220_{(-49)}^{(+32)}$ && $-8$&$.5_{(-64)}^{(+31)}$ && \cellcolor{prediction}$-11$&\cellcolor{prediction}$.43$ && \cellcolor{prediction}$-8$&\cellcolor{prediction}$.268_{(-38)}^{(+26)}$ && $-8$&$.482_{(-30)}^{(+26)}$ && $-8$&$.482(28)$ && Table~\ref{tab:fewnucleon:values}\\
\OEh && \cellcolor{prediction}$-10$&\cellcolor{prediction}$.39(4)$ && \cellcolor{prediction}$-7$&\cellcolor{prediction}$.474_{(-45)}^{(+29)}$ && $-7$&$.7_{(-62)}^{(+30)}$ && \cellcolor{prediction}$-10$&\cellcolor{prediction}$.43$ && \cellcolor{prediction}$-7$&\cellcolor{prediction}$.528_{(-31)}^{(+20)}$ && $-7$&$.717_{(-21)}^{(+17)}$ && $-7$&$.718(19)$ && Table~\ref{tab:fewnucleon:values}\\
\OEa && \cellcolor{prediction}$-40$&\cellcolor{prediction}$.27(13)$ && \cellcolor{prediction}$-27$&\cellcolor{prediction}$.56_{(-18)}^{(+14)}$ && \cellcolor{prediction}$-28$&\cellcolor{prediction}$_{(-18)}^{(+8)}$ && \cellcolor{prediction}$-40$&\cellcolor{prediction}$.38(1)$ && \cellcolor{prediction}$-27$&\cellcolor{prediction}$.44_{(-15)}^{(+13)}$ && \cellcolor{prediction}$-28$&\cellcolor{prediction}$.24_{(-11)}^{(+9)}$ && $-28$&$.30(11)$ && Table~\ref{tab:fewnucleon:values}\\
\Orpd && \cellcolor{prediction}$+1$&\cellcolor{prediction}$.916(5)$ && \cellcolor{prediction}$+1$&\cellcolor{prediction}$.977_{(-5)}^{(+2)}$ && \cellcolor{prediction}$+1$&\cellcolor{prediction}$.97_{(-52)}^{(+67)}$ && $+1$&$.912$ && $+1$&$.972_{(-2)}^{(+0)}$ && $+1$&$.966_{(-1)}^{(+0)}$ && $+1$&$.976(1)$ && Table~\ref{tab:fewnucleon:values}\\
\Orpt && \cellcolor{prediction}$+1$&\cellcolor{prediction}$.293(2)$ && \cellcolor{prediction}$+1$&\cellcolor{prediction}$.596(3)$ && $+1$&$.58_{(-30)}^{(+22)}$ && \cellcolor{prediction}$+1$&\cellcolor{prediction}$.292$ && \cellcolor{prediction}$+1$&\cellcolor{prediction}$.614_{(-3)}^{(+2)}$ && $+1$&$.581(2)$ && $+1$&$.587(41)$ && Table~\ref{tab:fewnucleon:values}\\
\Orph && \cellcolor{prediction}$+1$&\cellcolor{prediction}$.370(2)$ && \cellcolor{prediction}$+1$&\cellcolor{prediction}$.778_{(-4)}^{(+3)}$ && $+1$&$.76_{(-33)}^{(+23)}$ && \cellcolor{prediction}$+1$&\cellcolor{prediction}$.368$ && \cellcolor{prediction}$+1$&\cellcolor{prediction}$.791(3)$ && $+1$&$.761(2)$ && $+1$&$.766(13)$ && Table~\ref{tab:fewnucleon:values}\\
\Orpa && \cellcolor{prediction}$+1$&\cellcolor{prediction}$.081(1)$ && \cellcolor{prediction}$+1$&\cellcolor{prediction}$.459(4)$ && \cellcolor{prediction}$+1$&\cellcolor{prediction}$.44_{(-28)}^{(+15)}$ && \cellcolor{prediction}$+1$&\cellcolor{prediction}$.080$ && \cellcolor{prediction}$+1$&\cellcolor{prediction}$.482(3)$ && \cellcolor{prediction}$+1$&\cellcolor{prediction}$.445(3)$ && $+1$&$.455(7)$ && Table~\ref{tab:fewnucleon:values}\\
\OHL && \multicolumn{2}{c}{--} && \multicolumn{2}{c}{--} && $+0$&$.685_{(-50)}^{(+22)}$ && \multicolumn{2}{c}{--} && \multicolumn{2}{c}{--} && $+0$&$.6848(11)$ && $+0$&$.6848(11)$ && Table~\ref{tab:fewnucleon:values}\\
  \ODd && \cellcolor{prediction}$+7$&\cellcolor{prediction}$.794(17)$ && \cellcolor{prediction}$+2$&\cellcolor{prediction}$.942_{(-81)}^{(+85)}$ && \cellcolor{prediction}$+3$&\cellcolor{prediction}$.9_{(-12)}^{(+18)}$ && \cellcolor{prediction}$+7$&\cellcolor{prediction}$.807$ && \cellcolor{prediction}$+2$&\cellcolor{prediction}$.876_{(-82)}^{(+85)}$ && \cellcolor{prediction}$+3$&\cellcolor{prediction}$.381_{(-45)}^{(+46)}$ && \multicolumn{2}{c}{--} && \\
\OQd && \cellcolor{prediction}$+0$&\cellcolor{prediction}$.3035(7)$ && \cellcolor{prediction}$+0$&\cellcolor{prediction}$.2602_{(-20)}^{(+16)}$ && \cellcolor{prediction}$+0$&\cellcolor{prediction}$.270_{(-63)}^{(+60)}$ && $+0$&$.3030$ && $+0$&$.2589_{(-19)}^{(+17)}$ && $+0$&$.2623(8)$ && $+0$&$.270(11)$ && Table~\ref{tab:fewnucleon:values}\\
\Oann && \cellcolor{prediction}$-26$&\cellcolor{prediction}$.04(5)$ && $-18$&$.95_{(-47)}^{(+44)}$ && \cellcolor{prediction}$-19$&\cellcolor{prediction}$_{(-24)}^{(+7)}$ && \cellcolor{prediction}$-26$&\cellcolor{prediction}$.04(8)$ && $-18$&$.95_{(-41)}^{(+38)}$ && \cellcolor{prediction}$-19$&\cellcolor{prediction}$.28_{(-80)}^{(+74)}$ && $-18$&$.95(40)$ && \cite{machleidt2011}\\
\Oanp && \cellcolor{prediction}$-25$&\cellcolor{prediction}$.58(5)$ && \cellcolor{prediction}$-23$&\cellcolor{prediction}$.37_{(-19)}^{(+16)}$ && \cellcolor{prediction}$-24$&\cellcolor{prediction}$_{(-44)}^{(+11)}$ && \cellcolor{prediction}$-25$&\cellcolor{prediction}$.58(8)$ && \cellcolor{prediction}$-23$&\cellcolor{prediction}$.60_{(-13)}^{(+10)}$ && \cellcolor{prediction}$-23$&\cellcolor{prediction}$.83(11)$ && $-23$&$.71$ && \cite{hackenburg2006}\\
\OappC && \cellcolor{prediction}$-7$&\cellcolor{prediction}$.579(4)$ && \cellcolor{prediction}$-7$&\cellcolor{prediction}$.799_{(-3)}^{(+1)}$ && \cellcolor{prediction}$-7$&\cellcolor{prediction}$.8_{(-24)}^{(+10)}$ && \cellcolor{prediction}$-7$&\cellcolor{prediction}$.579(6)$ && \cellcolor{prediction}$-7$&\cellcolor{prediction}$.799_{(-3)}^{(+1)}$ && \cellcolor{prediction}$-7$&\cellcolor{prediction}$.811(1)$ && $-7$&$.820(3)$ && \cite{bergervoet1988}\\
\Ornn && \cellcolor{prediction}$+1$&\cellcolor{prediction}$.697$ && $+2$&$.752(7)$ && \cellcolor{prediction}$+2$&\cellcolor{prediction}$.85_{(-34)}^{(+21)}$ && \cellcolor{prediction}$+1$&\cellcolor{prediction}$.697(1)$ && $+2$&$.752_{(-8)}^{(+7)}$ && \cellcolor{prediction}$+2$&\cellcolor{prediction}$.793(14)$ && $+2$&$.75(11)$ && \cite{machleidt2011}\\
\Ornp && \cellcolor{prediction}$+1$&\cellcolor{prediction}$.700$ && \cellcolor{prediction}$+2$&\cellcolor{prediction}$.650_{(-4)}^{(+3)}$ && \cellcolor{prediction}$+2$&\cellcolor{prediction}$.74_{(-33)}^{(+20)}$ && \cellcolor{prediction}$+1$&\cellcolor{prediction}$.700(1)$ && \cellcolor{prediction}$+2$&\cellcolor{prediction}$.648(3)$ && \cellcolor{prediction}$+2$&\cellcolor{prediction}$.686(2)$ && $+2$&$.750(62)$ && \cite{hackenburg2006}\\
\OrppC && \cellcolor{prediction}$+1$&\cellcolor{prediction}$.812$ && \cellcolor{prediction}$+2$&\cellcolor{prediction}$.704(3)$ && \cellcolor{prediction}$+2$&\cellcolor{prediction}$.81_{(-30)}^{(+18)}$ && \cellcolor{prediction}$+1$&\cellcolor{prediction}$.812(1)$ && \cellcolor{prediction}$+2$&\cellcolor{prediction}$.704(3)$ && \cellcolor{prediction}$+2$&\cellcolor{prediction}$.758(2)$ && $+2$&$.790(14)$ && \cite{bergervoet1988}\\
        \end{tabular}
    \end{ruledtabular} 
\end{table*}

\begin{figure*}
    \centering
    \includegraphics[width=\textwidth]{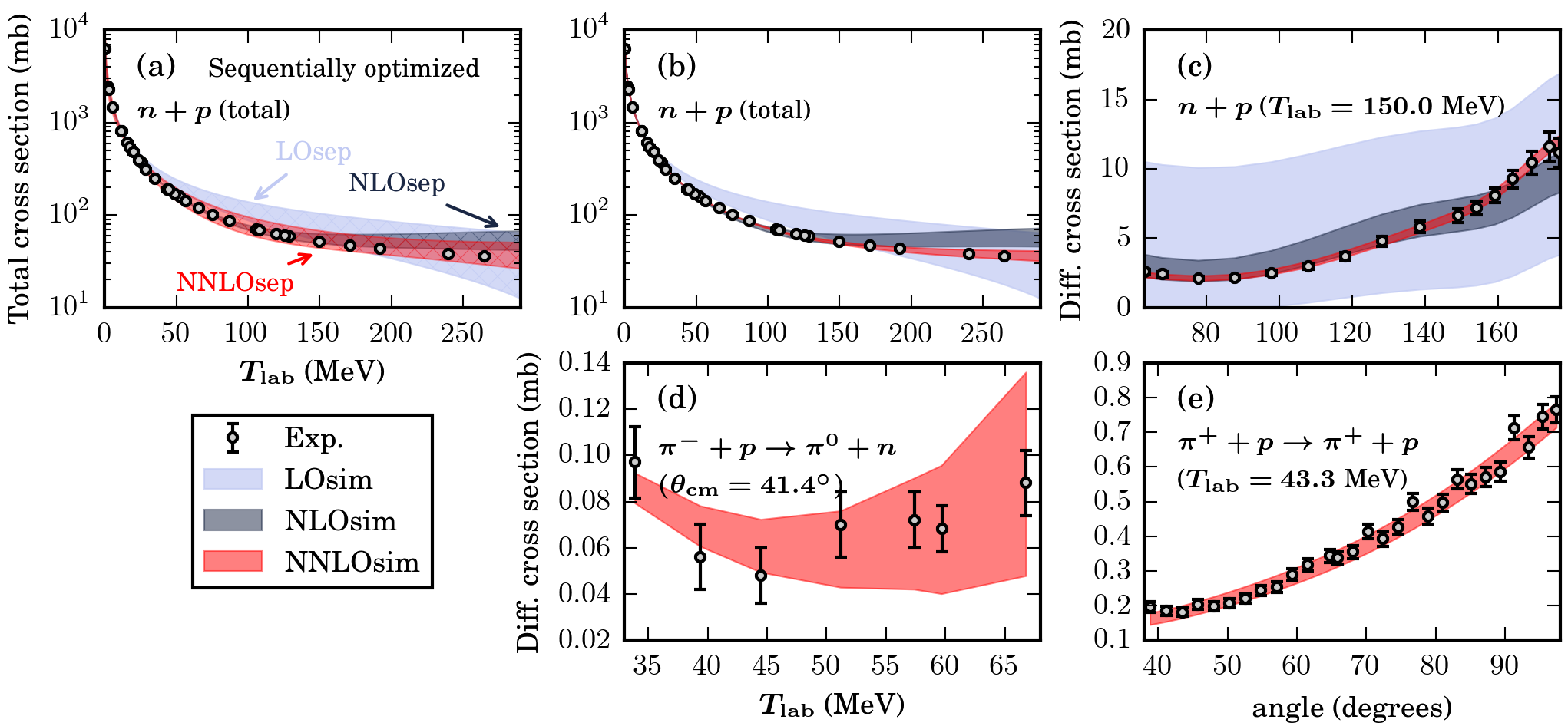}
    \caption{\label{fig:scatt_obs_xN}\co Comparison between selected
      \NN{} and \piN{} experimental data sets and theoretical
      calculations for chiral interactions at LO, NLO and NNLO. The
      bands indicate the total errors (statistical plus model
      errors). (a) $np$ total cross section for the sequentially
      optimized interactions with no clear signature of convergence
      with increasing chiral order. All other results are for the
      simultaneously optimized interactions: \LOsim{}, \NLOsim{} and
      \NNLOsim{}.  (b) $np$ total cross section; (c) $np$
      differential cross section; (d) \piN{} charge-exchange,
      differential cross section; (e) \piN{} elastic, differential
      cross section.}
\end{figure*}

Total uncertainties (statistical plus estimated model error from
the \xEFT{} truncation) are shown for scattering observables in
Fig.~\ref{fig:scatt_obs_xN}. The statistical errors are typically very
small compared to the model error for the \LOsim{}, \NLOsim{} and
\NNLOsim{} potentials, Fig.~\ref{fig:scatt_obs_xN}(b,c,d,e). Clear
signatures of an order-by-order convergence are seen in the \NN{}
scattering observables as illustrated by the $np$ total cross section
and the differential cross section that are shown in
Fig.~\ref{fig:scatt_obs_xN}(b,c). 
The same convergence is not seen when using the sequentially optimized
potentials as illustrated in Fig.~\ref{fig:scatt_obs_xN}(a). In this
case, the statistical errors are of the same order of magnitude as the
model errors, and the NNLO error band is even wider than the NLO band.
Note that \piN{} scattering is only described with the NNLOsep and
NNLOsim interactions.
%
%------------------------------
\subsection{\label{sec:optimizationprotocol}%
Optimization protocol}
%------------------------------
%
We have demonstrated that the statistical uncertainties of \xEFT{}, if
all correlations are accounted for, will induce rather small errors in
the predictions of observables. This reflects the fact that most of
the few-nucleon data is precise and diverse enough to constrain a
statistically meaningful \xEFT{} description of the nuclear
interaction. Also note that we only included experimentally observable
data in the objective function.

The existence of strong correlations between the LECs require a
complete determination of the corresponding covariance matrix, not
just the diagonal entries. For this, it is necessary to employ the
so-called \emph{simultaneous} optimization protocol. To further
demonstrate this point, we carried out error propagations with
\NNLOsim{} while neglecting the off-diagonal correlations between the
LECs. The statistical uncertainty of the binding energy in $^{4}$He
grew with a factor $\sim 90$ compared with the fully informed model.
Neglecting the statistical correlations will also obscure the desired
convergence pattern of \xEFT{}. Indeed, for the separately optimized
potentials there were no signs of convergence in the description of,
e.g., $np$ scattering data.

If the experimental database of \piN{} scattering cross sections would
be complete, then it would be possible to separately constrain, with
zero variances, the corresponding LECs. Only this scenario would
render it unnecessary to include the \piN{} scattering data in the
simultaneous objective function. Implicitly, this scenario also
assumes a perfect theory, i.e. that the employed \xEFT{} can account
for the dynamics of pionic interactions. Of course, reality lies
somewhere in between, and a simultaneous optimization approach is
preferable in the present situation. There exists ongoing efforts
where the $\pi N$ sector of \xEFT{} is extrapolated and fitted
separately in the unphysical kinematical region where it exhibits a
stronger curvature with respect to the data~\cite{hoferichter2015}.

Overall, the importance of applying simultaneous optimization is most
prominent at higher chiral orders, since the sub-leading \piN{} LECs
enter first at NNLO. In fact, the separately optimized \NNLOsep{}
potential contains a large systematic uncertainty by construction.
We find that the scaling factor for the \NN{} scattering
model error, $C_{\NN}$, decreases from $1.6$ to $\unit[1.0]{mb^{1/2}}$
when going from \NNLOsep{} to the simultaneously optimized
\NNLOsim. This implies that the separate, or sequential, optimization
protocol introduces additional artificial systematic errors not due to
the chiral expansion but due to incorrectly fitted LECs. This scenario
is avoided in a simultaneous optimization. The scaling factor for the
\piN{} scattering model error, $C_{\pi\text{N}}$, remains at
$\unit[3.6]{mb^{1/2}}$ for both \NNLOsep{} and \NNLOsim.

\begin{figure}
    \centering
    \includegraphics[width=\columnwidth]{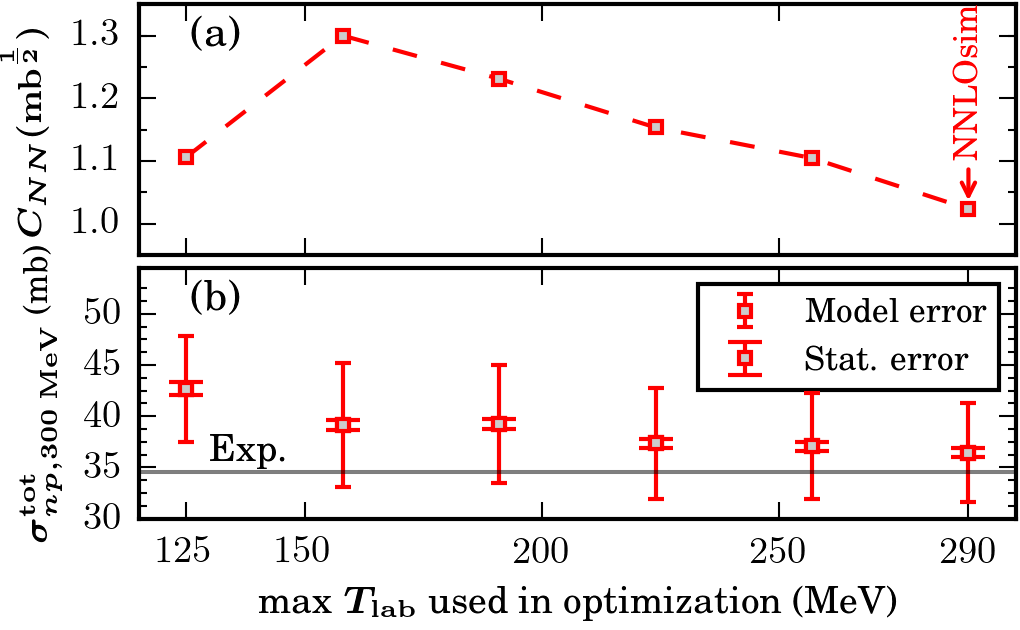}
    \caption{\label{fig:results:NNLOsim_serie_scatt}\co Predictions
      for the different re-optimizations of \NNLOsim. On the x-axis is
      the maximum $T_{\text{lab}}$ for the \NN{} scattering data used in
      the optimization. (a) Model error amplitude~\eqref{eq:NNscatt:erroramp} re-optimized so that
      $\chi^2/N_{\text{dof}} = 1$ for the respective data subset. (b)
      Model prediction for the $np$ total cross section at
      $T_\mathrm{lab} = 300$~MeV with error bars representing
      statistical and model errors for the different re-optimizations.}
\end{figure}

The size of the model error is determined such that the overall
scattering $\chi^2/N_{\text{dof}}$ is unity, which means that it
depends on the observables entering the optimization. We can explore
the stability of our approach by re-optimizing \NNLOsim{} with
respect to different truncations of the input \NN scattering data. To
this end we adjust the allowed $T_{\text{lab}}^{\text{max}}$ between
$\unit[125 \text{--} 290]{MeV}$ in six steps. It turns out that our
procedure for extracting the model error is very stable. The resulting
normalization constants $C_{\NN}$ vary between $\unit[1.0]{mb^{1/2}}$
and $\unit[1.3]{mb^{1/2}}$ as shown in
Fig.~\ref{fig:results:NNLOsim_serie_scatt}(a).

To see the corresponding effect on predicted observables we consider
the $np$ total cross section at laboratory scattering energy
$T_{\text{lab}} = \unit[300]{MeV}$. The model errors vary between
$\unit[4.8]{mb}$ and $\unit[6.1]{mb}$, and the calculated cross
sections vary between $\unit[36.5]{mb}$ and $\unit[42.7]{mb}$, see
Fig.~\ref{fig:results:NNLOsim_serie_scatt}(b). The measured value is
$\unit[34.563(174)]{mb}$~\cite{lisowski1982,arndt1999}. We note that
the size of the estimated model error is comparable with the variation
in the predictions due to changing $T_{\text{lab}}^{\text{max}}$.

Throughout the analysis, the model error for scattering observables
was assumed to scale with momentum $p$ according to
Eq.~\eqref{eq:NNscatt:erroramp}. However, the soft scale $Q$ in
\xEFT{} is set by $\max\{ p, m_{\pi}\}$ and it can be argued that the
model error should be implemented as
\begin{align}
\tilde{\sigma}_{\text{model,x}}^{(\text{amp})} =
C_{\text{x}}\left(\frac{\max\{p,
    m_\pi\}}{\Lambda_\chi}\right)^{\nu_{\text{x}}+1}. 
\end{align}
It turns out that resolving these two momentum scales has a
small impact on the estimated model errors. As an illustration, the
predictions of the \nuc{4}{He} binding energy changes by just $\unit[\sim
  20]{keV}$ (less than 0.1\%).
In fact, this effect is much smaller than the impact of changing the
$T_{\text{lab}}^{\max}$ cutoff in the experimental \NN{} scattering
database.
%
%------------------------------
\section{\label{sec:systematic}%
Extended analysis of systematic uncertainties}
%------------------------------
In nuclear physics the theoretical uncertainties very often dominate
over the experimental ones. In particular, this is true for the
systematic error. Therefore, it is crucial to establish a credible
program for assessing the error budget of any prediction or analysis
of experimental information. Thus, we focus our attention on the
convergence and missing physics in \xEFT{}. In particular, we discuss
consequences for predictions of bound-state observables in heavier
nuclei such as \nuc{4}{He} and \nuc{16}{O}. It would be valuable to
estimate the systematic uncertainty of predicted bound-state
observables --- due to the momentum-dependent \xEFT{} uncertainty
$\sigma_{\rm model}$ in Eq.~\eqref{eq:NNscatt:erroramp}. However, the
explicit momentum dependence is integrated over when solving the
non-relativistic Sch\"{o}dinger equation. Thus, a clear connection to
the momentum-expansion is lost.

As demonstrated already in Fig.~\ref{fig:results:NNLOsim_serie_scatt},
the variations in model predictions obtained from different
truncations of the input data (including only \NN{} scattering data
with $T_{\rm lab} \leq T_{\rm lab}^{\max}$) is a good first
approximation of the expected model uncertainty. To get a more
complete picture of the systematic uncertainty, we now also vary the
regulator cut-off parameter $\Lambda$ in the range
$\unit[450-600]{MeV}$ in steps of $\unit[25]{MeV}$.  For each
combination of $T_{\rm lab}^{\max}$ and $\Lambda$ we perform a
simultaneous optimization of the LECs, which results in a family of
$42$ NNLO interactions -- i.e., $42$ sets of LECs that each comes with
statistical uncertainties. It is clear from
Table~\ref{tab:optimresults:params} that the statistical uncertainties
of the LECs are smaller than the overall shifts induced by varying
$T_{\rm Lab}^{\max}$ and the cutoff $\Lambda$. All sets of LECs at LO,
NLO, NNLO that were obtained in this work are listed in the
Supplemental Material~\cite{supplemental}. Furthermore, each set is
accompanied by its own covariance matrix, also avaliable for
download. In the following discussion we use this family of potentials
to estimate the systematic uncertainty.
\begin{table}[htb!]
    \caption{\label{tab:optimresults:params}Ranges of LEC values and
      maximum statistical uncertainties among all $42$ simultaneously
      optimized NNLO potentials constructed in this work, see
      Sec.~\ref{sec:systematic} The first two columns show
      the global variation of the LEC values, in terms of minimum and
      maximum values, due to changes in $\Lambda$ and $T_{\rm
        Lab}^{\rm max}$. The third column shows the maximum
      statistical uncertainty of each LEC, which almost exclusively
      come from the $\Lambda=450$ MeV and $T_{\rm Lab}^{\rm max} =
      125$ MeV \NNLOsim{} potential. For a given LEC, the statistical
      uncertainty is rather similar for different
      potentials. $\tilde{C}_i$ are in units of
      $\unit[10^4]{GeV^{-2}}$, $C_i$ in units of
      $\unit[10^4]{GeV^{-4}}$, $c_D$ and $c_E$ are dimensionless while
      $c_i$, $d_i$ and $e_i$ are in units of $\unit{GeV^{-1}}$,
      $\unit{GeV^{-2}}$ and $\unit{GeV^{-3}}$, respectively.}
    \begin{ruledtabular}
        \begin{tabular}{c|c@{\phantom{\hspace{30pt}}}
  r@{$\,\,\,\,\ldots\,$}lc@{\phantom{\hspace{30pt}}}c}
            LEC && \multicolumn{2}{c}{range} && $\mathrm{max}(\sigma) $ \\\hline
            $\tilde{C}_{{}^1S_0}^{(np)}$&&    -0.1519&-0.1464  &&  $\pm    0.0020$   \\
            $\tilde{C}_{{}^1S_0}^{(pp)}$&&    -0.1512  &-0.1454  &&  $\pm    0.0020$   \\
            $\tilde{C}_{{}^1S_0}^{(nn)}$&&    -0.1518  &-0.1463  &&  $\pm    0.0021$   \\
            $C_{{}^1S_0}$               &&     2.4188  &\phantom{-}2.5476  &&  $\pm    0.0511$   \\
            $\tilde{C}_{{}^3S_1}$       &&    -0.1807  &-0.1348  &&  $\pm    0.0032$   \\
            $C_{{}^3S_1}$               &&     0.5037  &\phantom{-}0.7396  &&  $\pm    0.0521$   \\
            $C_{E_1}$                   &&     0.2792  &\phantom{-}0.6574  &&  $\pm    0.0253$   \\
            $C_{{}^3P_0}$               &&     0.9924  &\phantom{-}1.6343  &&  $\pm    0.0428$   \\
            $C_{{}^1P_1}$               &&     0.0618  &\phantom{-}0.6635  &&  $\pm    0.0438$   \\
            $C_{{}^3P_1}$               &&    -0.9666  &-0.4724  &&  $\pm    0.0416$   \\
            $C_{{}^3P_2}$               &&    -0.7941  &-0.6324  &&  $\pm    0.0327$   \\
            $c_D$                       &&    -0.5944  &\phantom{-}0.8348  &&  $\pm    0.0833$   \\
            $c_E$                       &&    -2.4019  &-0.0893  &&  $\pm    0.2282$   \\
            $c_1$                       &&    -0.8329  &\phantom{-}0.2784  &&  $\pm    0.3043$   \\
            $c_2$                       &&     2.7946  &\phantom{-}5.3258  &&  $\pm    1.0754$   \\
            $c_3$                       &&    -4.3601  &-3.4474  &&  $\pm    0.1506$   \\
            $c_4$                       &&     1.8999  &\phantom{-}4.2353  &&  $\pm    0.2179$   \\
            $d_1\!+\!d_2$               &&     4.4636  &\phantom{-}5.4505  &&  $\pm    0.1378$   \\
            $d_3$                       &&    -4.8549  &-4.4583  &&  $\pm    0.2302$   \\
            $d_5$                       &&    -0.2992  &\phantom{-}0.0233  &&  $\pm    0.1407$   \\
            $d_{14}\!-\!d_{15}$         &&   -10.3220  &-9.6902  &&  $\pm    0.2820$   \\
            $e_{14}$                    &&    -0.3700  &\phantom{-}0.9569  &&  $\pm    0.9079$   \\
            $e_{15}$                    &&   -11.9223  &-9.1307  &&  $\pm    2.4962$   \\
            $e_{16}$                    &&    -0.6847  &\phantom{-}7.4463  &&  $\pm    4.2436$   \\
            $e_{17}$                    &&     0.9322  &\phantom{-}1.4986  &&  $\pm    1.8143$   \\
            $e_{18}$                    &&    -2.5068  &\phantom{-}8.3777  &&  $\pm    1.9022$   \\    
        \end{tabular}
    \end{ruledtabular}
\end{table}

First, we would like to emphasize that all sets of simultaneously
optimized LECs provide an almost equally good description of all
$A\leq 4$ data.  Some of the $\pi$N LECs display large variations, but
the $\chi^2/N_{\rm dof}$ (without model error) for the $\pi$N data is
within $2.28(4)$ for all of these potentials.  The sub-leading $\pi N$
LECs become more positive when $NN$ scattering data at higher energies
is included, and $c_1$ in particular carries a larger (relative) statistical
uncertainty than the others. It is noteworthy that for a given
$T_{\rm Lab}^{\max}$, and up to $1\sigma$ precision, the $\pi N$ LECs
exhibit $\Lambda$-independence. The \NNN{} LECs, $c_D$ and $c_{E}$,
tend to depend less on $T_{\rm Lab}^{\max}$ at larger values of
$\Lambda$. However, they always remain natural. It is also interesting
to note that the tensor contact, $C_{E_1}$, is insensitive to
$\Lambda$-variations but strongly dependent on the
$T_{\rm Lab}^{\max}$ cut. It was shown in Fig.~\ref{fig:corr_par} that
$C_{E_1}$ and $c_4$ correlate strongly.  This effect can already be
expected from the structure of the underlying expression for the NNLO
interaction.

\begin{figure}
    \centering
    \includegraphics[width=\columnwidth]{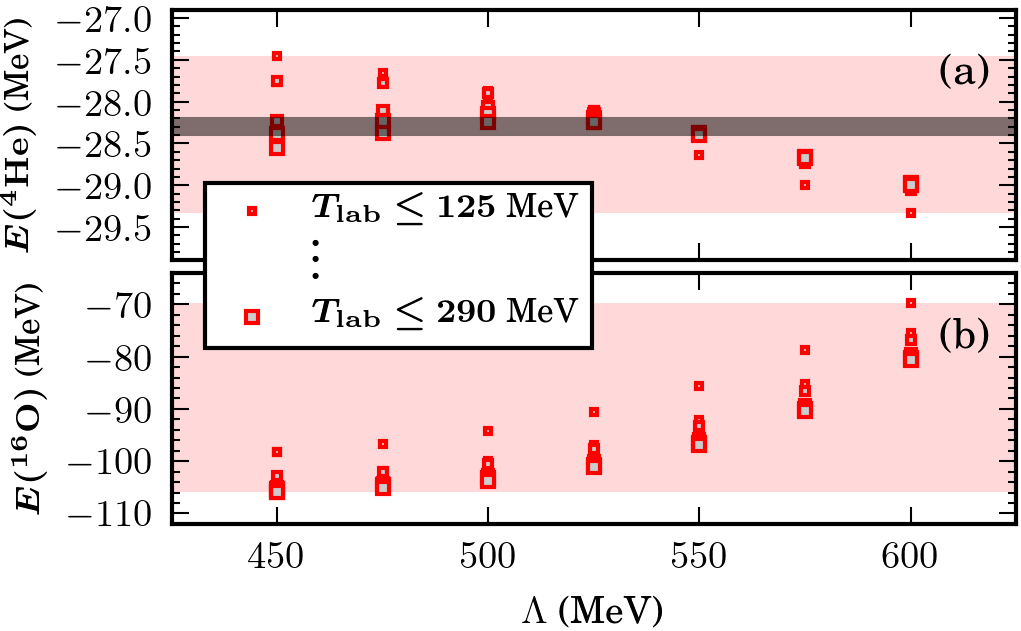}
    \caption{\label{fig:results:NNLOsim_serie_manybody_manymodels}\co
      Binding-energy predictions for (a) \nuc{4}{He} and (b)
      \nuc{16}{O} with the different re-optimizations of \NNLOsim. On
      the x-axis is the employed cutoff $\Lambda$. Vertically aligned
      red markers correspond to different $T_{\rm Lab}^{\rm max}$ for
    the \NN{} scattering data used in the optimization. The
    experimental binding energies are $\OEa{} \approx
    \unit[-28.30]{MeV}$, represented by a grey band in panel (a), and
    $\OEO \approx
    \unit[-127.6]{MeV}$~\cite{sukhoruchkin2009}. Statistical
      error bars on the theoretical results are smaller than the
      marker size on this energy scale.}
\end{figure}

To gauge the magnitude of model variations in heavier nuclei we
computed the binding energies of \nuc{4}{He} and \nuc{16}{O} using the
previously mentioned family of 42 NNLO potentials. The resulting
binding energies for \nuc{4}{He} and \nuc{16}{O}, computed in the NCSM
and CC, respectively, are shown in
Fig.~\ref{fig:results:NNLOsim_serie_manybody_manymodels}. The NCSM
calculations were carried out in a HO model space with
$N_{\rm max} =20 $ and $\hbar \omega = \unit[36]{MeV}$. The CC
calculations were carried out in the so-called $\Lambda-$CCSD(T)
approximation~\cite{hagen2013} in 15 major oscillator shells with
$\hbar \omega =\unit[22]{MeV}$. The largest energy difference when
going from 13 to 15 oscillator shells was 3.6 MeV (observed for
$\Lambda = \unit[600]{MeV}$). For our purposes, this provides
well-enough converged results. The \NNN{} force was truncated at the
normal-ordered two-body level in the Hartree-Fock basis.

The \OEa{} predictions vary within a $\sim \unit[2]{MeV}$ range. For
\OEO{} this variation increases dramatically to $\sim \unit[35]{MeV}$.
Irrespective of the discrepancy with the measured value, the spread of
the central values indicates the presence of a surprisingly large
systematic error when extrapolating to heavier systems.

The statistical uncertainties remain small: tens of keV for
\nuc{4}{He} and a few hundred keV for \nuc{16}{O}. These uncertainties
are obtained from the quadratic approximation with the computed
Jacobian and Hessian for \nuc{4}{He}, while a brute-force Monte Carlo
simulation with $2.5\times 10^4$ CC calculations was performed for
\nuc{16}{O}. This massive set of CC calculations employed the doubles
approximations in 9 major oscillator shells. We conclude that the
statistical uncertainties of the predictions for \OEa{} and \OEO{} at
NNLO are much smaller than the variations due to changing
$\Lambda$ or $T_{\rm Lab}^{\rm max}$.
However, this is only true for simultaneously optimized
potentials. For the separately optimized NNLO potential (\NNLOsep) the
statistical uncertainty of the \OEa{} prediction is five times larger
than the observed variations due to changing $\Lambda$ and
$T_{\rm Lab}^{\rm max}$.
%
%=======================================
\section{Outlook}\label{sec:outlook}
%-------------------
%
The extended analysis of systematic uncertainties presented above
suggests that large fluctuations are induced in heavier nuclei (see
Fig.~\ref{fig:results:NNLOsim_serie_manybody_manymodels}). Furthermore,
while predictions for \nuc{4}{He} are accurate over a rather wide
range of regulator parameters, the binding energy for \nuc{16}{O}
turns out to be underestimated for the entire range used in this
study. In fact, there is no overlap between the theoretical
predictions and the experimental results, even though the former ones
have large error bars.

Based on our findings we recommend that continued efforts towards an
\emph{ab initio} framework based on \xEFT{} should involve additional
work in, at least, three different directions:
\begin{enumerate}
\item Explore the alternative strategy of informing the model about low-energy
  many-body observables.
  \item Diversify and extend the statistical analysis and perform a
    sensitivity analysis of input data.
  \item Continue efforts towards higher orders of the
    chiral expansion, and possibly revisit the power counting.
\end{enumerate}
Let us comment briefly on these research directions. The poor
many-body scaling observed in
Fig.~\ref{fig:results:NNLOsim_serie_manybody_manymodels} was
pragmatically accounted for in the construction of the so-called
NNLO$_\mathrm{sat}$ potential presented in Ref.~\cite{ekstrom2015}
where also heavier nuclei were included in the fit. The accuracy of
many-body predictions was shown to be much improved, but the
uncertainty analysis is much more difficult within such a strategy.

Secondly, to get a handle on possible bias in the statistical analysis
due to the choice of statistical technique, it is important to apply
different types of optimization and uncertainty quantification
methods. Various choices exist, such as e.g.\ Lagrange multiplier
analysis~\cite{stump2001}, Bayesian methods~\cite{schindler2009}, or
Gaussian process modeling~\cite{habib2007,holsclaw2010}. In general,
stochastic modeling with Monte Carlo simulations offer a
straightforward and versatile approach. This tool is also
indispensable for computing the posterior probabilities in Bayesian
inference. The Monte Carlo results for $A\leq4$ observables that were
presented in this work consist of $10^5$ sampling points over a
multivariate Gaussian parameter space.  With our current
implementation, the computational cost for sampling all $A\leq 4$
observables presented in this work is very low --- less than 8000 CPU
hours. As such, the present work shows great promise also for future
stochastic applications. 

Furthermore, the computational framework that we have presented here,
and our present implementation, is not limited to any particular type
of regulator function or flavour of chiral expansion. Moreover, the
handling of a larger number of LECs, as would be the consequence of
working at a higher chiral order, should be relatively straightforward
and we don't foresee any computational bottlenecks.

Finally, the magnitude of the systematic uncertainties that were
observed in this work suggest the need to further explore and improve
the theoretical underpinnings of the chiral expansion of the nuclear
interaction.
%
%=======================================
\begin{acknowledgments}
  The authors thank D. Furnstahl, G. Hagen, M. Hjorth-Jensen,
  W. Nazarewicz, and T. Papenbrock for valuable comments and fruitful
  discussions. The research leading to these results has received
  funding from the European Research Council under the European
  Community's Seventh Framework Programme (FP7/2007-2013) / ERC Grant
  No.\ 240603 and the Swedish Foundation for International Cooperation
  in Research and Higher Education (STINT, Grant No.\ IG2012-5158).
  This material is based upon work supported by the
  U.S. Department of Energy, Office of Science, Office of Nuclear
  Physics under Awards No.\ DEFG02-96ER40963 (University of
  Tennessee) and No.\ DE-SC0008499 (NUCLEI SciDAC collaboration) and under
  Contract No.\  DE-AC05-00OR22725 (Oak Ridge National Laboratory). 
  The computations were performed on resources provided by the Swedish
  National Infrastructure for Computing at NSC, HPC2C, and
  C3SE\@. This research also used resources of the Oak Ridge
  Leadership Computing Facility located in the Oak Ridge National
  Laboratory.  One of us (AE) wants to acknowledge the hospitality of
  Chalmers University of Technology where the implementation of
  Automatic Differentiation was performed, while the hospitality of
  Oslo University is acknowledged by BC.
\end{acknowledgments}

\bibliography{chiral-error}

\begin{thebibliography}{102}%
\makeatletter
\providecommand \@ifxundefined [1]{%
 \@ifx{#1\undefined}
}%
\providecommand \@ifnum [1]{%
 \ifnum #1\expandafter \@firstoftwo
 \else \expandafter \@secondoftwo
 \fi
}%
\providecommand \@ifx [1]{%
 \ifx #1\expandafter \@firstoftwo
 \else \expandafter \@secondoftwo
 \fi
}%
\providecommand \natexlab [1]{#1}%
\providecommand \enquote  [1]{``#1''}%
\providecommand \bibnamefont  [1]{#1}%
\providecommand \bibfnamefont [1]{#1}%
\providecommand \citenamefont [1]{#1}%
\providecommand \href@noop [0]{\@secondoftwo}%
\providecommand \href [0]{\begingroup \@sanitize@url \@href}%
\providecommand \@href[1]{\@@startlink{#1}\@@href}%
\providecommand \@@href[1]{\endgroup#1\@@endlink}%
\providecommand \@sanitize@url [0]{\catcode `\\12\catcode `\$12\catcode
  `\&12\catcode `\#12\catcode `\^12\catcode `\_12\catcode `\%12\relax}%
\providecommand \@@startlink[1]{}%
\providecommand \@@endlink[0]{}%
\providecommand \url  [0]{\begingroup\@sanitize@url \@url }%
\providecommand \@url [1]{\endgroup\@href {#1}{\urlprefix }}%
\providecommand \urlprefix  [0]{URL }%
\providecommand \Eprint [0]{\href }%
\providecommand \doibase [0]{http://dx.doi.org/}%
\providecommand \selectlanguage [0]{\@gobble}%
\providecommand \bibinfo  [0]{\@secondoftwo}%
\providecommand \bibfield  [0]{\@secondoftwo}%
\providecommand \translation [1]{[#1]}%
\providecommand \BibitemOpen [0]{}%
\providecommand \bibitemStop [0]{}%
\providecommand \bibitemNoStop [0]{.\EOS\space}%
\providecommand \EOS [0]{\spacefactor3000\relax}%
\providecommand \BibitemShut  [1]{\csname bibitem#1\endcsname}%
\let\auto@bib@innerbib\@empty
%</preamble>
\bibitem [{\citenamefont {Torres~Valderrama}\ \emph {et~al.}(2015)\citenamefont
  {Torres~Valderrama}, \citenamefont {Witteveen}, \citenamefont {Navarro},\
  and\ \citenamefont {Blom}}]{valderrama2015b}%
  \BibitemOpen
  \bibfield  {author} {\bibinfo {author} {\bibfnamefont {A.}~\bibnamefont
  {Torres~Valderrama}}, \bibinfo {author} {\bibfnamefont {J.}~\bibnamefont
  {Witteveen}}, \bibinfo {author} {\bibfnamefont {M.}~\bibnamefont {Navarro}},
  \ and\ \bibinfo {author} {\bibfnamefont {J.}~\bibnamefont {Blom}},\ }\href
  {\doibase 10.1186/2190-8567-5-3} {\bibfield  {journal} {\bibinfo  {journal}
  {J. Math. Neurosci.}\ }\textbf {\bibinfo {volume} {5}},\ \bibinfo {pages} {3}
  (\bibinfo {year} {2015})}\BibitemShut {NoStop}%
\bibitem [{\citenamefont {Murphy}\ \emph {et~al.}(2004)\citenamefont {Murphy},
  \citenamefont {Sexton}, \citenamefont {Barnett}, \citenamefont {Jones},
  \citenamefont {Webb}, \citenamefont {Collins},\ and\ \citenamefont
  {Stainforth}}]{murphy2004}%
  \BibitemOpen
  \bibfield  {author} {\bibinfo {author} {\bibfnamefont {J.~M.}\ \bibnamefont
  {Murphy}}, \bibinfo {author} {\bibfnamefont {D.~M.~H.}\ \bibnamefont
  {Sexton}}, \bibinfo {author} {\bibfnamefont {D.~N.}\ \bibnamefont {Barnett}},
  \bibinfo {author} {\bibfnamefont {G.~S.}\ \bibnamefont {Jones}}, \bibinfo
  {author} {\bibfnamefont {M.~J.}\ \bibnamefont {Webb}}, \bibinfo {author}
  {\bibfnamefont {M.}~\bibnamefont {Collins}}, \ and\ \bibinfo {author}
  {\bibfnamefont {D.~A.}\ \bibnamefont {Stainforth}},\ }\href {\doibase
  10.1038/nature02771} {\bibfield  {journal} {\bibinfo  {journal} {Nature}\
  }\textbf {\bibinfo {volume} {430}},\ \bibinfo {pages} {768} (\bibinfo {year}
  {2004})}\BibitemShut {NoStop}%
\bibitem [{\citenamefont {Angelikopoulos}\ \emph {et~al.}(2012)\citenamefont
  {Angelikopoulos}, \citenamefont {Papadimitriou},\ and\ \citenamefont
  {Koumoutsakos}}]{angelikopoulos2012}%
  \BibitemOpen
  \bibfield  {author} {\bibinfo {author} {\bibfnamefont {P.}~\bibnamefont
  {Angelikopoulos}}, \bibinfo {author} {\bibfnamefont {C.}~\bibnamefont
  {Papadimitriou}}, \ and\ \bibinfo {author} {\bibfnamefont {P.}~\bibnamefont
  {Koumoutsakos}},\ }\href {\doibase 10.1063/1.4757266} {\bibfield  {journal}
  {\bibinfo  {journal} {J. Chem. Phys.}\ }\textbf {\bibinfo {volume} {137}},\
  \bibinfo {eid} {144103} (\bibinfo {year} {2012})}\BibitemShut {NoStop}%
\bibitem [{\citenamefont {Erler}\ \emph {et~al.}(2012)\citenamefont {Erler},
  \citenamefont {Birge}, \citenamefont {Kortelainen}, \citenamefont
  {Nazarewicz}, \citenamefont {Olsen}, \citenamefont {Perhac},\ and\
  \citenamefont {Stoitsov}}]{erler2012}%
  \BibitemOpen
  \bibfield  {author} {\bibinfo {author} {\bibfnamefont {J.}~\bibnamefont
  {Erler}}, \bibinfo {author} {\bibfnamefont {N.}~\bibnamefont {Birge}},
  \bibinfo {author} {\bibfnamefont {M.}~\bibnamefont {Kortelainen}}, \bibinfo
  {author} {\bibfnamefont {W.}~\bibnamefont {Nazarewicz}}, \bibinfo {author}
  {\bibfnamefont {E.}~\bibnamefont {Olsen}}, \bibinfo {author} {\bibfnamefont
  {A.~M.}\ \bibnamefont {Perhac}}, \ and\ \bibinfo {author} {\bibfnamefont
  {M.}~\bibnamefont {Stoitsov}},\ }\href {\doibase 10.1038/nature11188}
  {\bibfield  {journal} {\bibinfo  {journal} {Nature}\ }\textbf {\bibinfo
  {volume} {486}},\ \bibinfo {pages} {509} (\bibinfo {year}
  {2012})}\BibitemShut {NoStop}%
\bibitem [{\citenamefont {Cacciari}\ and\ \citenamefont
  {Houdeau}(2011)}]{cacciari2011}%
  \BibitemOpen
  \bibfield  {author} {\bibinfo {author} {\bibfnamefont {M.}~\bibnamefont
  {Cacciari}}\ and\ \bibinfo {author} {\bibfnamefont {N.}~\bibnamefont
  {Houdeau}},\ }\href {\doibase 10.1007/JHEP09(2011)039} {\bibfield  {journal}
  {\bibinfo  {journal} {J. High Energy Phys.}\ }\textbf {\bibinfo {volume}
  {2011}},\ \bibinfo {eid} {39} (\bibinfo {year} {2011}),\
  10.1007/JHEP09(2011)039}\BibitemShut {NoStop}%
\bibitem [{\citenamefont {Barrett}\ \emph {et~al.}(2013)\citenamefont
  {Barrett}, \citenamefont {Navr{\'a}til},\ and\ \citenamefont
  {Vary}}]{barrett2013}%
  \BibitemOpen
  \bibfield  {author} {\bibinfo {author} {\bibfnamefont {B.~R.}\ \bibnamefont
  {Barrett}}, \bibinfo {author} {\bibfnamefont {P.}~\bibnamefont
  {Navr{\'a}til}}, \ and\ \bibinfo {author} {\bibfnamefont {J.~P.}\
  \bibnamefont {Vary}},\ }\href {\doibase 10.1016/j.ppnp.2012.10.003}
  {\bibfield  {journal} {\bibinfo  {journal} {Prog. Part. Nucl. Phys.}\
  }\textbf {\bibinfo {volume} {69}},\ \bibinfo {pages} {131 } (\bibinfo {year}
  {2013})}\BibitemShut {NoStop}%
\bibitem [{\citenamefont {Hagen}\ \emph {et~al.}(2014)\citenamefont {Hagen},
  \citenamefont {Papenbrock}, \citenamefont {Hjorth-Jensen},\ and\
  \citenamefont {Dean}}]{hagen2013}%
  \BibitemOpen
  \bibfield  {author} {\bibinfo {author} {\bibfnamefont {G.}~\bibnamefont
  {Hagen}}, \bibinfo {author} {\bibfnamefont {T.}~\bibnamefont {Papenbrock}},
  \bibinfo {author} {\bibfnamefont {M.}~\bibnamefont {Hjorth-Jensen}}, \ and\
  \bibinfo {author} {\bibfnamefont {D.~J.}\ \bibnamefont {Dean}},\ }\href
  {\doibase 10.1088/0034-4885/77/9/096302} {\bibfield  {journal} {\bibinfo
  {journal} {Rep. Prog. Phys.}\ }\textbf {\bibinfo {volume} {77}},\ \bibinfo
  {pages} {096302} (\bibinfo {year} {2014})}\BibitemShut {NoStop}%
\bibitem [{\citenamefont {Pieper}\ and\ \citenamefont
  {Wiringa}(2001)}]{pieper2001}%
  \BibitemOpen
  \bibfield  {author} {\bibinfo {author} {\bibfnamefont {S.~C.}\ \bibnamefont
  {Pieper}}\ and\ \bibinfo {author} {\bibfnamefont {R.~B.}\ \bibnamefont
  {Wiringa}},\ }\href {\doibase 10.1146/annurev.nucl.51.101701.132506}
  {\bibfield  {journal} {\bibinfo  {journal} {Annu. Rev. Nucl. Part. S.}\
  }\textbf {\bibinfo {volume} {51}},\ \bibinfo {pages} {53} (\bibinfo {year}
  {2001})}\BibitemShut {NoStop}%
\bibitem [{\citenamefont {Lee}(2009)}]{lee2009}%
  \BibitemOpen
  \bibfield  {author} {\bibinfo {author} {\bibfnamefont {D.}~\bibnamefont
  {Lee}},\ }\href {\doibase 10.1016/j.ppnp.2008.12.001} {\bibfield  {journal}
  {\bibinfo  {journal} {Prog. Part. Nucl. Phys.}\ }\textbf {\bibinfo {volume}
  {63}},\ \bibinfo {pages} {117 } (\bibinfo {year} {2009})}\BibitemShut
  {NoStop}%
\bibitem [{\citenamefont {{Leidemann}}\ and\ \citenamefont
  {{Orlandini}}(2013)}]{leidemann2013}%
  \BibitemOpen
  \bibfield  {author} {\bibinfo {author} {\bibfnamefont {W.}~\bibnamefont
  {{Leidemann}}}\ and\ \bibinfo {author} {\bibfnamefont {G.}~\bibnamefont
  {{Orlandini}}},\ }\href {\doibase 10.1016/j.ppnp.2012.09.001} {\bibfield
  {journal} {\bibinfo  {journal} {Prog. Part. Nucl. Phys.}\ }\textbf {\bibinfo
  {volume} {68}},\ \bibinfo {pages} {158} (\bibinfo {year} {2013})}\BibitemShut
  {NoStop}%
\bibitem [{\citenamefont {Romero-Redondo}\ \emph {et~al.}(2014)\citenamefont
  {Romero-Redondo}, \citenamefont {Quaglioni}, \citenamefont {Navr\'atil},\
  and\ \citenamefont {Hupin}}]{romeroredondo2014}%
  \BibitemOpen
  \bibfield  {author} {\bibinfo {author} {\bibfnamefont {C.}~\bibnamefont
  {Romero-Redondo}}, \bibinfo {author} {\bibfnamefont {S.}~\bibnamefont
  {Quaglioni}}, \bibinfo {author} {\bibfnamefont {P.}~\bibnamefont
  {Navr\'atil}}, \ and\ \bibinfo {author} {\bibfnamefont {G.}~\bibnamefont
  {Hupin}},\ }\href {\doibase 10.1103/PhysRevLett.113.032503} {\bibfield
  {journal} {\bibinfo  {journal} {Phys. Rev. Lett.}\ }\textbf {\bibinfo
  {volume} {113}},\ \bibinfo {pages} {032503} (\bibinfo {year}
  {2014})}\BibitemShut {NoStop}%
\bibitem [{\citenamefont {Deltuva}\ and\ \citenamefont
  {Fonseca}(2015)}]{deltuva2015}%
  \BibitemOpen
  \bibfield  {author} {\bibinfo {author} {\bibfnamefont {A.}~\bibnamefont
  {Deltuva}}\ and\ \bibinfo {author} {\bibfnamefont {A.~C.}\ \bibnamefont
  {Fonseca}},\ }\href {\doibase 10.1103/PhysRevC.91.034001} {\bibfield
  {journal} {\bibinfo  {journal} {Phys. Rev. C}\ }\textbf {\bibinfo {volume}
  {91}},\ \bibinfo {pages} {034001} (\bibinfo {year} {2015})}\BibitemShut
  {NoStop}%
\bibitem [{\citenamefont {Mihaila}\ and\ \citenamefont
  {Heisenberg}(2000)}]{mihaila2000}%
  \BibitemOpen
  \bibfield  {author} {\bibinfo {author} {\bibfnamefont {B.}~\bibnamefont
  {Mihaila}}\ and\ \bibinfo {author} {\bibfnamefont {J.~H.}\ \bibnamefont
  {Heisenberg}},\ }\href {\doibase 10.1103/PhysRevLett.84.1403} {\bibfield
  {journal} {\bibinfo  {journal} {Phys. Rev. Lett.}\ }\textbf {\bibinfo
  {volume} {84}},\ \bibinfo {pages} {1403} (\bibinfo {year}
  {2000})}\BibitemShut {NoStop}%
\bibitem [{\citenamefont {L{\"a}hde}\ \emph {et~al.}(2014)\citenamefont
  {L{\"a}hde}, \citenamefont {Epelbaum}, \citenamefont {Krebs}, \citenamefont
  {Lee}, \citenamefont {Mei{\ss}ner},\ and\ \citenamefont {Rupak}}]{lahde2014}%
  \BibitemOpen
  \bibfield  {author} {\bibinfo {author} {\bibfnamefont {T.~A.}\ \bibnamefont
  {L{\"a}hde}}, \bibinfo {author} {\bibfnamefont {E.}~\bibnamefont {Epelbaum}},
  \bibinfo {author} {\bibfnamefont {H.}~\bibnamefont {Krebs}}, \bibinfo
  {author} {\bibfnamefont {D.}~\bibnamefont {Lee}}, \bibinfo {author}
  {\bibfnamefont {U.-G.}\ \bibnamefont {Mei{\ss}ner}}, \ and\ \bibinfo {author}
  {\bibfnamefont {G.}~\bibnamefont {Rupak}},\ }\href {\doibase
  10.1016/j.physletb.2014.03.023} {\bibfield  {journal} {\bibinfo  {journal}
  {Phys. Lett. B}\ }\textbf {\bibinfo {volume} {732}},\ \bibinfo {pages} {110 }
  (\bibinfo {year} {2014})}\BibitemShut {NoStop}%
\bibitem [{\citenamefont {Som\`a}\ \emph {et~al.}(2013)\citenamefont {Som\`a},
  \citenamefont {Barbieri},\ and\ \citenamefont {Duguet}}]{soma2013}%
  \BibitemOpen
  \bibfield  {author} {\bibinfo {author} {\bibfnamefont {V.}~\bibnamefont
  {Som\`a}}, \bibinfo {author} {\bibfnamefont {C.}~\bibnamefont {Barbieri}}, \
  and\ \bibinfo {author} {\bibfnamefont {T.}~\bibnamefont {Duguet}},\ }\href
  {\doibase 10.1103/PhysRevC.87.011303} {\bibfield  {journal} {\bibinfo
  {journal} {Phys. Rev. C}\ }\textbf {\bibinfo {volume} {87}},\ \bibinfo
  {pages} {011303} (\bibinfo {year} {2013})}\BibitemShut {NoStop}%
\bibitem [{\citenamefont {Hergert}\ \emph {et~al.}(2013)\citenamefont
  {Hergert}, \citenamefont {Bogner}, \citenamefont {Binder}, \citenamefont
  {Calci}, \citenamefont {Langhammer}, \citenamefont {Roth},\ and\
  \citenamefont {Schwenk}}]{hergert2013}%
  \BibitemOpen
  \bibfield  {author} {\bibinfo {author} {\bibfnamefont {H.}~\bibnamefont
  {Hergert}}, \bibinfo {author} {\bibfnamefont {S.~K.}\ \bibnamefont {Bogner}},
  \bibinfo {author} {\bibfnamefont {S.}~\bibnamefont {Binder}}, \bibinfo
  {author} {\bibfnamefont {A.}~\bibnamefont {Calci}}, \bibinfo {author}
  {\bibfnamefont {J.}~\bibnamefont {Langhammer}}, \bibinfo {author}
  {\bibfnamefont {R.}~\bibnamefont {Roth}}, \ and\ \bibinfo {author}
  {\bibfnamefont {A.}~\bibnamefont {Schwenk}},\ }\href {\doibase
  10.1103/PhysRevC.87.034307} {\bibfield  {journal} {\bibinfo  {journal} {Phys.
  Rev. C}\ }\textbf {\bibinfo {volume} {87}},\ \bibinfo {pages} {034307}
  (\bibinfo {year} {2013})}\BibitemShut {NoStop}%
\bibitem [{\citenamefont {Epelbaum}\ \emph {et~al.}(2009)\citenamefont
  {Epelbaum}, \citenamefont {Hammer},\ and\ \citenamefont
  {Mei\ss{}ner}}]{epelbaum2009}%
  \BibitemOpen
  \bibfield  {author} {\bibinfo {author} {\bibfnamefont {E.}~\bibnamefont
  {Epelbaum}}, \bibinfo {author} {\bibfnamefont {H.-W.}\ \bibnamefont
  {Hammer}}, \ and\ \bibinfo {author} {\bibfnamefont {U.-G.}\ \bibnamefont
  {Mei\ss{}ner}},\ }\href {\doibase 10.1103/RevModPhys.81.1773} {\bibfield
  {journal} {\bibinfo  {journal} {Rev. Mod. Phys.}\ }\textbf {\bibinfo {volume}
  {81}},\ \bibinfo {pages} {1773} (\bibinfo {year} {2009})}\BibitemShut
  {NoStop}%
\bibitem [{\citenamefont {Machleidt}\ and\ \citenamefont
  {Entem}(2011)}]{machleidt2011}%
  \BibitemOpen
  \bibfield  {author} {\bibinfo {author} {\bibfnamefont {R.}~\bibnamefont
  {Machleidt}}\ and\ \bibinfo {author} {\bibfnamefont {D.}~\bibnamefont
  {Entem}},\ }\href {\doibase 10.1016/j.physrep.2011.02.001} {\bibfield
  {journal} {\bibinfo  {journal} {Phys. Rep.}\ }\textbf {\bibinfo {volume}
  {503}},\ \bibinfo {pages} {1 } (\bibinfo {year} {2011})}\BibitemShut
  {NoStop}%
\bibitem [{\citenamefont {Barnea}\ \emph {et~al.}(2015)\citenamefont {Barnea},
  \citenamefont {Contessi}, \citenamefont {Gazit}, \citenamefont {Pederiva},\
  and\ \citenamefont {van Kolck}}]{barnea2015}%
  \BibitemOpen
  \bibfield  {author} {\bibinfo {author} {\bibfnamefont {N.}~\bibnamefont
  {Barnea}}, \bibinfo {author} {\bibfnamefont {L.}~\bibnamefont {Contessi}},
  \bibinfo {author} {\bibfnamefont {D.}~\bibnamefont {Gazit}}, \bibinfo
  {author} {\bibfnamefont {F.}~\bibnamefont {Pederiva}}, \ and\ \bibinfo
  {author} {\bibfnamefont {U.}~\bibnamefont {van Kolck}},\ }\href {\doibase
  10.1103/PhysRevLett.114.052501} {\bibfield  {journal} {\bibinfo  {journal}
  {Phys. Rev. Lett.}\ }\textbf {\bibinfo {volume} {114}},\ \bibinfo {pages}
  {052501} (\bibinfo {year} {2015})}\BibitemShut {NoStop}%
\bibitem [{\citenamefont {Stoks}\ \emph {et~al.}(1993)\citenamefont {Stoks},
  \citenamefont {Klomp}, \citenamefont {Rentmeester},\ and\ \citenamefont
  {de~Swart}}]{stoks1993}%
  \BibitemOpen
  \bibfield  {author} {\bibinfo {author} {\bibfnamefont {V.~G.~J.}\
  \bibnamefont {Stoks}}, \bibinfo {author} {\bibfnamefont {R.~A.~M.}\
  \bibnamefont {Klomp}}, \bibinfo {author} {\bibfnamefont {M.~C.~M.}\
  \bibnamefont {Rentmeester}}, \ and\ \bibinfo {author} {\bibfnamefont {J.~J.}\
  \bibnamefont {de~Swart}},\ }\href {\doibase 10.1103/PhysRevC.48.792}
  {\bibfield  {journal} {\bibinfo  {journal} {Phys. Rev. C}\ }\textbf {\bibinfo
  {volume} {48}},\ \bibinfo {pages} {792} (\bibinfo {year} {1993})}\BibitemShut
  {NoStop}%
\bibitem [{\citenamefont {Arndt}\ \emph {et~al.}(1999)\citenamefont {Arndt},
  \citenamefont {Strakovsky},\ and\ \citenamefont {Workman}}]{arndt1999}%
  \BibitemOpen
  \bibfield  {author} {\bibinfo {author} {\bibfnamefont {R.~A.}\ \bibnamefont
  {Arndt}}, \bibinfo {author} {\bibfnamefont {I.~I.}\ \bibnamefont
  {Strakovsky}}, \ and\ \bibinfo {author} {\bibfnamefont {R.~L.}\ \bibnamefont
  {Workman}},\ }\href@noop {} {\emph {\bibinfo {title} {{S}AID, Scattering
  Analysis Interactive Dial-in computer facility}}} (\bibinfo {year} {1999}),\
  \bibinfo {note} {{G}eorge {W}ashington {U}niversity (formerly {V}irginia
  {P}olytechnic {I}nstitute), solution SM99 (Summer 1999); for more information
  see, e.g., R. A. Arndt, I. I. Strakovsky, and R. L. Workman, Phys. Rev. C
  {\bf 50}, 2731 (1994).}\BibitemShut {Stop}%
\bibitem [{\citenamefont {Stoks}\ \emph {et~al.}(1994)\citenamefont {Stoks},
  \citenamefont {Klomp}, \citenamefont {Terheggen},\ and\ \citenamefont
  {de~Swart}}]{stoks1994}%
  \BibitemOpen
  \bibfield  {author} {\bibinfo {author} {\bibfnamefont {V.~G.~J.}\
  \bibnamefont {Stoks}}, \bibinfo {author} {\bibfnamefont {R.~A.~M.}\
  \bibnamefont {Klomp}}, \bibinfo {author} {\bibfnamefont {C.~P.~F.}\
  \bibnamefont {Terheggen}}, \ and\ \bibinfo {author} {\bibfnamefont {J.~J.}\
  \bibnamefont {de~Swart}},\ }\href {\doibase 10.1103/PhysRevC.49.2950}
  {\bibfield  {journal} {\bibinfo  {journal} {Phys. Rev. C}\ }\textbf {\bibinfo
  {volume} {49}},\ \bibinfo {pages} {2950} (\bibinfo {year}
  {1994})}\BibitemShut {NoStop}%
\bibitem [{\citenamefont {Wiringa}\ \emph {et~al.}(1995)\citenamefont
  {Wiringa}, \citenamefont {Stoks},\ and\ \citenamefont
  {Schiavilla}}]{wiringa1995}%
  \BibitemOpen
  \bibfield  {author} {\bibinfo {author} {\bibfnamefont {R.~B.}\ \bibnamefont
  {Wiringa}}, \bibinfo {author} {\bibfnamefont {V.~G.~J.}\ \bibnamefont
  {Stoks}}, \ and\ \bibinfo {author} {\bibfnamefont {R.}~\bibnamefont
  {Schiavilla}},\ }\href {\doibase 10.1103/PhysRevC.51.38} {\bibfield
  {journal} {\bibinfo  {journal} {Phys. Rev. C}\ }\textbf {\bibinfo {volume}
  {51}},\ \bibinfo {pages} {38} (\bibinfo {year} {1995})}\BibitemShut {NoStop}%
\bibitem [{\citenamefont {Machleidt}(2001)}]{machleidt2001}%
  \BibitemOpen
  \bibfield  {author} {\bibinfo {author} {\bibfnamefont {R.}~\bibnamefont
  {Machleidt}},\ }\href {\doibase 10.1103/PhysRevC.63.024001} {\bibfield
  {journal} {\bibinfo  {journal} {Phys. Rev. C}\ }\textbf {\bibinfo {volume}
  {63}},\ \bibinfo {pages} {024001} (\bibinfo {year} {2001})}\BibitemShut
  {NoStop}%
\bibitem [{\citenamefont {Pieper}\ \emph {et~al.}(2001)\citenamefont {Pieper},
  \citenamefont {Pandharipande}, \citenamefont {Wiringa},\ and\ \citenamefont
  {Carlson}}]{pieper2001b}%
  \BibitemOpen
  \bibfield  {author} {\bibinfo {author} {\bibfnamefont {S.~C.}\ \bibnamefont
  {Pieper}}, \bibinfo {author} {\bibfnamefont {V.~R.}\ \bibnamefont
  {Pandharipande}}, \bibinfo {author} {\bibfnamefont {R.~B.}\ \bibnamefont
  {Wiringa}}, \ and\ \bibinfo {author} {\bibfnamefont {J.}~\bibnamefont
  {Carlson}},\ }\href {\doibase 10.1103/PhysRevC.64.014001} {\bibfield
  {journal} {\bibinfo  {journal} {Phys. Rev. C}\ }\textbf {\bibinfo {volume}
  {64}},\ \bibinfo {pages} {014001} (\bibinfo {year} {2001})}\BibitemShut
  {NoStop}%
\bibitem [{\citenamefont {Entem}\ and\ \citenamefont
  {Machleidt}(2003)}]{entem2003}%
  \BibitemOpen
  \bibfield  {author} {\bibinfo {author} {\bibfnamefont {D.~R.}\ \bibnamefont
  {Entem}}\ and\ \bibinfo {author} {\bibfnamefont {R.}~\bibnamefont
  {Machleidt}},\ }\href {\doibase 10.1103/PhysRevC.68.041001} {\bibfield
  {journal} {\bibinfo  {journal} {Phys. Rev. C}\ }\textbf {\bibinfo {volume}
  {68}},\ \bibinfo {pages} {041001} (\bibinfo {year} {2003})}\BibitemShut
  {NoStop}%
\bibitem [{\citenamefont {Epelbaum}\ \emph {et~al.}(2005)\citenamefont
  {Epelbaum}, \citenamefont {Gl{\"o}ckle},\ and\ \citenamefont
  {Mei{\ss}ner}}]{epelbaum2005}%
  \BibitemOpen
  \bibfield  {author} {\bibinfo {author} {\bibfnamefont {E.}~\bibnamefont
  {Epelbaum}}, \bibinfo {author} {\bibfnamefont {W.}~\bibnamefont
  {Gl{\"o}ckle}}, \ and\ \bibinfo {author} {\bibfnamefont {U.-G.}\ \bibnamefont
  {Mei{\ss}ner}},\ }\href {\doibase 10.1016/j.nuclphysa.2004.09.107} {\bibfield
   {journal} {\bibinfo  {journal} {Nucl. Phys. A}\ }\textbf {\bibinfo {volume}
  {747}},\ \bibinfo {pages} {362 } (\bibinfo {year} {2005})}\BibitemShut
  {NoStop}%
\bibitem [{\citenamefont {Epelbaum}\ \emph
  {et~al.}(2015{\natexlab{a}})\citenamefont {Epelbaum}, \citenamefont {Krebs},\
  and\ \citenamefont {Mei{\ss}ner}}]{epelbaum2014}%
  \BibitemOpen
  \bibfield  {author} {\bibinfo {author} {\bibfnamefont {E.}~\bibnamefont
  {Epelbaum}}, \bibinfo {author} {\bibfnamefont {H.}~\bibnamefont {Krebs}}, \
  and\ \bibinfo {author} {\bibfnamefont {U.-G.}\ \bibnamefont {Mei{\ss}ner}},\
  }\href {\doibase 10.1140/epja/i2015-15053-8} {\bibfield  {journal} {\bibinfo
  {journal} {Eur. Phys. J. A}\ }\textbf {\bibinfo {volume} {51}},\ \bibinfo
  {eid} {53} (\bibinfo {year} {2015}{\natexlab{a}}),\
  10.1140/epja/i2015-15053-8}\BibitemShut {NoStop}%
\bibitem [{\citenamefont {Ekstr\"om}\ \emph {et~al.}(2013)\citenamefont
  {Ekstr\"om}, \citenamefont {Baardsen}, \citenamefont {Forss\'en},
  \citenamefont {Hagen}, \citenamefont {Hjorth-Jensen}, \citenamefont {Jansen},
  \citenamefont {Machleidt}, \citenamefont {Nazarewicz}, \citenamefont
  {Papenbrock}, \citenamefont {Sarich},\ and\ \citenamefont
  {Wild}}]{ekstrom2013}%
  \BibitemOpen
  \bibfield  {author} {\bibinfo {author} {\bibfnamefont {A.}~\bibnamefont
  {Ekstr\"om}}, \bibinfo {author} {\bibfnamefont {G.}~\bibnamefont {Baardsen}},
  \bibinfo {author} {\bibfnamefont {C.}~\bibnamefont {Forss\'en}}, \bibinfo
  {author} {\bibfnamefont {G.}~\bibnamefont {Hagen}}, \bibinfo {author}
  {\bibfnamefont {M.}~\bibnamefont {Hjorth-Jensen}}, \bibinfo {author}
  {\bibfnamefont {G.~R.}\ \bibnamefont {Jansen}}, \bibinfo {author}
  {\bibfnamefont {R.}~\bibnamefont {Machleidt}}, \bibinfo {author}
  {\bibfnamefont {W.}~\bibnamefont {Nazarewicz}}, \bibinfo {author}
  {\bibfnamefont {T.}~\bibnamefont {Papenbrock}}, \bibinfo {author}
  {\bibfnamefont {J.}~\bibnamefont {Sarich}}, \ and\ \bibinfo {author}
  {\bibfnamefont {S.~M.}\ \bibnamefont {Wild}},\ }\href {\doibase
  10.1103/PhysRevLett.110.192502} {\bibfield  {journal} {\bibinfo  {journal}
  {Phys. Rev. Lett.}\ }\textbf {\bibinfo {volume} {110}},\ \bibinfo {pages}
  {192502} (\bibinfo {year} {2013})}\BibitemShut {NoStop}%
\bibitem [{\citenamefont {Navarro~Perez}\ \emph {et~al.}(2013)\citenamefont
  {Navarro~Perez}, \citenamefont {Amaro},\ and\ \citenamefont
  {Ruiz~Arriola}}]{navarro2013}%
  \BibitemOpen
  \bibfield  {author} {\bibinfo {author} {\bibfnamefont {R.}~\bibnamefont
  {Navarro~Perez}}, \bibinfo {author} {\bibfnamefont {J.~E.}\ \bibnamefont
  {Amaro}}, \ and\ \bibinfo {author} {\bibfnamefont {E.}~\bibnamefont
  {Ruiz~Arriola}},\ }\href {\doibase 10.1016/j.physletb.2013.05.066} {\bibfield
   {journal} {\bibinfo  {journal} {Phys. Lett. B}\ }\textbf {\bibinfo {volume}
  {724}},\ \bibinfo {pages} {138} (\bibinfo {year} {2013})}\BibitemShut
  {NoStop}%
\bibitem [{\citenamefont {P\'erez}\ \emph {et~al.}(2013)\citenamefont
  {P\'erez}, \citenamefont {Amaro},\ and\ \citenamefont
  {Arriola}}]{navarro2013b}%
  \BibitemOpen
  \bibfield  {author} {\bibinfo {author} {\bibfnamefont {R.~N.}\ \bibnamefont
  {P\'erez}}, \bibinfo {author} {\bibfnamefont {J.~E.}\ \bibnamefont {Amaro}},
  \ and\ \bibinfo {author} {\bibfnamefont {E.~R.}\ \bibnamefont {Arriola}},\
  }\href {\doibase 10.1103/PhysRevC.88.064002} {\bibfield  {journal} {\bibinfo
  {journal} {Phys. Rev. C}\ }\textbf {\bibinfo {volume} {88}},\ \bibinfo
  {pages} {064002} (\bibinfo {year} {2013})}\BibitemShut {NoStop}%
\bibitem [{\citenamefont {Navarro~P\'erez}\ \emph
  {et~al.}(2014{\natexlab{a}})\citenamefont {Navarro~P\'erez}, \citenamefont
  {Amaro},\ and\ \citenamefont {Arriola}}]{navarro2014}%
  \BibitemOpen
  \bibfield  {author} {\bibinfo {author} {\bibfnamefont {R.}~\bibnamefont
  {Navarro~P\'erez}}, \bibinfo {author} {\bibfnamefont {J.~E.}\ \bibnamefont
  {Amaro}}, \ and\ \bibinfo {author} {\bibfnamefont {E.~R.}\ \bibnamefont
  {Arriola}},\ }\href {\doibase 10.1103/PhysRevC.89.024004} {\bibfield
  {journal} {\bibinfo  {journal} {Phys. Rev. C}\ }\textbf {\bibinfo {volume}
  {89}},\ \bibinfo {pages} {024004} (\bibinfo {year}
  {2014}{\natexlab{a}})}\BibitemShut {NoStop}%
\bibitem [{\citenamefont {Navarro~P\'erez}\ \emph
  {et~al.}(2014{\natexlab{b}})\citenamefont {Navarro~P\'erez}, \citenamefont
  {Amaro},\ and\ \citenamefont {Ruiz~Arriola}}]{navarro2014b}%
  \BibitemOpen
  \bibfield  {author} {\bibinfo {author} {\bibfnamefont {R.}~\bibnamefont
  {Navarro~P\'erez}}, \bibinfo {author} {\bibfnamefont {J.~E.}\ \bibnamefont
  {Amaro}}, \ and\ \bibinfo {author} {\bibfnamefont {E.}~\bibnamefont
  {Ruiz~Arriola}},\ }\href {\doibase 10.1103/PhysRevC.89.064006} {\bibfield
  {journal} {\bibinfo  {journal} {Phys. Rev. C}\ }\textbf {\bibinfo {volume}
  {89}},\ \bibinfo {pages} {064006} (\bibinfo {year}
  {2014}{\natexlab{b}})}\BibitemShut {NoStop}%
\bibitem [{\citenamefont {P\'erez}\ \emph
  {et~al.}(2015{\natexlab{a}})\citenamefont {P\'erez}, \citenamefont {Amaro},\
  and\ \citenamefont {Arriola}}]{navarro2015b}%
  \BibitemOpen
  \bibfield  {author} {\bibinfo {author} {\bibfnamefont {R.~N.}\ \bibnamefont
  {P\'erez}}, \bibinfo {author} {\bibfnamefont {J.~E.}\ \bibnamefont {Amaro}},
  \ and\ \bibinfo {author} {\bibfnamefont {E.~R.}\ \bibnamefont {Arriola}},\
  }\href {\doibase 10.1103/PhysRevC.91.054002} {\bibfield  {journal} {\bibinfo
  {journal} {Phys. Rev. C}\ }\textbf {\bibinfo {volume} {91}},\ \bibinfo
  {pages} {054002} (\bibinfo {year} {2015}{\natexlab{a}})}\BibitemShut
  {NoStop}%
\bibitem [{\citenamefont {Ekstr{\"o}m}\ \emph {et~al.}(2015)\citenamefont
  {Ekstr{\"o}m}, \citenamefont {Carlsson}, \citenamefont {Wendt}, \citenamefont
  {Forss{\'e}n}, \citenamefont {Jensen}, \citenamefont {Machleidt},\ and\
  \citenamefont {Wild}}]{ekstrom2015}%
  \BibitemOpen
  \bibfield  {author} {\bibinfo {author} {\bibfnamefont {A.}~\bibnamefont
  {Ekstr{\"o}m}}, \bibinfo {author} {\bibfnamefont {B.~D.}\ \bibnamefont
  {Carlsson}}, \bibinfo {author} {\bibfnamefont {K.~A.}\ \bibnamefont {Wendt}},
  \bibinfo {author} {\bibfnamefont {C.}~\bibnamefont {Forss{\'e}n}}, \bibinfo
  {author} {\bibfnamefont {M.~H.}\ \bibnamefont {Jensen}}, \bibinfo {author}
  {\bibfnamefont {R.}~\bibnamefont {Machleidt}}, \ and\ \bibinfo {author}
  {\bibfnamefont {S.~M.}\ \bibnamefont {Wild}},\ }\href {\doibase
  10.1088/0954-3899/42/3/034003} {\bibfield  {journal} {\bibinfo  {journal} {J.
  Phys. G}\ }\textbf {\bibinfo {volume} {42}},\ \bibinfo {pages} {034003}
  (\bibinfo {year} {2015})}\BibitemShut {NoStop}%
\bibitem [{\citenamefont {P\'erez}\ \emph
  {et~al.}(2015{\natexlab{b}})\citenamefont {P\'erez}, \citenamefont {Amaro},\
  and\ \citenamefont {Arriola}}]{navarro2015}%
  \BibitemOpen
  \bibfield  {author} {\bibinfo {author} {\bibfnamefont {R.~N.}\ \bibnamefont
  {P\'erez}}, \bibinfo {author} {\bibfnamefont {J.~E.}\ \bibnamefont {Amaro}},
  \ and\ \bibinfo {author} {\bibfnamefont {E.~R.}\ \bibnamefont {Arriola}},\
  }\href {\doibase 10.1088/0954-3899/42/3/034013} {\bibfield  {journal}
  {\bibinfo  {journal} {J. Phys. G}\ }\textbf {\bibinfo {volume} {42}},\
  \bibinfo {pages} {034013} (\bibinfo {year} {2015}{\natexlab{b}})}\BibitemShut
  {NoStop}%
\bibitem [{\citenamefont {Navr\'atil}\ \emph {et~al.}(2007)\citenamefont
  {Navr\'atil}, \citenamefont {Gueorguiev}, \citenamefont {Vary}, \citenamefont
  {Ormand},\ and\ \citenamefont {Nogga}}]{navratil2007}%
  \BibitemOpen
  \bibfield  {author} {\bibinfo {author} {\bibfnamefont {P.}~\bibnamefont
  {Navr\'atil}}, \bibinfo {author} {\bibfnamefont {V.~G.}\ \bibnamefont
  {Gueorguiev}}, \bibinfo {author} {\bibfnamefont {J.~P.}\ \bibnamefont
  {Vary}}, \bibinfo {author} {\bibfnamefont {W.~E.}\ \bibnamefont {Ormand}}, \
  and\ \bibinfo {author} {\bibfnamefont {A.}~\bibnamefont {Nogga}},\ }\href
  {\doibase 10.1103/PhysRevLett.99.042501} {\bibfield  {journal} {\bibinfo
  {journal} {Phys. Rev. Lett.}\ }\textbf {\bibinfo {volume} {99}},\ \bibinfo
  {pages} {042501} (\bibinfo {year} {2007})}\BibitemShut {NoStop}%
\bibitem [{\citenamefont {Hammer}\ \emph {et~al.}(2013)\citenamefont {Hammer},
  \citenamefont {Nogga},\ and\ \citenamefont {Schwenk}}]{hammer2013}%
  \BibitemOpen
  \bibfield  {author} {\bibinfo {author} {\bibfnamefont {H.-W.}\ \bibnamefont
  {Hammer}}, \bibinfo {author} {\bibfnamefont {A.}~\bibnamefont {Nogga}}, \
  and\ \bibinfo {author} {\bibfnamefont {A.}~\bibnamefont {Schwenk}},\ }\href
  {\doibase 10.1103/RevModPhys.85.197} {\bibfield  {journal} {\bibinfo
  {journal} {Rev. Mod. Phys.}\ }\textbf {\bibinfo {volume} {85}},\ \bibinfo
  {pages} {197} (\bibinfo {year} {2013})}\BibitemShut {NoStop}%
\bibitem [{\citenamefont {Weinberg}(1979)}]{weinberg1979}%
  \BibitemOpen
  \bibfield  {author} {\bibinfo {author} {\bibfnamefont {S.}~\bibnamefont
  {Weinberg}},\ }\href {\doibase 10.1016/0378-4371(79)90223-1} {\bibfield
  {journal} {\bibinfo  {journal} {Physica A}\ }\textbf {\bibinfo {volume}
  {96}},\ \bibinfo {pages} {327 } (\bibinfo {year} {1979})}\BibitemShut
  {NoStop}%
\bibitem [{\citenamefont {Weinberg}(1990)}]{weinberg1990}%
  \BibitemOpen
  \bibfield  {author} {\bibinfo {author} {\bibfnamefont {S.}~\bibnamefont
  {Weinberg}},\ }\href {\doibase 10.1016/0370-2693(90)90938-3} {\bibfield
  {journal} {\bibinfo  {journal} {Phys. Lett. B}\ }\textbf {\bibinfo {volume}
  {251}},\ \bibinfo {pages} {288 } (\bibinfo {year} {1990})}\BibitemShut
  {NoStop}%
\bibitem [{\citenamefont {Entem}\ \emph {et~al.}(2015)\citenamefont {Entem},
  \citenamefont {Kaiser}, \citenamefont {Machleidt},\ and\ \citenamefont
  {Nosyk}}]{machleidt2015}%
  \BibitemOpen
  \bibfield  {author} {\bibinfo {author} {\bibfnamefont {D.~R.}\ \bibnamefont
  {Entem}}, \bibinfo {author} {\bibfnamefont {N.}~\bibnamefont {Kaiser}},
  \bibinfo {author} {\bibfnamefont {R.}~\bibnamefont {Machleidt}}, \ and\
  \bibinfo {author} {\bibfnamefont {Y.}~\bibnamefont {Nosyk}},\ }\href
  {\doibase 10.1103/PhysRevC.91.014002} {\bibfield  {journal} {\bibinfo
  {journal} {Phys. Rev. C}\ }\textbf {\bibinfo {volume} {91}},\ \bibinfo
  {pages} {014002} (\bibinfo {year} {2015})}\BibitemShut {NoStop}%
\bibitem [{\citenamefont {Epelbaum}\ \emph
  {et~al.}(2015{\natexlab{b}})\citenamefont {Epelbaum}, \citenamefont {Krebs},\
  and\ \citenamefont {Mei\ss{}ner}}]{epelbaum2015}%
  \BibitemOpen
  \bibfield  {author} {\bibinfo {author} {\bibfnamefont {E.}~\bibnamefont
  {Epelbaum}}, \bibinfo {author} {\bibfnamefont {H.}~\bibnamefont {Krebs}}, \
  and\ \bibinfo {author} {\bibfnamefont {U.-G.}\ \bibnamefont {Mei\ss{}ner}},\
  }\href {\doibase 10.1103/PhysRevLett.115.122301} {\bibfield  {journal}
  {\bibinfo  {journal} {Phys. Rev. Lett.}\ }\textbf {\bibinfo {volume} {115}},\
  \bibinfo {pages} {122301} (\bibinfo {year} {2015}{\natexlab{b}})}\BibitemShut
  {NoStop}%
\bibitem [{\citenamefont {Epelbaum}\ \emph {et~al.}(2002)\citenamefont
  {Epelbaum}, \citenamefont {Nogga}, \citenamefont {Gl\"ockle}, \citenamefont
  {Kamada}, \citenamefont {Mei\ss{}ner},\ and\ \citenamefont
  {Wita\l{}a}}]{epelbaum2002}%
  \BibitemOpen
  \bibfield  {author} {\bibinfo {author} {\bibfnamefont {E.}~\bibnamefont
  {Epelbaum}}, \bibinfo {author} {\bibfnamefont {A.}~\bibnamefont {Nogga}},
  \bibinfo {author} {\bibfnamefont {W.}~\bibnamefont {Gl\"ockle}}, \bibinfo
  {author} {\bibfnamefont {H.}~\bibnamefont {Kamada}}, \bibinfo {author}
  {\bibfnamefont {U.-G.}\ \bibnamefont {Mei\ss{}ner}}, \ and\ \bibinfo {author}
  {\bibfnamefont {H.}~\bibnamefont {Wita\l{}a}},\ }\href {\doibase
  10.1103/PhysRevC.66.064001} {\bibfield  {journal} {\bibinfo  {journal} {Phys.
  Rev. C}\ }\textbf {\bibinfo {volume} {66}},\ \bibinfo {pages} {064001}
  (\bibinfo {year} {2002})}\BibitemShut {NoStop}%
\bibitem [{\citenamefont {Hebeler}\ \emph {et~al.}(2015)\citenamefont
  {Hebeler}, \citenamefont {Krebs}, \citenamefont {Epelbaum}, \citenamefont
  {Golak},\ and\ \citenamefont {Skibi\ifmmode~\acute{n}\else
  \'{n}\fi{}ski}}]{hebeler2015}%
  \BibitemOpen
  \bibfield  {author} {\bibinfo {author} {\bibfnamefont {K.}~\bibnamefont
  {Hebeler}}, \bibinfo {author} {\bibfnamefont {H.}~\bibnamefont {Krebs}},
  \bibinfo {author} {\bibfnamefont {E.}~\bibnamefont {Epelbaum}}, \bibinfo
  {author} {\bibfnamefont {J.}~\bibnamefont {Golak}}, \ and\ \bibinfo {author}
  {\bibfnamefont {R.}~\bibnamefont {Skibi\ifmmode~\acute{n}\else
  \'{n}\fi{}ski}},\ }\href {\doibase 10.1103/PhysRevC.91.044001} {\bibfield
  {journal} {\bibinfo  {journal} {Phys. Rev. C}\ }\textbf {\bibinfo {volume}
  {91}},\ \bibinfo {pages} {044001} (\bibinfo {year} {2015})}\BibitemShut
  {NoStop}%
\bibitem [{\citenamefont {Liu}\ \emph {et~al.}(2010)\citenamefont {Liu},
  \citenamefont {Mendenhall}, \citenamefont {Holley}, \citenamefont {Back},
  \citenamefont {Bowles}, \citenamefont {Broussard}, \citenamefont {Carr},
  \citenamefont {Clayton}, \citenamefont {Currie}, \citenamefont {Filippone},
  \citenamefont {Garc\'ia}, \citenamefont {Geltenbort}, \citenamefont
  {Hickerson}, \citenamefont {Hoagland}, \citenamefont {Hogan}, \citenamefont
  {Hona}, \citenamefont {Ito}, \citenamefont {Liu}, \citenamefont {Makela},
  \citenamefont {Mammei}, \citenamefont {Martin}, \citenamefont {Melconian},
  \citenamefont {Morris}, \citenamefont {Pattie}, \citenamefont
  {P\'erez~Galv\'an}, \citenamefont {Pitt}, \citenamefont {Plaster},
  \citenamefont {Ramsey}, \citenamefont {Rios}, \citenamefont {Russell},
  \citenamefont {Saunders}, \citenamefont {Seestrom}, \citenamefont {Sondheim},
  \citenamefont {Tatar}, \citenamefont {Vogelaar}, \citenamefont {VornDick},
  \citenamefont {Wrede}, \citenamefont {Yan},\ and\ \citenamefont
  {Young}}]{Liu2010}%
  \BibitemOpen
  \bibfield  {author} {\bibinfo {author} {\bibfnamefont {J.}~\bibnamefont
  {Liu}}, \bibinfo {author} {\bibfnamefont {M.}~\bibnamefont {Mendenhall}},
  \bibinfo {author} {\bibfnamefont {A.}~\bibnamefont {Holley}}, \bibinfo
  {author} {\bibfnamefont {H.}~\bibnamefont {Back}}, \bibinfo {author}
  {\bibfnamefont {T.}~\bibnamefont {Bowles}}, \bibinfo {author} {\bibfnamefont
  {L.}~\bibnamefont {Broussard}}, \bibinfo {author} {\bibfnamefont
  {R.}~\bibnamefont {Carr}}, \bibinfo {author} {\bibfnamefont {S.}~\bibnamefont
  {Clayton}}, \bibinfo {author} {\bibfnamefont {S.}~\bibnamefont {Currie}},
  \bibinfo {author} {\bibfnamefont {B.}~\bibnamefont {Filippone}}, \bibinfo
  {author} {\bibfnamefont {A.}~\bibnamefont {Garc\'ia}}, \bibinfo {author}
  {\bibfnamefont {P.}~\bibnamefont {Geltenbort}}, \bibinfo {author}
  {\bibfnamefont {K.}~\bibnamefont {Hickerson}}, \bibinfo {author}
  {\bibfnamefont {J.}~\bibnamefont {Hoagland}}, \bibinfo {author}
  {\bibfnamefont {G.}~\bibnamefont {Hogan}}, \bibinfo {author} {\bibfnamefont
  {B.}~\bibnamefont {Hona}}, \bibinfo {author} {\bibfnamefont {T.}~\bibnamefont
  {Ito}}, \bibinfo {author} {\bibfnamefont {C.-Y.}\ \bibnamefont {Liu}},
  \bibinfo {author} {\bibfnamefont {M.}~\bibnamefont {Makela}}, \bibinfo
  {author} {\bibfnamefont {R.}~\bibnamefont {Mammei}}, \bibinfo {author}
  {\bibfnamefont {J.}~\bibnamefont {Martin}}, \bibinfo {author} {\bibfnamefont
  {D.}~\bibnamefont {Melconian}}, \bibinfo {author} {\bibfnamefont
  {C.}~\bibnamefont {Morris}}, \bibinfo {author} {\bibfnamefont
  {R.}~\bibnamefont {Pattie}}, \bibinfo {author} {\bibfnamefont
  {A.}~\bibnamefont {P\'erez~Galv\'an}}, \bibinfo {author} {\bibfnamefont
  {M.}~\bibnamefont {Pitt}}, \bibinfo {author} {\bibfnamefont {B.}~\bibnamefont
  {Plaster}}, \bibinfo {author} {\bibfnamefont {J.}~\bibnamefont {Ramsey}},
  \bibinfo {author} {\bibfnamefont {R.}~\bibnamefont {Rios}}, \bibinfo {author}
  {\bibfnamefont {R.}~\bibnamefont {Russell}}, \bibinfo {author} {\bibfnamefont
  {A.}~\bibnamefont {Saunders}}, \bibinfo {author} {\bibfnamefont
  {S.}~\bibnamefont {Seestrom}}, \bibinfo {author} {\bibfnamefont
  {W.}~\bibnamefont {Sondheim}}, \bibinfo {author} {\bibfnamefont
  {E.}~\bibnamefont {Tatar}}, \bibinfo {author} {\bibfnamefont
  {R.}~\bibnamefont {Vogelaar}}, \bibinfo {author} {\bibfnamefont
  {B.}~\bibnamefont {VornDick}}, \bibinfo {author} {\bibfnamefont
  {C.}~\bibnamefont {Wrede}}, \bibinfo {author} {\bibfnamefont
  {H.}~\bibnamefont {Yan}}, \ and\ \bibinfo {author} {\bibfnamefont
  {A.}~\bibnamefont {Young}} (\bibinfo {collaboration} {UCNA Collaboration}),\
  }\href {\doibase 10.1103/PhysRevLett.105.181803} {\bibfield  {journal}
  {\bibinfo  {journal} {Phys. Rev. Lett.}\ }\textbf {\bibinfo {volume} {105}},\
  \bibinfo {pages} {181803} (\bibinfo {year} {2010})}\BibitemShut {NoStop}%
\bibitem [{\citenamefont {Mohr}\ \emph {et~al.}(2012)\citenamefont {Mohr},
  \citenamefont {Taylor},\ and\ \citenamefont {Newell}}]{mohr2012}%
  \BibitemOpen
  \bibfield  {author} {\bibinfo {author} {\bibfnamefont {P.~J.}\ \bibnamefont
  {Mohr}}, \bibinfo {author} {\bibfnamefont {B.~N.}\ \bibnamefont {Taylor}}, \
  and\ \bibinfo {author} {\bibfnamefont {D.~B.}\ \bibnamefont {Newell}},\
  }\href {\doibase 10.1103/RevModPhys.84.1527} {\bibfield  {journal} {\bibinfo
  {journal} {Rev. Mod. Phys.}\ }\textbf {\bibinfo {volume} {84}},\ \bibinfo
  {pages} {1527} (\bibinfo {year} {2012})}\BibitemShut {NoStop}%
\bibitem [{\citenamefont {Beringer}\ \emph {et~al.}(2012)\citenamefont
  {Beringer} \emph {et~al.}}]{beringer2012}%
  \BibitemOpen
  \bibfield  {author} {\bibinfo {author} {\bibfnamefont {J.}~\bibnamefont
  {Beringer}} \emph {et~al.},\ }\href {\doibase 10.1103/PhysRevD.86.010001}
  {\bibfield  {journal} {\bibinfo  {journal} {Phys. Rev. D}\ }\textbf {\bibinfo
  {volume} {86}},\ \bibinfo {pages} {010001} (\bibinfo {year}
  {2012})}\BibitemShut {NoStop}%
\bibitem [{\citenamefont {Austen}\ and\ \citenamefont
  {de~Swart}(1983)}]{austen1983}%
  \BibitemOpen
  \bibfield  {author} {\bibinfo {author} {\bibfnamefont {G.~J.~M.}\
  \bibnamefont {Austen}}\ and\ \bibinfo {author} {\bibfnamefont {J.~J.}\
  \bibnamefont {de~Swart}},\ }\href {\doibase 10.1103/PhysRevLett.50.2039}
  {\bibfield  {journal} {\bibinfo  {journal} {Phys. Rev. Lett.}\ }\textbf
  {\bibinfo {volume} {50}},\ \bibinfo {pages} {2039} (\bibinfo {year}
  {1983})}\BibitemShut {NoStop}%
\bibitem [{\citenamefont {Durand}(1957)}]{durand1957}%
  \BibitemOpen
  \bibfield  {author} {\bibinfo {author} {\bibfnamefont {L.}~\bibnamefont
  {Durand}},\ }\href {\doibase 10.1103/PhysRev.108.1597} {\bibfield  {journal}
  {\bibinfo  {journal} {Phys. Rev.}\ }\textbf {\bibinfo {volume} {108}},\
  \bibinfo {pages} {1597} (\bibinfo {year} {1957})}\BibitemShut {NoStop}%
\bibitem [{\citenamefont {Stoks}\ and\ \citenamefont
  {de~Swart}(1990)}]{stoks1990}%
  \BibitemOpen
  \bibfield  {author} {\bibinfo {author} {\bibfnamefont {V.~G.~J.}\
  \bibnamefont {Stoks}}\ and\ \bibinfo {author} {\bibfnamefont {J.~J.}\
  \bibnamefont {de~Swart}},\ }\href {\doibase 10.1103/PhysRevC.42.1235}
  {\bibfield  {journal} {\bibinfo  {journal} {Phys. Rev. C}\ }\textbf {\bibinfo
  {volume} {42}},\ \bibinfo {pages} {1235} (\bibinfo {year}
  {1990})}\BibitemShut {NoStop}%
\bibitem [{\citenamefont {Weinberg}(1991)}]{weinberg1991}%
  \BibitemOpen
  \bibfield  {author} {\bibinfo {author} {\bibfnamefont {S.}~\bibnamefont
  {Weinberg}},\ }\href {\doibase 10.1016/0550-3213(91)90231-L} {\bibfield
  {journal} {\bibinfo  {journal} {Nucl. Phys. B}\ }\textbf {\bibinfo {volume}
  {363}},\ \bibinfo {pages} {3 } (\bibinfo {year} {1991})}\BibitemShut
  {NoStop}%
\bibitem [{\citenamefont {Epelbaum}(2006)}]{epelbaum2006}%
  \BibitemOpen
  \bibfield  {author} {\bibinfo {author} {\bibfnamefont {E.}~\bibnamefont
  {Epelbaum}},\ }\href {\doibase 10.1016/j.ppnp.2005.09.002} {\bibfield
  {journal} {\bibinfo  {journal} {Prog. Part. Nucl. Phys.}\ }\textbf {\bibinfo
  {volume} {57}},\ \bibinfo {pages} {654 } (\bibinfo {year}
  {2006})}\BibitemShut {NoStop}%
\bibitem [{\citenamefont {Ord\'o\~nez}\ \emph {et~al.}(1994)\citenamefont
  {Ord\'o\~nez}, \citenamefont {Ray},\ and\ \citenamefont {van
  Kolck}}]{ordonez1994}%
  \BibitemOpen
  \bibfield  {author} {\bibinfo {author} {\bibfnamefont {C.}~\bibnamefont
  {Ord\'o\~nez}}, \bibinfo {author} {\bibfnamefont {L.}~\bibnamefont {Ray}}, \
  and\ \bibinfo {author} {\bibfnamefont {U.}~\bibnamefont {van Kolck}},\ }\href
  {\doibase 10.1103/PhysRevLett.72.1982} {\bibfield  {journal} {\bibinfo
  {journal} {Phys. Rev. Lett.}\ }\textbf {\bibinfo {volume} {72}},\ \bibinfo
  {pages} {1982} (\bibinfo {year} {1994})}\BibitemShut {NoStop}%
\bibitem [{\citenamefont {Brown}\ \emph {et~al.}(1969)\citenamefont {Brown},
  \citenamefont {Jackson},\ and\ \citenamefont {Kuo}}]{brown1969}%
  \BibitemOpen
  \bibfield  {author} {\bibinfo {author} {\bibfnamefont {G.~E.}\ \bibnamefont
  {Brown}}, \bibinfo {author} {\bibfnamefont {A.~D.}\ \bibnamefont {Jackson}},
  \ and\ \bibinfo {author} {\bibfnamefont {T.~T.~S.}\ \bibnamefont {Kuo}},\
  }\href {\doibase 10.1016/0375-9474(69)90549-1} {\bibfield  {journal}
  {\bibinfo  {journal} {Nucl. Phys. A}\ }\textbf {\bibinfo {volume} {133}},\
  \bibinfo {pages} {481 } (\bibinfo {year} {1969})}\BibitemShut {NoStop}%
\bibitem [{\citenamefont {Nogga}\ \emph {et~al.}(2005)\citenamefont {Nogga},
  \citenamefont {Timmermans},\ and\ \citenamefont {Kolck}}]{nogga2005}%
  \BibitemOpen
  \bibfield  {author} {\bibinfo {author} {\bibfnamefont {A.}~\bibnamefont
  {Nogga}}, \bibinfo {author} {\bibfnamefont {R.~G.~E.}\ \bibnamefont
  {Timmermans}}, \ and\ \bibinfo {author} {\bibfnamefont {U.~v.}\ \bibnamefont
  {Kolck}},\ }\href {\doibase 10.1103/PhysRevC.72.054006} {\bibfield  {journal}
  {\bibinfo  {journal} {Phys. Rev. C}\ }\textbf {\bibinfo {volume} {72}},\
  \bibinfo {pages} {054006} (\bibinfo {year} {2005})}\BibitemShut {NoStop}%
\bibitem [{\citenamefont {Valderrama}\ and\ \citenamefont
  {Phillips}(2015)}]{valderrama2015a}%
  \BibitemOpen
  \bibfield  {author} {\bibinfo {author} {\bibfnamefont {M.~P.}\ \bibnamefont
  {Valderrama}}\ and\ \bibinfo {author} {\bibfnamefont {D.~R.}\ \bibnamefont
  {Phillips}},\ }\href {\doibase 10.1103/PhysRevLett.114.082502} {\bibfield
  {journal} {\bibinfo  {journal} {Phys. Rev. Lett.}\ }\textbf {\bibinfo
  {volume} {114}},\ \bibinfo {pages} {082502} (\bibinfo {year}
  {2015})}\BibitemShut {NoStop}%
\bibitem [{\citenamefont {Bystricky}\ \emph {et~al.}(1978)\citenamefont
  {Bystricky}, \citenamefont {Lehar},\ and\ \citenamefont
  {Winternitz}}]{bystricky1978}%
  \BibitemOpen
  \bibfield  {author} {\bibinfo {author} {\bibfnamefont {J.}~\bibnamefont
  {Bystricky}}, \bibinfo {author} {\bibfnamefont {F.}~\bibnamefont {Lehar}}, \
  and\ \bibinfo {author} {\bibfnamefont {P.}~\bibnamefont {Winternitz}},\
  }\href {\doibase 10.1051/jphys:019780039010100} {\bibfield  {journal}
  {\bibinfo  {journal} {J. Phys. France}\ }\textbf {\bibinfo {volume} {39}},\
  \bibinfo {pages} {1} (\bibinfo {year} {1978})}\BibitemShut {NoStop}%
\bibitem [{\citenamefont {La~France}\ and\ \citenamefont
  {Winternitz}(1980)}]{lafrance1980}%
  \BibitemOpen
  \bibfield  {author} {\bibinfo {author} {\bibfnamefont {P.}~\bibnamefont
  {La~France}}\ and\ \bibinfo {author} {\bibfnamefont {P.}~\bibnamefont
  {Winternitz}},\ }\href {\doibase 10.1051/jphys:0198000410120139100}
  {\bibfield  {journal} {\bibinfo  {journal} {J. Phys. France}\ }\textbf
  {\bibinfo {volume} {41}},\ \bibinfo {pages} {1391} (\bibinfo {year}
  {1980})}\BibitemShut {NoStop}%
\bibitem [{\citenamefont {Stapp}\ \emph {et~al.}(1957)\citenamefont {Stapp},
  \citenamefont {Ypsilantis},\ and\ \citenamefont {Metropolis}}]{stapp1957}%
  \BibitemOpen
  \bibfield  {author} {\bibinfo {author} {\bibfnamefont {H.~P.}\ \bibnamefont
  {Stapp}}, \bibinfo {author} {\bibfnamefont {T.~J.}\ \bibnamefont
  {Ypsilantis}}, \ and\ \bibinfo {author} {\bibfnamefont {N.}~\bibnamefont
  {Metropolis}},\ }\href {\doibase 10.1103/PhysRev.105.302} {\bibfield
  {journal} {\bibinfo  {journal} {Phys. Rev.}\ }\textbf {\bibinfo {volume}
  {105}},\ \bibinfo {pages} {302} (\bibinfo {year} {1957})}\BibitemShut
  {NoStop}%
\bibitem [{\citenamefont {Gersten}(1977)}]{gersten1977}%
  \BibitemOpen
  \bibfield  {author} {\bibinfo {author} {\bibfnamefont {A.}~\bibnamefont
  {Gersten}},\ }\href {\doibase 10.1016/0375-9474(77)90447-X} {\bibfield
  {journal} {\bibinfo  {journal} {Nucl. Phys. A}\ }\textbf {\bibinfo {volume}
  {290}},\ \bibinfo {pages} {445 } (\bibinfo {year} {1977})}\BibitemShut
  {NoStop}%
\bibitem [{\citenamefont {Vincent}\ and\ \citenamefont
  {Phatak}(1974)}]{vincent1974}%
  \BibitemOpen
  \bibfield  {author} {\bibinfo {author} {\bibfnamefont {C.~M.}\ \bibnamefont
  {Vincent}}\ and\ \bibinfo {author} {\bibfnamefont {S.~C.}\ \bibnamefont
  {Phatak}},\ }\href {\doibase 10.1103/PhysRevC.10.391} {\bibfield  {journal}
  {\bibinfo  {journal} {Phys. Rev. C}\ }\textbf {\bibinfo {volume} {10}},\
  \bibinfo {pages} {391} (\bibinfo {year} {1974})}\BibitemShut {NoStop}%
\bibitem [{\citenamefont {Bergervoet}\ \emph {et~al.}(1988)\citenamefont
  {Bergervoet}, \citenamefont {van Campen}, \citenamefont {van~der Sanden},\
  and\ \citenamefont {de~Swart}}]{bergervoet1988}%
  \BibitemOpen
  \bibfield  {author} {\bibinfo {author} {\bibfnamefont {J.~R.}\ \bibnamefont
  {Bergervoet}}, \bibinfo {author} {\bibfnamefont {P.~C.}\ \bibnamefont {van
  Campen}}, \bibinfo {author} {\bibfnamefont {W.~A.}\ \bibnamefont {van~der
  Sanden}}, \ and\ \bibinfo {author} {\bibfnamefont {J.~J.}\ \bibnamefont
  {de~Swart}},\ }\href {\doibase 10.1103/PhysRevC.38.15} {\bibfield  {journal}
  {\bibinfo  {journal} {Phys. Rev. C}\ }\textbf {\bibinfo {volume} {38}},\
  \bibinfo {pages} {15} (\bibinfo {year} {1988})}\BibitemShut {NoStop}%
\bibitem [{\citenamefont {Calogero}(1967)}]{calogero1967}%
  \BibitemOpen
  \bibfield  {author} {\bibinfo {author} {\bibfnamefont {F.}~\bibnamefont
  {Calogero}},\ }\href@noop {} {\emph {\bibinfo {title} {Variable Phase
  Approach to Potential Scattering}}}\ (\bibinfo  {publisher} {Academic
  Press},\ \bibinfo {year} {1967})\BibitemShut {NoStop}%
\bibitem [{\citenamefont {Erkelenz}\ \emph {et~al.}(1971)\citenamefont
  {Erkelenz}, \citenamefont {Alzetta},\ and\ \citenamefont
  {Holinde}}]{erkelenz1971}%
  \BibitemOpen
  \bibfield  {author} {\bibinfo {author} {\bibfnamefont {K.}~\bibnamefont
  {Erkelenz}}, \bibinfo {author} {\bibfnamefont {R.}~\bibnamefont {Alzetta}}, \
  and\ \bibinfo {author} {\bibfnamefont {K.}~\bibnamefont {Holinde}},\ }\href
  {\doibase 10.1016/0375-9474(71)90279-X} {\bibfield  {journal} {\bibinfo
  {journal} {Nucl. Phys. A}\ }\textbf {\bibinfo {volume} {176}},\ \bibinfo
  {pages} {413 } (\bibinfo {year} {1971})}\BibitemShut {NoStop}%
\bibitem [{\citenamefont {Wendt}\ \emph {et~al.}()\citenamefont {Wendt},
  \citenamefont {Carlsson},\ and\ \citenamefont {Ekstr\"om}}]{wendt2014arxiv}%
  \BibitemOpen
  \bibfield  {author} {\bibinfo {author} {\bibfnamefont {K.}~\bibnamefont
  {Wendt}}, \bibinfo {author} {\bibfnamefont {B.}~\bibnamefont {Carlsson}}, \
  and\ \bibinfo {author} {\bibfnamefont {A.}~\bibnamefont {Ekstr\"om}},\
  }\href@noop {} {\ }\Eprint {http://arxiv.org/abs/1410.0646} {arXiv:1410.0646
  [nucl-th]} \BibitemShut {NoStop}%
%%CITATION = ARXIV:1410.0646;%%
\bibitem [{\citenamefont {Krebs}\ \emph {et~al.}(2012)\citenamefont {Krebs},
  \citenamefont {Gasparyan},\ and\ \citenamefont {Epelbaum}}]{krebs2012}%
  \BibitemOpen
  \bibfield  {author} {\bibinfo {author} {\bibfnamefont {H.}~\bibnamefont
  {Krebs}}, \bibinfo {author} {\bibfnamefont {A.}~\bibnamefont {Gasparyan}}, \
  and\ \bibinfo {author} {\bibfnamefont {E.}~\bibnamefont {Epelbaum}},\ }\href
  {\doibase 10.1103/PhysRevC.85.054006} {\bibfield  {journal} {\bibinfo
  {journal} {Phys. Rev. C}\ }\textbf {\bibinfo {volume} {85}},\ \bibinfo
  {pages} {054006} (\bibinfo {year} {2012})}\BibitemShut {NoStop}%
\bibitem [{\citenamefont {Tromborg}\ \emph {et~al.}(1978)\citenamefont
  {Tromborg}, \citenamefont {Waldenstr{\o}m},\ and\ \citenamefont
  {{\O}verb{\o}}}]{tromborg1978}%
  \BibitemOpen
  \bibfield  {author} {\bibinfo {author} {\bibfnamefont {B.}~\bibnamefont
  {Tromborg}}, \bibinfo {author} {\bibfnamefont {S.}~\bibnamefont
  {Waldenstr{\o}m}}, \ and\ \bibinfo {author} {\bibfnamefont {I.}~\bibnamefont
  {{\O}verb{\o}}},\ }\href {\doibase 10.5169/seals-114961} {\bibfield
  {journal} {\bibinfo  {journal} {Helv. Phys. Acta}\ }\textbf {\bibinfo
  {volume} {51}},\ \bibinfo {pages} {584} (\bibinfo {year} {1978})}\BibitemShut
  {NoStop}%
\bibitem [{\citenamefont {Tromborg}\ and\ \citenamefont
  {Hamilton}(1974)}]{tromborg1974}%
  \BibitemOpen
  \bibfield  {author} {\bibinfo {author} {\bibfnamefont {B.}~\bibnamefont
  {Tromborg}}\ and\ \bibinfo {author} {\bibfnamefont {J.}~\bibnamefont
  {Hamilton}},\ }\href {\doibase 10.1016/0550-3213(74)90538-0} {\bibfield
  {journal} {\bibinfo  {journal} {Nucl. Phys. B}\ }\textbf {\bibinfo {volume}
  {76}},\ \bibinfo {pages} {483 } (\bibinfo {year} {1974})}\BibitemShut
  {NoStop}%
\bibitem [{\citenamefont {Tromborg}\ \emph {et~al.}(1977)\citenamefont
  {Tromborg}, \citenamefont {Waldenstr\o{}m},\ and\ \citenamefont
  {\O{}verb\o{}}}]{tromborg1977}%
  \BibitemOpen
  \bibfield  {author} {\bibinfo {author} {\bibfnamefont {B.}~\bibnamefont
  {Tromborg}}, \bibinfo {author} {\bibfnamefont {S.}~\bibnamefont
  {Waldenstr\o{}m}}, \ and\ \bibinfo {author} {\bibfnamefont {I.}~\bibnamefont
  {\O{}verb\o{}}},\ }\href {\doibase 10.1103/PhysRevD.15.725} {\bibfield
  {journal} {\bibinfo  {journal} {Phys. Rev. D}\ }\textbf {\bibinfo {volume}
  {15}},\ \bibinfo {pages} {725} (\bibinfo {year} {1977})}\BibitemShut
  {NoStop}%
\bibitem [{\citenamefont {Bugg}(1973)}]{bugg1973}%
  \BibitemOpen
  \bibfield  {author} {\bibinfo {author} {\bibfnamefont {D.}~\bibnamefont
  {Bugg}},\ }\href {\doibase 10.1016/0550-3213(73)90591-9} {\bibfield
  {journal} {\bibinfo  {journal} {Nucl. Phys. B}\ }\textbf {\bibinfo {volume}
  {58}},\ \bibinfo {pages} {397 } (\bibinfo {year} {1973})}\BibitemShut
  {NoStop}%
\bibitem [{\citenamefont {Bethe}(1949)}]{bethe1949}%
  \BibitemOpen
  \bibfield  {author} {\bibinfo {author} {\bibfnamefont {H.~A.}\ \bibnamefont
  {Bethe}},\ }\href {\doibase 10.1103/PhysRev.76.38} {\bibfield  {journal}
  {\bibinfo  {journal} {Phys. Rev.}\ }\textbf {\bibinfo {volume} {76}},\
  \bibinfo {pages} {38} (\bibinfo {year} {1949})}\BibitemShut {NoStop}%
\bibitem [{\citenamefont {Navr\'atil}\ \emph {et~al.}(2000)\citenamefont
  {Navr\'atil}, \citenamefont {Kamuntavi\ifmmode~\check{c}\else
  \v{c}\fi{}ius},\ and\ \citenamefont {Barrett}}]{navratil2000}%
  \BibitemOpen
  \bibfield  {author} {\bibinfo {author} {\bibfnamefont {P.}~\bibnamefont
  {Navr\'atil}}, \bibinfo {author} {\bibfnamefont {G.~P.}\ \bibnamefont
  {Kamuntavi\ifmmode~\check{c}\else \v{c}\fi{}ius}}, \ and\ \bibinfo {author}
  {\bibfnamefont {B.~R.}\ \bibnamefont {Barrett}},\ }\href {\doibase
  10.1103/PhysRevC.61.044001} {\bibfield  {journal} {\bibinfo  {journal} {Phys.
  Rev. C}\ }\textbf {\bibinfo {volume} {61}},\ \bibinfo {pages} {044001}
  (\bibinfo {year} {2000})}\BibitemShut {NoStop}%
\bibitem [{\citenamefont {Kamuntavi\ifmmode~\check{c}\else \v{c}\fi{}ius}\
  \emph {et~al.}(1999)\citenamefont {Kamuntavi\ifmmode~\check{c}\else
  \v{c}\fi{}ius}, \citenamefont {Navr\'atil}, \citenamefont {Barrett},
  \citenamefont {Sapragonaite},\ and\ \citenamefont
  {Kalinauskas}}]{kamuntavicius1999}%
  \BibitemOpen
  \bibfield  {author} {\bibinfo {author} {\bibfnamefont {G.~P.}\ \bibnamefont
  {Kamuntavi\ifmmode~\check{c}\else \v{c}\fi{}ius}}, \bibinfo {author}
  {\bibfnamefont {P.}~\bibnamefont {Navr\'atil}}, \bibinfo {author}
  {\bibfnamefont {B.~R.}\ \bibnamefont {Barrett}}, \bibinfo {author}
  {\bibfnamefont {G.}~\bibnamefont {Sapragonaite}}, \ and\ \bibinfo {author}
  {\bibfnamefont {R.~K.}\ \bibnamefont {Kalinauskas}},\ }\href {\doibase
  10.1103/PhysRevC.60.044304} {\bibfield  {journal} {\bibinfo  {journal} {Phys.
  Rev. C}\ }\textbf {\bibinfo {volume} {60}},\ \bibinfo {pages} {044304}
  (\bibinfo {year} {1999})}\BibitemShut {NoStop}%
\bibitem [{\citenamefont {Friar}\ and\ \citenamefont
  {Negele}(1975)}]{friar1976}%
  \BibitemOpen
  \bibfield  {author} {\bibinfo {author} {\bibfnamefont {J.~L.}\ \bibnamefont
  {Friar}}\ and\ \bibinfo {author} {\bibfnamefont {J.~W.}\ \bibnamefont
  {Negele}},\ }\href@noop {} {\emph {\bibinfo {title} {Advances in Nucl.
  Phys.}}},\ edited by\ \bibinfo {editor} {\bibfnamefont {M.}~\bibnamefont
  {Baranger}}\ and\ \bibinfo {editor} {\bibfnamefont {E.}~\bibnamefont
  {Vogt}},\ Vol.~\bibinfo {volume} {8}\ (\bibinfo  {publisher} {Springer US,
  New York},\ \bibinfo {year} {1975})\ Chap.~\bibinfo {chapter} {3}, pp.\
  \bibinfo {pages} {219--376}\BibitemShut {NoStop}%
\bibitem [{\citenamefont {Jentschura}(2011)}]{jentschura2011}%
  \BibitemOpen
  \bibfield  {author} {\bibinfo {author} {\bibfnamefont {U.~D.}\ \bibnamefont
  {Jentschura}},\ }\href {\doibase 10.1140/epjd/e2010-10414-6} {\bibfield
  {journal} {\bibinfo  {journal} {Eur. Phys. J. D}\ }\textbf {\bibinfo {volume}
  {61}},\ \bibinfo {pages} {7} (\bibinfo {year} {2011})}\BibitemShut {NoStop}%
\bibitem [{\citenamefont {Angeli}\ and\ \citenamefont
  {Marinova}(2013)}]{angeli2013}%
  \BibitemOpen
  \bibfield  {author} {\bibinfo {author} {\bibfnamefont {I.}~\bibnamefont
  {Angeli}}\ and\ \bibinfo {author} {\bibfnamefont {K.}~\bibnamefont
  {Marinova}},\ }\href {\doibase 10.1016/j.adt.2011.12.006} {\bibfield
  {journal} {\bibinfo  {journal} {At. Data Nucl. Data Tables}\ }\textbf
  {\bibinfo {volume} {99}},\ \bibinfo {pages} {69 } (\bibinfo {year}
  {2013})}\BibitemShut {NoStop}%
\bibitem [{\citenamefont {Gazit}\ \emph {et~al.}(2009)\citenamefont {Gazit},
  \citenamefont {Quaglioni},\ and\ \citenamefont {Navr\'atil}}]{gazit2009}%
  \BibitemOpen
  \bibfield  {author} {\bibinfo {author} {\bibfnamefont {D.}~\bibnamefont
  {Gazit}}, \bibinfo {author} {\bibfnamefont {S.}~\bibnamefont {Quaglioni}}, \
  and\ \bibinfo {author} {\bibfnamefont {P.}~\bibnamefont {Navr\'atil}},\
  }\href {\doibase 10.1103/PhysRevLett.103.102502} {\bibfield  {journal}
  {\bibinfo  {journal} {Phys. Rev. Lett.}\ }\textbf {\bibinfo {volume} {103}},\
  \bibinfo {pages} {102502} (\bibinfo {year} {2009})}\BibitemShut {NoStop}%
\bibitem [{\citenamefont {Akulov}\ and\ \citenamefont
  {Mamyrin}(2005)}]{akulov2005}%
  \BibitemOpen
  \bibfield  {author} {\bibinfo {author} {\bibfnamefont {Y.}~\bibnamefont
  {Akulov}}\ and\ \bibinfo {author} {\bibfnamefont {B.}~\bibnamefont
  {Mamyrin}},\ }\href {\doibase 10.1016/j.physletb.2005.01.094} {\bibfield
  {journal} {\bibinfo  {journal} {Phys. Lett. B}\ }\textbf {\bibinfo {volume}
  {610}},\ \bibinfo {pages} {45 } (\bibinfo {year} {2005})}\BibitemShut
  {NoStop}%
\bibitem [{\citenamefont {Dobaczewski}\ \emph {et~al.}(2014)\citenamefont
  {Dobaczewski}, \citenamefont {Nazarewicz},\ and\ \citenamefont
  {Reinhard}}]{dobaczewski2014}%
  \BibitemOpen
  \bibfield  {author} {\bibinfo {author} {\bibfnamefont {J.}~\bibnamefont
  {Dobaczewski}}, \bibinfo {author} {\bibfnamefont {W.}~\bibnamefont
  {Nazarewicz}}, \ and\ \bibinfo {author} {\bibfnamefont {P.-G.}\ \bibnamefont
  {Reinhard}},\ }\href {http://stacks.iop.org/0954-3899/41/i=7/a=074001}
  {\bibfield  {journal} {\bibinfo  {journal} {J. Phys. G}\ }\textbf {\bibinfo
  {volume} {41}},\ \bibinfo {pages} {074001} (\bibinfo {year}
  {2014})}\BibitemShut {NoStop}%
\bibitem [{\citenamefont {Miller}\ \emph {et~al.}(1990)\citenamefont {Miller},
  \citenamefont {Nefkens},\ and\ \citenamefont {\v{S}laus}}]{miller1990}%
  \BibitemOpen
  \bibfield  {author} {\bibinfo {author} {\bibfnamefont {G.}~\bibnamefont
  {Miller}}, \bibinfo {author} {\bibfnamefont {B.}~\bibnamefont {Nefkens}}, \
  and\ \bibinfo {author} {\bibfnamefont {I.}~\bibnamefont {\v{S}laus}},\ }\href
  {\doibase 10.1016/0370-1573(90)90102-8} {\bibfield  {journal} {\bibinfo
  {journal} {Phys. Rep.}\ }\textbf {\bibinfo {volume} {194}},\ \bibinfo {pages}
  {1 } (\bibinfo {year} {1990})}\BibitemShut {NoStop}%
\bibitem [{\citenamefont {Workman}\ \emph {et~al.}(2012)\citenamefont
  {Workman}, \citenamefont {Arndt}, \citenamefont {Briscoe}, \citenamefont
  {Paris},\ and\ \citenamefont {Strakovsky}}]{workman2012}%
  \BibitemOpen
  \bibfield  {author} {\bibinfo {author} {\bibfnamefont {R.~L.}\ \bibnamefont
  {Workman}}, \bibinfo {author} {\bibfnamefont {R.~A.}\ \bibnamefont {Arndt}},
  \bibinfo {author} {\bibfnamefont {W.~J.}\ \bibnamefont {Briscoe}}, \bibinfo
  {author} {\bibfnamefont {M.~W.}\ \bibnamefont {Paris}}, \ and\ \bibinfo
  {author} {\bibfnamefont {I.~I.}\ \bibnamefont {Strakovsky}},\ }\href
  {\doibase 10.1103/PhysRevC.86.035202} {\bibfield  {journal} {\bibinfo
  {journal} {Phys. Rev. C}\ }\textbf {\bibinfo {volume} {86}},\ \bibinfo
  {pages} {035202} (\bibinfo {year} {2012})}\BibitemShut {NoStop}%
\bibitem [{\citenamefont {Huber}\ \emph {et~al.}(1998)\citenamefont {Huber},
  \citenamefont {Udem}, \citenamefont {Gross}, \citenamefont {Reichert},
  \citenamefont {Kourogi}, \citenamefont {Pachucki}, \citenamefont {Weitz},\
  and\ \citenamefont {H\"ansch}}]{huber1998}%
  \BibitemOpen
  \bibfield  {author} {\bibinfo {author} {\bibfnamefont {A.}~\bibnamefont
  {Huber}}, \bibinfo {author} {\bibfnamefont {T.}~\bibnamefont {Udem}},
  \bibinfo {author} {\bibfnamefont {B.}~\bibnamefont {Gross}}, \bibinfo
  {author} {\bibfnamefont {J.}~\bibnamefont {Reichert}}, \bibinfo {author}
  {\bibfnamefont {M.}~\bibnamefont {Kourogi}}, \bibinfo {author} {\bibfnamefont
  {K.}~\bibnamefont {Pachucki}}, \bibinfo {author} {\bibfnamefont
  {M.}~\bibnamefont {Weitz}}, \ and\ \bibinfo {author} {\bibfnamefont {T.~W.}\
  \bibnamefont {H\"ansch}},\ }\href {\doibase 10.1103/PhysRevLett.80.468}
  {\bibfield  {journal} {\bibinfo  {journal} {Phys. Rev. Lett.}\ }\textbf
  {\bibinfo {volume} {80}},\ \bibinfo {pages} {468} (\bibinfo {year}
  {1998})}\BibitemShut {NoStop}%
\bibitem [{\citenamefont {Wild}(2014)}]{wild2014}%
  \BibitemOpen
  \bibfield  {author} {\bibinfo {author} {\bibfnamefont {S.}~\bibnamefont
  {Wild}},\ }\href
  {http://www.mcs.anl.gov/publication/solving-derivative-free-nonlinear-least-squares-pounders}
  {\bibfield  {journal} {\bibinfo  {journal} {Preprint ANL/MCS-P5120-0414,
  Argonne Nat. Lab., Argonne, IL}\ } (\bibinfo {year} {2014})}\BibitemShut
  {NoStop}%
\bibitem [{\citenamefont {Munson}\ \emph {et~al.}(2012)\citenamefont {Munson},
  \citenamefont {Sarich}, \citenamefont {Wild}, \citenamefont {Benson},\ and\
  \citenamefont {McInnes}}]{munson2012}%
  \BibitemOpen
  \bibfield  {author} {\bibinfo {author} {\bibfnamefont {T.}~\bibnamefont
  {Munson}}, \bibinfo {author} {\bibfnamefont {J.}~\bibnamefont {Sarich}},
  \bibinfo {author} {\bibfnamefont {S.}~\bibnamefont {Wild}}, \bibinfo {author}
  {\bibfnamefont {S.}~\bibnamefont {Benson}}, \ and\ \bibinfo {author}
  {\bibfnamefont {L.~C.}\ \bibnamefont {McInnes}},\ }\href
  {http://www.mcs.anl.gov/tao} {\emph {\bibinfo {title} {TAO 2.0 Users
  Manual}}},\ \bibinfo {type} {Tech. Rep.}\ \bibinfo {number} {ANL/MCS-TM-322}\
  (\bibinfo  {institution} {Mathematics and Computer Science Division, Argonne
  National Laboratory},\ \bibinfo {year} {2012})\ \bibinfo {note}
  {\url{http://www.mcs.anl.gov/tao}}\BibitemShut {NoStop}%
\bibitem [{\citenamefont {Kortelainen}\ \emph {et~al.}(2010)\citenamefont
  {Kortelainen}, \citenamefont {Lesinski}, \citenamefont {Mor\'e},
  \citenamefont {Nazarewicz}, \citenamefont {Sarich}, \citenamefont {Schunck},
  \citenamefont {Stoitsov},\ and\ \citenamefont {Wild}}]{kortelainen2010}%
  \BibitemOpen
  \bibfield  {author} {\bibinfo {author} {\bibfnamefont {M.}~\bibnamefont
  {Kortelainen}}, \bibinfo {author} {\bibfnamefont {T.}~\bibnamefont
  {Lesinski}}, \bibinfo {author} {\bibfnamefont {J.}~\bibnamefont {Mor\'e}},
  \bibinfo {author} {\bibfnamefont {W.}~\bibnamefont {Nazarewicz}}, \bibinfo
  {author} {\bibfnamefont {J.}~\bibnamefont {Sarich}}, \bibinfo {author}
  {\bibfnamefont {N.}~\bibnamefont {Schunck}}, \bibinfo {author} {\bibfnamefont
  {M.~V.}\ \bibnamefont {Stoitsov}}, \ and\ \bibinfo {author} {\bibfnamefont
  {S.}~\bibnamefont {Wild}},\ }\href {\doibase 10.1103/PhysRevC.82.024313}
  {\bibfield  {journal} {\bibinfo  {journal} {Phys. Rev. C}\ }\textbf {\bibinfo
  {volume} {82}},\ \bibinfo {pages} {024313} (\bibinfo {year}
  {2010})}\BibitemShut {NoStop}%
\bibitem [{\citenamefont {Ekstr\"om}\ \emph {et~al.}(2015)\citenamefont
  {Ekstr\"om}, \citenamefont {Jansen}, \citenamefont {Wendt}, \citenamefont
  {Hagen}, \citenamefont {Papenbrock}, \citenamefont {Carlsson}, \citenamefont
  {Forss\'en}, \citenamefont {Hjorth-Jensen}, \citenamefont {Navr\'atil},\ and\
  \citenamefont {Nazarewicz}}]{ekstrom2015b}%
  \BibitemOpen
  \bibfield  {author} {\bibinfo {author} {\bibfnamefont {A.}~\bibnamefont
  {Ekstr\"om}}, \bibinfo {author} {\bibfnamefont {G.~R.}\ \bibnamefont
  {Jansen}}, \bibinfo {author} {\bibfnamefont {K.~A.}\ \bibnamefont {Wendt}},
  \bibinfo {author} {\bibfnamefont {G.}~\bibnamefont {Hagen}}, \bibinfo
  {author} {\bibfnamefont {T.}~\bibnamefont {Papenbrock}}, \bibinfo {author}
  {\bibfnamefont {B.~D.}\ \bibnamefont {Carlsson}}, \bibinfo {author}
  {\bibfnamefont {C.}~\bibnamefont {Forss\'en}}, \bibinfo {author}
  {\bibfnamefont {M.}~\bibnamefont {Hjorth-Jensen}}, \bibinfo {author}
  {\bibfnamefont {P.}~\bibnamefont {Navr\'atil}}, \ and\ \bibinfo {author}
  {\bibfnamefont {W.}~\bibnamefont {Nazarewicz}},\ }\href {\doibase
  10.1103/PhysRevC.91.051301} {\bibfield  {journal} {\bibinfo  {journal} {Phys.
  Rev. C}\ }\textbf {\bibinfo {volume} {91}},\ \bibinfo {pages} {051301}
  (\bibinfo {year} {2015})}\BibitemShut {NoStop}%
\bibitem [{\citenamefont {Charpentier}\ and\ \citenamefont
  {Utke}()}]{charpentier2014}%
  \BibitemOpen
  \bibfield  {author} {\bibinfo {author} {\bibfnamefont {I.}~\bibnamefont
  {Charpentier}}\ and\ \bibinfo {author} {\bibfnamefont {J.}~\bibnamefont
  {Utke}},\ }\href@noop {} {\emph {\bibinfo {title} {Rapsodia: {U}ser
  manual}}},\ \bibinfo {type} {Tech. Rep.}\ (\bibinfo  {institution} {Argonne
  National Laboratory})\ \bibinfo {note} {latest version available online at
  \url{http://www.mcs.anl.gov/Rapsodia/userManual.pdf}}\BibitemShut {NoStop}%
\bibitem [{\citenamefont {Haftel}\ and\ \citenamefont
  {Tabakin}(1970)}]{haftel1970}%
  \BibitemOpen
  \bibfield  {author} {\bibinfo {author} {\bibfnamefont {M.~I.}\ \bibnamefont
  {Haftel}}\ and\ \bibinfo {author} {\bibfnamefont {F.}~\bibnamefont
  {Tabakin}},\ }\href {\doibase 10.1016/0375-9474(70)90047-3} {\bibfield
  {journal} {\bibinfo  {journal} {Nucl. Phys. A}\ }\textbf {\bibinfo {volume}
  {158}},\ \bibinfo {pages} {1 } (\bibinfo {year} {1970})}\BibitemShut
  {NoStop}%
\bibitem [{\citenamefont {Carlsson}(2015)}]{carlsson2015}%
  \BibitemOpen
  \bibfield  {author} {\bibinfo {author} {\bibfnamefont {B.~D.}\ \bibnamefont
  {Carlsson}},\ }\emph {\bibinfo {title} {Making predictions using
  $\chi$EFT}},\ \href@noop {} {\bibinfo {type} {Licentiate thesis}},\ \bibinfo
  {school} {Chalmers University of Technology, Gothenburg, Sweden} (\bibinfo
  {year} {2015})\BibitemShut {NoStop}%
\bibitem [{\citenamefont {Mehen}\ \emph {et~al.}(1999)\citenamefont {Mehen},
  \citenamefont {Stewart},\ and\ \citenamefont {Wise}}]{mehen1999}%
  \BibitemOpen
  \bibfield  {author} {\bibinfo {author} {\bibfnamefont {T.}~\bibnamefont
  {Mehen}}, \bibinfo {author} {\bibfnamefont {I.~W.}\ \bibnamefont {Stewart}},
  \ and\ \bibinfo {author} {\bibfnamefont {M.~B.}\ \bibnamefont {Wise}},\
  }\href {\doibase 10.1103/PhysRevLett.83.931} {\bibfield  {journal} {\bibinfo
  {journal} {Phys. Rev. Lett.}\ }\textbf {\bibinfo {volume} {83}},\ \bibinfo
  {pages} {931} (\bibinfo {year} {1999})}\BibitemShut {NoStop}%
\bibitem [{\citenamefont {Friar}(1997)}]{friar1997b}%
  \BibitemOpen
  \bibfield  {author} {\bibinfo {author} {\bibfnamefont {J.~L.}\ \bibnamefont
  {Friar}},\ }\href {\doibase 10.1007/s006010050059} {\bibfield  {journal}
  {\bibinfo  {journal} {Few-Body Syst.}\ }\textbf {\bibinfo {volume} {22}},\
  \bibinfo {pages} {161} (\bibinfo {year} {1997})}\BibitemShut {NoStop}%
\bibitem [{sup()}]{supplemental}%
  \BibitemOpen
  \href@noop {} {}\bibinfo {note} {See Supplemental Material at [URL will be
  inserted by publisher] for a tabulation of numerical values for LECs,
  statistical uncertainties, and covariance matrices.}\BibitemShut {Stop}%
\bibitem [{\citenamefont {Klarsfeld}\ \emph {et~al.}(1986)\citenamefont
  {Klarsfeld}, \citenamefont {Martorell}, \citenamefont {Oteo}, \citenamefont
  {Nishimura},\ and\ \citenamefont {Sprung}}]{klarsfeld1986}%
  \BibitemOpen
  \bibfield  {author} {\bibinfo {author} {\bibfnamefont {S.}~\bibnamefont
  {Klarsfeld}}, \bibinfo {author} {\bibfnamefont {J.}~\bibnamefont
  {Martorell}}, \bibinfo {author} {\bibfnamefont {J.}~\bibnamefont {Oteo}},
  \bibinfo {author} {\bibfnamefont {M.}~\bibnamefont {Nishimura}}, \ and\
  \bibinfo {author} {\bibfnamefont {D.}~\bibnamefont {Sprung}},\ }\href
  {\doibase 10.1016/0375-9474(86)90400-8} {\bibfield  {journal} {\bibinfo
  {journal} {Nucl. Phys. A}\ }\textbf {\bibinfo {volume} {456}},\ \bibinfo
  {pages} {373 } (\bibinfo {year} {1986})}\BibitemShut {NoStop}%
\bibitem [{\citenamefont {Friar}\ \emph {et~al.}(1997)\citenamefont {Friar},
  \citenamefont {Martorell},\ and\ \citenamefont {Sprung}}]{friar1997a}%
  \BibitemOpen
  \bibfield  {author} {\bibinfo {author} {\bibfnamefont {J.~L.}\ \bibnamefont
  {Friar}}, \bibinfo {author} {\bibfnamefont {J.}~\bibnamefont {Martorell}}, \
  and\ \bibinfo {author} {\bibfnamefont {D.~W.~L.}\ \bibnamefont {Sprung}},\
  }\href {\doibase 10.1103/PhysRevA.56.4579} {\bibfield  {journal} {\bibinfo
  {journal} {Phys. Rev. A}\ }\textbf {\bibinfo {volume} {56}},\ \bibinfo
  {pages} {4579} (\bibinfo {year} {1997})}\BibitemShut {NoStop}%
\bibitem [{\citenamefont {Hackenburg}(2006)}]{hackenburg2006}%
  \BibitemOpen
  \bibfield  {author} {\bibinfo {author} {\bibfnamefont {R.~W.}\ \bibnamefont
  {Hackenburg}},\ }\href {\doibase 10.1103/PhysRevC.73.044002} {\bibfield
  {journal} {\bibinfo  {journal} {Phys. Rev. C}\ }\textbf {\bibinfo {volume}
  {73}},\ \bibinfo {pages} {044002} (\bibinfo {year} {2006})}\BibitemShut
  {NoStop}%
\bibitem [{\citenamefont {Hoferichter}\ \emph {et~al.}(2015)\citenamefont
  {Hoferichter}, \citenamefont {Ruiz~de Elvira}, \citenamefont {Kubis},\ and\
  \citenamefont {Mei\ss{}ner}}]{hoferichter2015}%
  \BibitemOpen
  \bibfield  {author} {\bibinfo {author} {\bibfnamefont {M.}~\bibnamefont
  {Hoferichter}}, \bibinfo {author} {\bibfnamefont {J.}~\bibnamefont {Ruiz~de
  Elvira}}, \bibinfo {author} {\bibfnamefont {B.}~\bibnamefont {Kubis}}, \ and\
  \bibinfo {author} {\bibfnamefont {U.-G.}\ \bibnamefont {Mei\ss{}ner}},\
  }\href {\doibase 10.1103/PhysRevLett.115.092301} {\bibfield  {journal}
  {\bibinfo  {journal} {Phys. Rev. Lett.}\ }\textbf {\bibinfo {volume} {115}},\
  \bibinfo {pages} {092301} (\bibinfo {year} {2015})}\BibitemShut {NoStop}%
\bibitem [{\citenamefont {Lisowski}\ \emph {et~al.}(1982)\citenamefont
  {Lisowski}, \citenamefont {Shamu}, \citenamefont {Auchampaugh}, \citenamefont
  {King}, \citenamefont {Moore}, \citenamefont {Morgan},\ and\ \citenamefont
  {Singleton}}]{lisowski1982}%
  \BibitemOpen
  \bibfield  {author} {\bibinfo {author} {\bibfnamefont {P.}~\bibnamefont
  {Lisowski}}, \bibinfo {author} {\bibfnamefont {R.}~\bibnamefont {Shamu}},
  \bibinfo {author} {\bibfnamefont {G.}~\bibnamefont {Auchampaugh}}, \bibinfo
  {author} {\bibfnamefont {N.}~\bibnamefont {King}}, \bibinfo {author}
  {\bibfnamefont {M.}~\bibnamefont {Moore}}, \bibinfo {author} {\bibfnamefont
  {G.}~\bibnamefont {Morgan}}, \ and\ \bibinfo {author} {\bibfnamefont
  {T.}~\bibnamefont {Singleton}},\ }\href {\doibase 10.1103/physrevlett.49.255}
  {\bibfield  {journal} {\bibinfo  {journal} {Phys. Rev. Lett.}\ }\textbf
  {\bibinfo {volume} {49}},\ \bibinfo {pages} {255} (\bibinfo {year}
  {1982})}\BibitemShut {NoStop}%
\bibitem [{\citenamefont {Sukhoruchkin}\ and\ \citenamefont
  {Soroko}(2009)}]{sukhoruchkin2009}%
  \BibitemOpen
  \bibfield  {author} {\bibinfo {author} {\bibfnamefont {S.}~\bibnamefont
  {Sukhoruchkin}}\ and\ \bibinfo {author} {\bibfnamefont {Z.}~\bibnamefont
  {Soroko}},\ }in\ \href {\doibase 10.1007/978-3-540-69945-3_127} {\emph
  {\bibinfo {booktitle} {Nuclei with Z = 1 - 54}}},\ \bibinfo {series}
  {Landolt-B\"ornstein - Group I Elementary Particles, Nuclei and Atoms}, Vol.\
  \bibinfo {volume} {22A},\ \bibinfo {editor} {edited by\ \bibinfo {editor}
  {\bibfnamefont {H.}~\bibnamefont {Schopper}}}\ (\bibinfo  {publisher}
  {Springer Berlin Heidelberg},\ \bibinfo {year} {2009})\ pp.\ \bibinfo {pages}
  {320--322}\BibitemShut {NoStop}%
\bibitem [{\citenamefont {Stump}\ \emph {et~al.}(2001)\citenamefont {Stump},
  \citenamefont {Pumplin}, \citenamefont {Brock}, \citenamefont {Casey},
  \citenamefont {Huston}, \citenamefont {Kalk}, \citenamefont {Lai},\ and\
  \citenamefont {Tung}}]{stump2001}%
  \BibitemOpen
  \bibfield  {author} {\bibinfo {author} {\bibfnamefont {D.}~\bibnamefont
  {Stump}}, \bibinfo {author} {\bibfnamefont {J.}~\bibnamefont {Pumplin}},
  \bibinfo {author} {\bibfnamefont {R.}~\bibnamefont {Brock}}, \bibinfo
  {author} {\bibfnamefont {D.}~\bibnamefont {Casey}}, \bibinfo {author}
  {\bibfnamefont {J.}~\bibnamefont {Huston}}, \bibinfo {author} {\bibfnamefont
  {J.}~\bibnamefont {Kalk}}, \bibinfo {author} {\bibfnamefont {H.~L.}\
  \bibnamefont {Lai}}, \ and\ \bibinfo {author} {\bibfnamefont {W.~K.}\
  \bibnamefont {Tung}},\ }\href {\doibase 10.1103/PhysRevD.65.014012}
  {\bibfield  {journal} {\bibinfo  {journal} {Phys. Rev. D}\ }\textbf {\bibinfo
  {volume} {65}},\ \bibinfo {pages} {014012} (\bibinfo {year}
  {2001})}\BibitemShut {NoStop}%
\bibitem [{\citenamefont {Schindler}\ and\ \citenamefont
  {Phillips}(2009)}]{schindler2009}%
  \BibitemOpen
  \bibfield  {author} {\bibinfo {author} {\bibfnamefont {M.}~\bibnamefont
  {Schindler}}\ and\ \bibinfo {author} {\bibfnamefont {D.}~\bibnamefont
  {Phillips}},\ }\href {\doibase 10.1016/j.aop.2008.09.003} {\bibfield
  {journal} {\bibinfo  {journal} {Ann. Phys.}\ }\textbf {\bibinfo {volume}
  {324}},\ \bibinfo {pages} {682 } (\bibinfo {year} {2009})}\BibitemShut
  {NoStop}%
\bibitem [{\citenamefont {Habib}\ \emph {et~al.}(2007)\citenamefont {Habib},
  \citenamefont {Heitmann}, \citenamefont {Higdon}, \citenamefont {Nakhleh},\
  and\ \citenamefont {Williams}}]{habib2007}%
  \BibitemOpen
  \bibfield  {author} {\bibinfo {author} {\bibfnamefont {S.}~\bibnamefont
  {Habib}}, \bibinfo {author} {\bibfnamefont {K.}~\bibnamefont {Heitmann}},
  \bibinfo {author} {\bibfnamefont {D.}~\bibnamefont {Higdon}}, \bibinfo
  {author} {\bibfnamefont {C.}~\bibnamefont {Nakhleh}}, \ and\ \bibinfo
  {author} {\bibfnamefont {B.}~\bibnamefont {Williams}},\ }\href {\doibase
  10.1103/PhysRevD.76.083503} {\bibfield  {journal} {\bibinfo  {journal} {Phys.
  Rev. D}\ }\textbf {\bibinfo {volume} {76}},\ \bibinfo {pages} {083503}
  (\bibinfo {year} {2007})}\BibitemShut {NoStop}%
\bibitem [{\citenamefont {Holsclaw}\ \emph {et~al.}(2010)\citenamefont
  {Holsclaw}, \citenamefont {Alam}, \citenamefont {Sans\'o}, \citenamefont
  {Lee}, \citenamefont {Heitmann}, \citenamefont {Habib},\ and\ \citenamefont
  {Higdon}}]{holsclaw2010}%
  \BibitemOpen
  \bibfield  {author} {\bibinfo {author} {\bibfnamefont {T.}~\bibnamefont
  {Holsclaw}}, \bibinfo {author} {\bibfnamefont {U.}~\bibnamefont {Alam}},
  \bibinfo {author} {\bibfnamefont {B.}~\bibnamefont {Sans\'o}}, \bibinfo
  {author} {\bibfnamefont {H.}~\bibnamefont {Lee}}, \bibinfo {author}
  {\bibfnamefont {K.}~\bibnamefont {Heitmann}}, \bibinfo {author}
  {\bibfnamefont {S.}~\bibnamefont {Habib}}, \ and\ \bibinfo {author}
  {\bibfnamefont {D.}~\bibnamefont {Higdon}},\ }\href {\doibase
  10.1103/PhysRevLett.105.241302} {\bibfield  {journal} {\bibinfo  {journal}
  {Phys. Rev. Lett.}\ }\textbf {\bibinfo {volume} {105}},\ \bibinfo {pages}
  {241302} (\bibinfo {year} {2010})}\BibitemShut {NoStop}%
\end{thebibliography}%

\appendix*

\section{Supplemental material for\\``Uncertainty analysis and order-by-order optimization of chiral
  nuclear interactions''}

Here we present all optimized central values and statistical
uncertainties for the LECs in the \piN{}, \NN{} and the \NNN{} sector for all potentials referred to in the main text.
The $c_i$, $d_i$ and $e_i$ are in units of $\unit{GeV^{-1}}$,
$\unit{GeV^{-2}}$ and $\unit{GeV^{-3}}$ respectively. $\tilde{C}_i$
are in units of $\unit[10^4]{GeV^{-2}}$, $C_i$ in units of
$\unit[10^4]{GeV^{-4}}$ and $c_D$ and $c_E$ are dimensionless.

The covariance matrices for all potentials and numerical tables with LECs
can be downloaded as a tarball from the supplemental material page at
the published version of the manuscript. They are also available
upon request from the authors.
The ordering of the LECs follow the ordering in the tables below.

\begin{table}[htb]\caption{\label{LO_500_290}\LOsep{} and \LOsim{} ($\Lambda=500, T_{\max}=290$)}
\begin{ruledtabular}\begin{tabular}{ccc}
    LEC & \LOsep & \LOsim\\\hline
    $\tilde{C}_{{}^1S_0}$ & $-0.1076841(50)$ & $-0.1076845(80)$\\
    $\tilde{C}_{{}^3S_1}$ & $-0.07172(11)$ & $-0.0718086(27)$\\
\end{tabular}\end{ruledtabular}
\end{table}
\begin{table}[htb]\caption{\label{NLO_500_290}\NLOsep{} and \NLOsim{} ($\Lambda=500, T_{\max}=290$)}
\begin{ruledtabular}\begin{tabular}{ccc}
    LEC & \NLOsep & \NLOsim\\\hline
    $\tilde{C}_{{}^1S_0}^{(np)}$ & $-0.150533(96)$ & $-0.150623(79)$\\
    $\tilde{C}_{{}^1S_0}^{(pp)}$ & $-0.14893(11)$ & $-0.14891(11)$\\
    $\tilde{C}_{{}^1S_0}^{(nn)}$ & $-0.14992(27)$ & $-0.14991(27)$\\
    $C_{{}^1S_0}$ & $+1.6926(82)$ & $+1.6935(83)$\\
    $\tilde{C}_{{}^3S_1}$ & $-0.1742(20)$ & $-0.1843(16)$\\
    $C_{{}^3S_1}$ & $-0.408(23)$ & $-0.218(14)$\\
    $C_{E_1}$ & $+0.238(14)$ & $+0.263(16)$\\
    $C_{{}^3P_0}$ & $+1.3085(86)$ & $+1.2998(85)$\\
    $C_{{}^1P_1}$ & $+0.849(47)$ & $+1.025(59)$\\
    $C_{{}^3P_1}$ & $-0.3409(98)$ & $-0.336(10)$\\
    $C_{{}^3P_2}$ & $-0.2011(15)$ & $-0.2029(15)$\\
\end{tabular}\end{ruledtabular}
\end{table}
\begin{table}[htb]\caption{\label{NNLO_500_290}\NNLOsep{} and \NNLOsim{} ($\Lambda=500, T_{\max}=290$)}
\begin{ruledtabular}\begin{tabular}{ccc}
    LEC & \NNLOsep & \NNLOsim\\\hline
    $\tilde{C}_{{}^1S_0}^{(np)}$ & $-0.15387(10)$ & $-0.1474(20)$\\
    $\tilde{C}_{{}^1S_0}^{(pp)}$ & $-0.152935(72)$ & $-0.1465(20)$\\
    $\tilde{C}_{{}^1S_0}^{(nn)}$ & $-0.15354(43)$ & $-0.1471(20)$\\
    $C_{{}^1S_0}$ & $+2.7442(19)$ & $+2.548(47)$\\
    $\tilde{C}_{{}^3S_1}$ & $-0.1671(10)$ & $-0.1687(21)$\\
    $C_{{}^3S_1}$ & $+0.8738(64)$ & $+0.705(47)$\\
    $C_{E_1}$ & $+0.6899(67)$ & $+0.597(11)$\\
    $C_{{}^3P_0}$ & $+1.2782(66)$ & $+1.161(31)$\\
    $C_{{}^1P_1}$ & $+0.521(12)$ & $+0.520(33)$\\
    $C_{{}^3P_1}$ & $-0.9378(69)$ & $-0.955(31)$\\
    $C_{{}^3P_2}$ & $-0.68645(76)$ & $-0.658(30)$\\
    $c_D$ & $-0.581(28)$ & $-0.325(51)$\\
    $c_E$ & $-0.6666(99)$ & $-0.521(17)$\\
    $c_1$ & $-0.69(50)$ & $+0.22(30)$\\
    $c_2$ & $+3.0(14)$ & $+5.1(10)$\\
    $c_3$ & $-4.12(32)$ & $-3.56(13)$\\
    $c_4$ & $+5.35(81)$ & $+3.933(85)$\\
    $d_1\!+\!d_2$ & $+6.22(44)$ & $+5.320(94)$\\
    $d_3$ & $-5.31(30)$ & $-4.83(22)$\\
    $d_5$ & $-0.46(18)$ & $-0.24(14)$\\
    $d_{14}\!-\!d_{15}$ & $-11.00(42)$ & $-10.23(27)$\\
    $e_{14}$ & $-0.63(95)$ & $-0.26(89)$\\
    $e_{15}$ & $-7.7(26)$ & $-9.3(24)$\\
    $e_{16}$ & $+5.9(49)$ & $-0.0(41)$\\
    $e_{17}$ & $+2.1(18)$ & $+1.5(18)$\\
    $e_{18}$ & $-8.1(42)$ & $-1.2(16)$\\
\end{tabular}\end{ruledtabular}
\end{table}
\begin{table*}[htb]\caption{\label{NNLOsim_450}\NNLOsim{} ($\Lambda=450$)}
\begin{ruledtabular}\begin{tabular}{ccccccc}
    LEC & $T_{\rm lab}^{\max}\!=\!125$ & $T_{\rm lab}^{\max}\!=\!158$ & $T_{\rm lab}^{\max}\!=\!191$ & $T_{\rm lab}^{\max}\!=\!224$ & $T_{\rm lab}^{\max}\!=\!257$ & $T_{\rm lab}^{\max}\!=\!290$ \\\hline
    $\tilde{C}_{{}^1S_0}^{(np)}$ & $-0.1519(20)$ & $-0.1512(20)$ & $-0.1510(20)$ & $-0.1501(20)$ & $-0.1499(20)$ & $-0.1496(20)$ \\
    $\tilde{C}_{{}^1S_0}^{(pp)}$ & $-0.1512(20)$ & $-0.1504(20)$ & $-0.1502(20)$ & $-0.1493(20)$ & $-0.1491(20)$ & $-0.1488(20)$ \\
    $\tilde{C}_{{}^1S_0}^{(nn)}$ & $-0.1518(21)$ & $-0.1510(21)$ & $-0.1508(21)$ & $-0.1498(20)$ & $-0.1496(20)$ & $-0.1493(20)$ \\
    $C_{{}^1S_0}$ & $+2.498(51)$ & $+2.480(50)$ & $+2.477(49)$ & $+2.511(48)$ & $+2.518(48)$ & $+2.527(48)$ \\
    $\tilde{C}_{{}^3S_1}$ & $-0.1743(23)$ & $-0.1780(22)$ & $-0.1785(22)$ & $-0.1801(21)$ & $-0.1806(21)$ & $-0.1807(21)$ \\
    $C_{{}^3S_1}$ & $+0.676(52)$ & $+0.682(51)$ & $+0.677(50)$ & $+0.722(49)$ & $+0.730(48)$ & $+0.740(48)$ \\
    $C_{E_1}$ & $+0.313(25)$ & $+0.475(20)$ & $+0.478(18)$ & $+0.600(15)$ & $+0.629(14)$ & $+0.657(13)$ \\
    $C_{{}^3P_0}$ & $+0.992(37)$ & $+1.122(34)$ & $+1.111(33)$ & $+1.164(32)$ & $+1.171(32)$ & $+1.183(31)$ \\
    $C_{{}^1P_1}$ & $+0.064(44)$ & $+0.333(40)$ & $+0.355(39)$ & $+0.551(36)$ & $+0.610(36)$ & $+0.664(35)$ \\
    $C_{{}^3P_1}$ & $-0.893(34)$ & $-0.940(33)$ & $-0.943(33)$ & $-0.960(32)$ & $-0.961(32)$ & $-0.967(31)$ \\
    $C_{{}^3P_2}$ & $-0.794(33)$ & $-0.706(32)$ & $-0.708(32)$ & $-0.663(31)$ & $-0.651(31)$ & $-0.639(31)$ \\
    $c_D$ & $+0.359(83)$ & $-0.109(68)$ & $-0.089(63)$ & $-0.432(56)$ & $-0.508(54)$ & $-0.594(52)$ \\
    $c_E$ & $-0.089(29)$ & $-0.281(28)$ & $-0.276(26)$ & $-0.443(24)$ & $-0.483(23)$ & $-0.528(22)$ \\
    $c_1$ & $-0.83(30)$ & $-0.53(30)$ & $-0.50(30)$ & $-0.21(30)$ & $-0.13(30)$ & $-0.05(30)$ \\
    $c_2$ & $+2.8(11)$ & $+3.1(11)$ & $+3.2(11)$ & $+3.8(11)$ & $+4.0(11)$ & $+4.2(11)$ \\
    $c_3$ & $-4.36(15)$ & $-3.82(14)$ & $-3.82(14)$ & $-3.57(14)$ & $-3.51(14)$ & $-3.45(14)$ \\
    $c_4$ & $+1.90(22)$ & $+3.02(16)$ & $+2.95(15)$ & $+3.84(12)$ & $+4.02(11)$ & $+4.235(98)$ \\
    $d_1\!+\!d_2$ & $+4.47(14)$ & $+4.91(12)$ & $+4.88(11)$ & $+5.28(10)$ & $+5.36(10)$ & $+5.450(97)$ \\
    $d_3$ & $-4.49(23)$ & $-4.63(23)$ & $-4.61(23)$ & $-4.78(22)$ & $-4.82(22)$ & $-4.85(22)$ \\
    $d_5$ & $+0.02(14)$ & $-0.14(14)$ & $-0.13(14)$ & $-0.25(14)$ & $-0.27(14)$ & $-0.30(14)$ \\
    $d_{14}\!-\!d_{15}$ & $-9.71(28)$ & $-9.98(28)$ & $-9.95(27)$ & $-10.21(27)$ & $-10.26(27)$ & $-10.32(27)$ \\
    $e_{14}$ & $+0.96(91)$ & $+0.33(90)$ & $+0.35(90)$ & $-0.14(89)$ & $-0.25(89)$ & $-0.37(89)$ \\
    $e_{15}$ & $-10.0(25)$ & $-11.0(25)$ & $-11.0(25)$ & $-10.6(24)$ & $-10.5(24)$ & $-10.4(24)$ \\
    $e_{16}$ & $+7.4(42)$ & $+7.0(42)$ & $+6.7(42)$ & $+4.6(42)$ & $+4.1(42)$ & $+3.5(41)$ \\
    $e_{17}$ & $+1.2(18)$ & $+1.2(18)$ & $+1.1(18)$ & $+1.3(18)$ & $+1.4(18)$ & $+1.4(18)$ \\
    $e_{18}$ & $+8.4(19)$ & $+3.3(18)$ & $+3.6(17)$ & $-0.6(17)$ & $-1.5(17)$ & $-2.5(16)$ \\
\end{tabular}\end{ruledtabular}
\end{table*}
\begin{table*}[htb]\caption{\label{NNLOsim_475}\NNLOsim{} ($\Lambda=475$)}
\begin{ruledtabular}\begin{tabular}{ccccccc}
    LEC & $T_{\rm lab}^{\max}\!=\!125$ & $T_{\rm lab}^{\max}\!=\!158$ & $T_{\rm lab}^{\max}\!=\!191$ & $T_{\rm lab}^{\max}\!=\!224$ & $T_{\rm lab}^{\max}\!=\!257$ & $T_{\rm lab}^{\max}\!=\!290$ \\\hline
    $\tilde{C}_{{}^1S_0}^{(np)}$ & $-0.1513(20)$ & $-0.1507(20)$ & $-0.1503(20)$ & $-0.1493(20)$ & $-0.1489(20)$ & $-0.1483(20)$ \\
    $\tilde{C}_{{}^1S_0}^{(pp)}$ & $-0.1506(20)$ & $-0.1500(20)$ & $-0.1496(20)$ & $-0.1485(20)$ & $-0.1481(20)$ & $-0.1475(20)$ \\
    $\tilde{C}_{{}^1S_0}^{(nn)}$ & $-0.1512(21)$ & $-0.1506(21)$ & $-0.1502(21)$ & $-0.1491(20)$ & $-0.1486(20)$ & $-0.1481(20)$ \\
    $C_{{}^1S_0}$ & $+2.492(51)$ & $+2.464(50)$ & $+2.470(49)$ & $+2.508(48)$ & $+2.524(48)$ & $+2.541(47)$ \\
    $\tilde{C}_{{}^3S_1}$ & $-0.1673(23)$ & $-0.1722(22)$ & $-0.1727(22)$ & $-0.1743(22)$ & $-0.1746(22)$ & $-0.1743(21)$ \\
    $C_{{}^3S_1}$ & $+0.639(52)$ & $+0.638(50)$ & $+0.644(50)$ & $+0.690(48)$ & $+0.707(48)$ & $+0.723(47)$ \\
    $C_{E_1}$ & $+0.297(24)$ & $+0.443(18)$ & $+0.455(17)$ & $+0.570(14)$ & $+0.599(13)$ & $+0.627(12)$ \\
    $C_{{}^3P_0}$ & $+1.036(36)$ & $+1.139(33)$ & $+1.130(33)$ & $+1.161(32)$ & $+1.162(31)$ & $+1.167(31)$ \\
    $C_{{}^1P_1}$ & $+0.062(42)$ & $+0.299(39)$ & $+0.318(38)$ & $+0.489(35)$ & $+0.540(35)$ & $+0.583(34)$ \\
    $C_{{}^3P_1}$ & $-0.857(34)$ & $-0.913(33)$ & $-0.925(33)$ & $-0.946(32)$ & $-0.954(31)$ & $-0.967(31)$ \\
    $C_{{}^3P_2}$ & $-0.784(33)$ & $-0.700(32)$ & $-0.703(31)$ & $-0.663(31)$ & $-0.656(31)$ & $-0.649(30)$ \\
    $c_D$ & $+0.404(78)$ & $-0.003(64)$ & $-0.003(60)$ & $-0.317(54)$ & $-0.391(53)$ & $-0.471(51)$ \\
    $c_E$ & $-0.175(22)$ & $-0.301(23)$ & $-0.304(21)$ & $-0.440(20)$ & $-0.475(20)$ & $-0.515(19)$ \\
    $c_1$ & $-0.75(30)$ & $-0.49(30)$ & $-0.42(30)$ & $-0.12(30)$ & $-0.01(30)$ & $+0.11(30)$ \\
    $c_2$ & $+3.0(11)$ & $+3.2(11)$ & $+3.4(11)$ & $+4.1(11)$ & $+4.4(11)$ & $+4.7(10)$ \\
    $c_3$ & $-4.30(15)$ & $-3.78(14)$ & $-3.80(14)$ & $-3.58(14)$ & $-3.54(14)$ & $-3.51(13)$ \\
    $c_4$ & $+1.91(20)$ & $+2.85(15)$ & $+2.85(13)$ & $+3.68(11)$ & $+3.877(99)$ & $+4.092(90)$ \\
    $d_1\!+\!d_2$ & $+4.46(13)$ & $+4.81(11)$ & $+4.82(11)$ & $+5.20(10)$ & $+5.288(98)$ & $+5.391(95)$ \\
    $d_3$ & $-4.48(23)$ & $-4.58(23)$ & $-4.59(23)$ & $-4.75(22)$ & $-4.80(22)$ & $-4.85(22)$ \\
    $d_5$ & $+0.02(14)$ & $-0.11(14)$ & $-0.11(14)$ & $-0.22(14)$ & $-0.24(14)$ & $-0.27(14)$ \\
    $d_{14}\!-\!d_{15}$ & $-9.69(28)$ & $-9.90(27)$ & $-9.90(27)$ & $-10.15(27)$ & $-10.21(27)$ & $-10.28(27)$ \\
    $e_{14}$ & $+0.93(91)$ & $+0.39(90)$ & $+0.38(90)$ & $-0.08(89)$ & $-0.20(89)$ & $-0.32(89)$ \\
    $e_{15}$ & $-10.0(25)$ & $-11.2(25)$ & $-10.9(25)$ & $-10.4(24)$ & $-10.1(24)$ & $-9.7(24)$ \\
    $e_{16}$ & $+6.9(42)$ & $+6.8(42)$ & $+6.0(42)$ & $+3.7(42)$ & $+2.7(41)$ & $+1.4(41)$ \\
    $e_{17}$ & $+1.1(18)$ & $+1.1(18)$ & $+1.1(18)$ & $+1.3(18)$ & $+1.4(18)$ & $+1.5(18)$ \\
    $e_{18}$ & $+8.3(19)$ & $+4.1(17)$ & $+4.1(17)$ & $+0.1(17)$ & $-0.8(16)$ & $-1.9(16)$ \\
\end{tabular}\end{ruledtabular}
\end{table*}
\begin{table*}[htb]\caption{\label{NNLOsim_500}\NNLOsim{} ($\Lambda=500$)}
\begin{ruledtabular}\begin{tabular}{ccccccc}
    LEC & $T_{\rm lab}^{\max}\!=\!125$ & $T_{\rm lab}^{\max}\!=\!158$ & $T_{\rm lab}^{\max}\!=\!191$ & $T_{\rm lab}^{\max}\!=\!224$ & $T_{\rm lab}^{\max}\!=\!257$ & $T_{\rm lab}^{\max}\!=\!290$ \\\hline
    $\tilde{C}_{{}^1S_0}^{(np)}$ & $-0.1507(20)$ & $-0.1503(20)$ & $-0.1497(20)$ & $-0.1488(20)$ & $-0.1481(20)$ & $-0.1474(20)$ \\
    $\tilde{C}_{{}^1S_0}^{(pp)}$ & $-0.1500(20)$ & $-0.1496(20)$ & $-0.1490(20)$ & $-0.1480(20)$ & $-0.1473(20)$ & $-0.1465(20)$ \\
    $\tilde{C}_{{}^1S_0}^{(nn)}$ & $-0.1508(21)$ & $-0.1503(21)$ & $-0.1497(21)$ & $-0.1486(20)$ & $-0.1479(20)$ & $-0.1471(20)$ \\
    $C_{{}^1S_0}$ & $+2.488(50)$ & $+2.451(49)$ & $+2.465(49)$ & $+2.502(48)$ & $+2.524(47)$ & $+2.548(47)$ \\
    $\tilde{C}_{{}^3S_1}$ & $-0.1605(24)$ & $-0.1668(23)$ & $-0.1673(23)$ & $-0.1691(22)$ & $-0.1693(22)$ & $-0.1687(21)$ \\
    $C_{{}^3S_1}$ & $+0.608(51)$ & $+0.601(50)$ & $+0.615(49)$ & $+0.660(48)$ & $+0.682(48)$ & $+0.705(47)$ \\
    $C_{E_1}$ & $+0.287(22)$ & $+0.418(17)$ & $+0.436(16)$ & $+0.542(13)$ & $+0.572(12)$ & $+0.597(11)$ \\
    $C_{{}^3P_0}$ & $+1.096(35)$ & $+1.170(33)$ & $+1.161(32)$ & $+1.169(31)$ & $+1.164(31)$ & $+1.161(31)$ \\
    $C_{{}^1P_1}$ & $+0.071(42)$ & $+0.281(39)$ & $+0.294(37)$ & $+0.445(35)$ & $+0.487(34)$ & $+0.520(33)$ \\
    $C_{{}^3P_1}$ & $-0.813(34)$ & $-0.882(33)$ & $-0.900(32)$ & $-0.923(31)$ & $-0.936(31)$ & $-0.955(31)$ \\
    $C_{{}^3P_2}$ & $-0.772(32)$ & $-0.694(32)$ & $-0.699(31)$ & $-0.663(31)$ & $-0.659(30)$ & $-0.658(30)$ \\
    $c_D$ & $+0.450(74)$ & $+0.102(62)$ & $+0.091(59)$ & $-0.189(54)$ & $-0.256(52)$ & $-0.325(51)$ \\
    $c_E$ & $-0.297(15)$ & $-0.357(17)$ & $-0.362(17)$ & $-0.460(17)$ & $-0.490(17)$ & $-0.521(17)$ \\
    $c_1$ & $-0.66(30)$ & $-0.45(30)$ & $-0.36(30)$ & $-0.07(30)$ & $+0.07(30)$ & $+0.22(30)$ \\
    $c_2$ & $+3.2(11)$ & $+3.3(11)$ & $+3.6(11)$ & $+4.3(11)$ & $+4.7(11)$ & $+5.1(10)$ \\
    $c_3$ & $-4.23(15)$ & $-3.75(14)$ & $-3.78(14)$ & $-3.58(14)$ & $-3.57(13)$ & $-3.56(13)$ \\
    $c_4$ & $+1.97(19)$ & $+2.73(14)$ & $+2.78(12)$ & $+3.527(99)$ & $+3.727(92)$ & $+3.933(85)$ \\
    $d_1\!+\!d_2$ & $+4.48(12)$ & $+4.74(11)$ & $+4.77(10)$ & $+5.115(98)$ & $+5.215(96)$ & $+5.320(94)$ \\
    $d_3$ & $-4.48(23)$ & $-4.54(23)$ & $-4.57(23)$ & $-4.72(22)$ & $-4.77(22)$ & $-4.83(22)$ \\
    $d_5$ & $+0.02(14)$ & $-0.10(14)$ & $-0.10(14)$ & $-0.19(14)$ & $-0.22(14)$ & $-0.24(14)$ \\
    $d_{14}\!-\!d_{15}$ & $-9.69(28)$ & $-9.84(27)$ & $-9.86(27)$ & $-10.09(27)$ & $-10.16(27)$ & $-10.23(27)$ \\
    $e_{14}$ & $+0.88(90)$ & $+0.43(90)$ & $+0.40(90)$ & $-0.02(89)$ & $-0.14(89)$ & $-0.26(89)$ \\
    $e_{15}$ & $-10.1(25)$ & $-11.4(25)$ & $-10.9(25)$ & $-10.4(24)$ & $-9.9(24)$ & $-9.3(24)$ \\
    $e_{16}$ & $+6.3(42)$ & $+6.7(42)$ & $+5.5(42)$ & $+3.1(42)$ & $+1.7(41)$ & $-0.0(41)$ \\
    $e_{17}$ & $+1.1(18)$ & $+1.0(18)$ & $+1.1(18)$ & $+1.3(18)$ & $+1.4(18)$ & $+1.5(18)$ \\
    $e_{18}$ & $+8.1(18)$ & $+4.7(17)$ & $+4.4(17)$ & $+0.9(16)$ & $-0.1(16)$ & $-1.2(16)$ \\
\end{tabular}\end{ruledtabular}
\end{table*}
\begin{table*}[htb]\caption{\label{NNLOsim_525}\NNLOsim{} ($\Lambda=525$)}
\begin{ruledtabular}\begin{tabular}{ccccccc}
    LEC & $T_{\rm lab}^{\max}\!=\!125$ & $T_{\rm lab}^{\max}\!=\!158$ & $T_{\rm lab}^{\max}\!=\!191$ & $T_{\rm lab}^{\max}\!=\!224$ & $T_{\rm lab}^{\max}\!=\!257$ & $T_{\rm lab}^{\max}\!=\!290$ \\\hline
    $\tilde{C}_{{}^1S_0}^{(np)}$ & $-0.1502(20)$ & $-0.1499(20)$ & $-0.1493(20)$ & $-0.1484(20)$ & $-0.1477(20)$ & $-0.1467(20)$ \\
    $\tilde{C}_{{}^1S_0}^{(pp)}$ & $-0.1495(20)$ & $-0.1492(20)$ & $-0.1485(20)$ & $-0.1476(20)$ & $-0.1468(20)$ & $-0.1459(20)$ \\
    $\tilde{C}_{{}^1S_0}^{(nn)}$ & $-0.1504(21)$ & $-0.1500(21)$ & $-0.1493(21)$ & $-0.1483(20)$ & $-0.1475(20)$ & $-0.1466(20)$ \\
    $C_{{}^1S_0}$ & $+2.484(50)$ & $+2.441(49)$ & $+2.460(49)$ & $+2.492(48)$ & $+2.517(47)$ & $+2.545(47)$ \\
    $\tilde{C}_{{}^3S_1}$ & $-0.1540(26)$ & $-0.1616(24)$ & $-0.1625(23)$ & $-0.1646(22)$ & $-0.1648(22)$ & $-0.1639(22)$ \\
    $C_{{}^3S_1}$ & $+0.581(51)$ & $+0.570(50)$ & $+0.590(49)$ & $+0.629(48)$ & $+0.655(47)$ & $+0.681(47)$ \\
    $C_{E_1}$ & $+0.283(22)$ & $+0.399(17)$ & $+0.422(15)$ & $+0.519(12)$ & $+0.548(12)$ & $+0.571(11)$ \\
    $C_{{}^3P_0}$ & $+1.176(34)$ & $+1.219(33)$ & $+1.209(32)$ & $+1.193(31)$ & $+1.182(31)$ & $+1.171(30)$ \\
    $C_{{}^1P_1}$ & $+0.087(41)$ & $+0.273(38)$ & $+0.283(37)$ & $+0.416(34)$ & $+0.451(34)$ & $+0.475(33)$ \\
    $C_{{}^3P_1}$ & $-0.759(34)$ & $-0.844(33)$ & $-0.866(32)$ & $-0.890(31)$ & $-0.906(31)$ & $-0.930(31)$ \\
    $C_{{}^3P_2}$ & $-0.760(32)$ & $-0.688(31)$ & $-0.694(31)$ & $-0.660(30)$ & $-0.659(30)$ & $-0.661(30)$ \\
    $c_D$ & $+0.508(72)$ & $+0.213(62)$ & $+0.193(59)$ & $-0.052(54)$ & $-0.111(53)$ & $-0.166(52)$ \\
    $c_E$ & $-0.473(15)$ & $-0.462(14)$ & $-0.464(13)$ & $-0.521(14)$ & $-0.543(14)$ & $-0.566(14)$ \\
    $c_1$ & $-0.59(30)$ & $-0.42(30)$ & $-0.31(30)$ & $-0.05(30)$ & $+0.10(30)$ & $+0.28(29)$ \\
    $c_2$ & $+3.3(11)$ & $+3.4(11)$ & $+3.7(11)$ & $+4.3(11)$ & $+4.8(10)$ & $+5.3(10)$ \\
    $c_3$ & $-4.17(15)$ & $-3.72(14)$ & $-3.75(14)$ & $-3.57(13)$ & $-3.57(13)$ & $-3.58(13)$ \\
    $c_4$ & $+2.06(17)$ & $+2.65(13)$ & $+2.73(11)$ & $+3.393(93)$ & $+3.589(88)$ & $+3.781(81)$ \\
    $d_1\!+\!d_2$ & $+4.50(12)$ & $+4.69(11)$ & $+4.74(10)$ & $+5.041(96)$ & $+5.143(95)$ & $+5.245(93)$ \\
    $d_3$ & $-4.49(23)$ & $-4.51(23)$ & $-4.55(23)$ & $-4.68(22)$ & $-4.74(22)$ & $-4.80(22)$ \\
    $d_5$ & $+0.01(14)$ & $-0.08(14)$ & $-0.09(14)$ & $-0.17(14)$ & $-0.19(14)$ & $-0.21(14)$ \\
    $d_{14}\!-\!d_{15}$ & $-9.70(28)$ & $-9.79(27)$ & $-9.83(27)$ & $-10.03(27)$ & $-10.10(27)$ & $-10.18(27)$ \\
    $e_{14}$ & $+0.82(90)$ & $+0.46(90)$ & $+0.41(90)$ & $+0.03(89)$ & $-0.09(89)$ & $-0.20(89)$ \\
    $e_{15}$ & $-10.2(25)$ & $-11.5(25)$ & $-10.9(25)$ & $-10.5(24)$ & $-9.9(24)$ & $-9.1(24)$ \\
    $e_{16}$ & $+5.9(42)$ & $+6.5(42)$ & $+5.1(42)$ & $+3.0(41)$ & $+1.4(41)$ & $-0.7(41)$ \\
    $e_{17}$ & $+1.1(18)$ & $+1.0(18)$ & $+1.1(18)$ & $+1.2(18)$ & $+1.4(18)$ & $+1.5(18)$ \\
    $e_{18}$ & $+7.7(18)$ & $+5.1(17)$ & $+4.7(17)$ & $+1.5(16)$ & $+0.5(16)$ & $-0.4(16)$ \\
\end{tabular}\end{ruledtabular}
\end{table*}
\begin{table*}[htb]\caption{\label{NNLOsim_550}\NNLOsim{} ($\Lambda=550$)}
\begin{ruledtabular}\begin{tabular}{ccccccc}
    LEC & $T_{\rm lab}^{\max}\!=\!125$ & $T_{\rm lab}^{\max}\!=\!158$ & $T_{\rm lab}^{\max}\!=\!191$ & $T_{\rm lab}^{\max}\!=\!224$ & $T_{\rm lab}^{\max}\!=\!257$ & $T_{\rm lab}^{\max}\!=\!290$ \\\hline
    $\tilde{C}_{{}^1S_0}^{(np)}$ & $-0.1498(20)$ & $-0.1495(20)$ & $-0.1489(20)$ & $-0.1482(20)$ & $-0.1474(20)$ & $-0.1464(20)$ \\
    $\tilde{C}_{{}^1S_0}^{(pp)}$ & $-0.1491(20)$ & $-0.1488(20)$ & $-0.1481(20)$ & $-0.1473(20)$ & $-0.1465(20)$ & $-0.1455(20)$ \\
    $\tilde{C}_{{}^1S_0}^{(nn)}$ & $-0.1501(21)$ & $-0.1497(21)$ & $-0.1490(21)$ & $-0.1482(20)$ & $-0.1474(20)$ & $-0.1463(20)$ \\
    $C_{{}^1S_0}$ & $+2.480(50)$ & $+2.433(49)$ & $+2.454(48)$ & $+2.478(47)$ & $+2.504(47)$ & $+2.533(46)$ \\
    $\tilde{C}_{{}^3S_1}$ & $-0.1476(27)$ & $-0.1568(24)$ & $-0.1580(24)$ & $-0.1608(23)$ & $-0.1610(22)$ & $-0.1601(22)$ \\
    $C_{{}^3S_1}$ & $+0.557(51)$ & $+0.544(49)$ & $+0.567(49)$ & $+0.598(48)$ & $+0.626(47)$ & $+0.652(46)$ \\
    $C_{E_1}$ & $+0.280(21)$ & $+0.385(16)$ & $+0.411(14)$ & $+0.500(12)$ & $+0.529(11)$ & $+0.550(11)$ \\
    $C_{{}^3P_0}$ & $+1.282(35)$ & $+1.292(33)$ & $+1.281(33)$ & $+1.237(31)$ & $+1.220(31)$ & $+1.203(30)$ \\
    $C_{{}^1P_1}$ & $+0.106(41)$ & $+0.274(38)$ & $+0.281(37)$ & $+0.400(34)$ & $+0.429(34)$ & $+0.447(33)$ \\
    $C_{{}^3P_1}$ & $-0.689(35)$ & $-0.796(33)$ & $-0.821(33)$ & $-0.844(31)$ & $-0.863(31)$ & $-0.890(31)$ \\
    $C_{{}^3P_2}$ & $-0.748(32)$ & $-0.682(31)$ & $-0.687(31)$ & $-0.654(30)$ & $-0.654(30)$ & $-0.658(30)$ \\
    $c_D$ & $+0.587(71)$ & $+0.334(62)$ & $+0.307(60)$ & $+0.094(56)$ & $+0.043(55)$ & $+0.000(54)$ \\
    $c_E$ & $-0.744(33)$ & $-0.646(19)$ & $-0.638(18)$ & $-0.645(15)$ & $-0.658(15)$ & $-0.673(14)$ \\
    $c_1$ & $-0.54(30)$ & $-0.39(30)$ & $-0.28(30)$ & $-0.06(30)$ & $+0.09(30)$ & $+0.27(29)$ \\
    $c_2$ & $+3.4(11)$ & $+3.4(11)$ & $+3.8(11)$ & $+4.3(10)$ & $+4.7(10)$ & $+5.3(10)$ \\
    $c_3$ & $-4.10(15)$ & $-3.69(14)$ & $-3.72(14)$ & $-3.54(13)$ & $-3.55(13)$ & $-3.56(13)$ \\
    $c_4$ & $+2.15(17)$ & $+2.60(12)$ & $+2.70(11)$ & $+3.281(90)$ & $+3.467(84)$ & $+3.644(78)$ \\
    $d_1\!+\!d_2$ & $+4.54(12)$ & $+4.66(10)$ & $+4.72(10)$ & $+4.975(96)$ & $+5.073(94)$ & $+5.170(92)$ \\
    $d_3$ & $-4.49(23)$ & $-4.49(22)$ & $-4.53(22)$ & $-4.64(22)$ & $-4.70(22)$ & $-4.76(22)$ \\
    $d_5$ & $-0.01(14)$ & $-0.08(14)$ & $-0.08(14)$ & $-0.16(14)$ & $-0.18(14)$ & $-0.19(14)$ \\
    $d_{14}\!-\!d_{15}$ & $-9.72(27)$ & $-9.76(27)$ & $-9.81(27)$ & $-9.98(27)$ & $-10.05(27)$ & $-10.12(27)$ \\
    $e_{14}$ & $+0.76(90)$ & $+0.47(90)$ & $+0.41(90)$ & $+0.08(89)$ & $-0.03(89)$ & $-0.14(89)$ \\
    $e_{15}$ & $-10.3(25)$ & $-11.6(25)$ & $-11.0(24)$ & $-10.8(24)$ & $-10.1(24)$ & $-9.3(24)$ \\
    $e_{16}$ & $+5.7(42)$ & $+6.4(42)$ & $+5.0(42)$ & $+3.4(41)$ & $+1.6(41)$ & $-0.5(41)$ \\
    $e_{17}$ & $+1.1(18)$ & $+1.0(18)$ & $+1.1(18)$ & $+1.2(18)$ & $+1.3(18)$ & $+1.4(18)$ \\
    $e_{18}$ & $+7.3(18)$ & $+5.3(17)$ & $+4.8(17)$ & $+2.0(16)$ & $+1.1(16)$ & $+0.2(16)$ \\
\end{tabular}\end{ruledtabular}
\end{table*}
\begin{table*}[htb]\caption{\label{NNLOsim_575}\NNLOsim{} ($\Lambda=575$)}
\begin{ruledtabular}\begin{tabular}{ccccccc}
    LEC & $T_{\rm lab}^{\max}\!=\!125$ & $T_{\rm lab}^{\max}\!=\!158$ & $T_{\rm lab}^{\max}\!=\!191$ & $T_{\rm lab}^{\max}\!=\!224$ & $T_{\rm lab}^{\max}\!=\!257$ & $T_{\rm lab}^{\max}\!=\!290$ \\\hline
    $\tilde{C}_{{}^1S_0}^{(np)}$ & $-0.1495(20)$ & $-0.1492(20)$ & $-0.1486(20)$ & $-0.1481(20)$ & $-0.1474(20)$ & $-0.1464(20)$ \\
    $\tilde{C}_{{}^1S_0}^{(pp)}$ & $-0.1487(20)$ & $-0.1483(20)$ & $-0.1478(20)$ & $-0.1472(20)$ & $-0.1464(20)$ & $-0.1454(20)$ \\
    $\tilde{C}_{{}^1S_0}^{(nn)}$ & $-0.1498(21)$ & $-0.1494(21)$ & $-0.1488(21)$ & $-0.1482(20)$ & $-0.1474(20)$ & $-0.1464(20)$ \\
    $C_{{}^1S_0}$ & $+2.475(49)$ & $+2.426(49)$ & $+2.446(48)$ & $+2.461(47)$ & $+2.485(46)$ & $+2.512(46)$ \\
    $\tilde{C}_{{}^3S_1}$ & $-0.1413(30)$ & $-0.1522(25)$ & $-0.1539(24)$ & $-0.1575(23)$ & $-0.1579(23)$ & $-0.1571(22)$ \\
    $C_{{}^3S_1}$ & $+0.534(51)$ & $+0.522(49)$ & $+0.546(49)$ & $+0.568(47)$ & $+0.594(46)$ & $+0.619(46)$ \\
    $C_{E_1}$ & $+0.280(21)$ & $+0.375(16)$ & $+0.404(14)$ & $+0.486(12)$ & $+0.513(11)$ & $+0.533(10)$ \\
    $C_{{}^3P_0}$ & $+1.426(37)$ & $+1.397(35)$ & $+1.386(34)$ & $+1.308(32)$ & $+1.284(31)$ & $+1.260(31)$ \\
    $C_{{}^1P_1}$ & $+0.127(41)$ & $+0.282(39)$ & $+0.286(37)$ & $+0.396(35)$ & $+0.421(34)$ & $+0.433(33)$ \\
    $C_{{}^3P_1}$ & $-0.598(37)$ & $-0.734(34)$ & $-0.762(33)$ & $-0.783(32)$ & $-0.804(31)$ & $-0.833(31)$ \\
    $C_{{}^3P_2}$ & $-0.735(32)$ & $-0.676(31)$ & $-0.680(31)$ & $-0.645(30)$ & $-0.645(30)$ & $-0.649(30)$ \\
    $c_D$ & $+0.694(72)$ & $+0.472(64)$ & $+0.438(62)$ & $+0.251(59)$ & $+0.208(58)$ & $+0.176(57)$ \\
    $c_E$ & $-1.232(78)$ & $-0.990(45)$ & $-0.955(40)$ & $-0.887(30)$ & $-0.887(28)$ & $-0.893(27)$ \\
    $c_1$ & $-0.49(30)$ & $-0.37(30)$ & $-0.27(30)$ & $-0.10(30)$ & $+0.04(29)$ & $+0.21(29)$ \\
    $c_2$ & $+3.5(11)$ & $+3.5(11)$ & $+3.8(11)$ & $+4.1(10)$ & $+4.5(10)$ & $+5.1(10)$ \\
    $c_3$ & $-4.03(14)$ & $-3.66(14)$ & $-3.69(14)$ & $-3.50(13)$ & $-3.51(13)$ & $-3.52(13)$ \\
    $c_4$ & $+2.25(16)$ & $+2.57(12)$ & $+2.69(10)$ & $+3.188(87)$ & $+3.363(81)$ & $+3.523(76)$ \\
    $d_1\!+\!d_2$ & $+4.57(12)$ & $+4.63(10)$ & $+4.704(99)$ & $+4.914(95)$ & $+5.007(93)$ & $+5.096(92)$ \\
    $d_3$ & $-4.50(23)$ & $-4.47(22)$ & $-4.52(22)$ & $-4.60(22)$ & $-4.66(22)$ & $-4.72(22)$ \\
    $d_5$ & $-0.02(14)$ & $-0.07(14)$ & $-0.08(14)$ & $-0.15(14)$ & $-0.17(14)$ & $-0.18(14)$ \\
    $d_{14}\!-\!d_{15}$ & $-9.73(27)$ & $-9.74(27)$ & $-9.79(27)$ & $-9.93(27)$ & $-9.99(27)$ & $-10.06(27)$ \\
    $e_{14}$ & $+0.70(90)$ & $+0.48(90)$ & $+0.41(89)$ & $+0.12(89)$ & $+0.02(89)$ & $-0.08(89)$ \\
    $e_{15}$ & $-10.4(25)$ & $-11.8(25)$ & $-11.2(24)$ & $-11.2(24)$ & $-10.6(24)$ & $-9.8(24)$ \\
    $e_{16}$ & $+5.5(42)$ & $+6.4(42)$ & $+5.1(42)$ & $+4.1(41)$ & $+2.5(40)$ & $+0.5(41)$ \\
    $e_{17}$ & $+1.1(18)$ & $+1.0(18)$ & $+1.0(18)$ & $+1.1(18)$ & $+1.2(18)$ & $+1.3(18)$ \\
    $e_{18}$ & $+6.9(17)$ & $+5.5(17)$ & $+4.9(16)$ & $+2.5(16)$ & $+1.7(16)$ & $+0.8(16)$ \\
\end{tabular}\end{ruledtabular}
\end{table*}
\begin{table*}[htb]\caption{\label{NNLOsim_600}\NNLOsim{} ($\Lambda=600$)}
\begin{ruledtabular}\begin{tabular}{ccccccc}
    LEC & $T_{\rm lab}^{\max}\!=\!125$ & $T_{\rm lab}^{\max}\!=\!158$ & $T_{\rm lab}^{\max}\!=\!191$ & $T_{\rm lab}^{\max}\!=\!224$ & $T_{\rm lab}^{\max}\!=\!257$ & $T_{\rm lab}^{\max}\!=\!290$ \\\hline
    $\tilde{C}_{{}^1S_0}^{(np)}$ & $-0.1491(20)$ & $-0.1487(20)$ & $-0.1483(20)$ & $-0.1481(20)$ & $-0.1474(20)$ & $-0.1466(20)$ \\
    $\tilde{C}_{{}^1S_0}^{(pp)}$ & $-0.1483(20)$ & $-0.1479(20)$ & $-0.1474(20)$ & $-0.1471(20)$ & $-0.1464(20)$ & $-0.1456(20)$ \\
    $\tilde{C}_{{}^1S_0}^{(nn)}$ & $-0.1495(21)$ & $-0.1491(21)$ & $-0.1486(21)$ & $-0.1483(20)$ & $-0.1476(20)$ & $-0.1467(20)$ \\
    $C_{{}^1S_0}$ & $+2.469(49)$ & $+2.419(48)$ & $+2.437(48)$ & $+2.441(47)$ & $+2.461(46)$ & $+2.485(46)$ \\
    $\tilde{C}_{{}^3S_1}$ & $-0.1348(32)$ & $-0.1477(27)$ & $-0.1500(25)$ & $-0.1547(24)$ & $-0.1554(23)$ & $-0.1547(23)$ \\
    $C_{{}^3S_1}$ & $+0.512(50)$ & $+0.504(49)$ & $+0.527(48)$ & $+0.538(47)$ & $+0.563(47)$ & $+0.583(46)$ \\
    $C_{E_1}$ & $+0.279(21)$ & $+0.368(15)$ & $+0.399(14)$ & $+0.476(12)$ & $+0.502(11)$ & $+0.520(10)$ \\
    $C_{{}^3P_0}$ & $+1.634(43)$ & $+1.556(38)$ & $+1.541(37)$ & $+1.416(33)$ & $+1.382(32)$ & $+1.351(31)$ \\
    $C_{{}^1P_1}$ & $+0.149(42)$ & $+0.296(40)$ & $+0.296(38)$ & $+0.400(35)$ & $+0.423(34)$ & $+0.432(33)$ \\
    $C_{{}^3P_1}$ & $-0.472(42)$ & $-0.653(35)$ & $-0.683(34)$ & $-0.703(32)$ & $-0.725(32)$ & $-0.756(31)$ \\
    $C_{{}^3P_2}$ & $-0.723(32)$ & $-0.669(31)$ & $-0.671(31)$ & $-0.633(30)$ & $-0.632(30)$ & $-0.635(29)$ \\
    $c_D$ & $+0.835(74)$ & $+0.632(67)$ & $+0.592(65)$ & $+0.424(62)$ & $+0.388(62)$ & $+0.363(61)$ \\
    $c_E$ & $-2.40(23)$ & $-1.76(13)$ & $-1.64(11)$ & $-1.409(77)$ & $-1.383(72)$ & $-1.372(68)$ \\
    $c_1$ & $-0.46(30)$ & $-0.36(30)$ & $-0.28(30)$ & $-0.17(30)$ & $-0.04(29)$ & $+0.10(29)$ \\
    $c_2$ & $+3.5(11)$ & $+3.5(11)$ & $+3.7(11)$ & $+3.8(10)$ & $+4.2(10)$ & $+4.7(10)$ \\
    $c_3$ & $-3.97(14)$ & $-3.63(14)$ & $-3.65(14)$ & $-3.45(13)$ & $-3.45(13)$ & $-3.46(13)$ \\
    $c_4$ & $+2.33(15)$ & $+2.55(11)$ & $+2.69(10)$ & $+3.112(85)$ & $+3.274(80)$ & $+3.417(74)$ \\
    $d_1\!+\!d_2$ & $+4.60(11)$ & $+4.61(10)$ & $+4.692(99)$ & $+4.860(94)$ & $+4.944(93)$ & $+5.023(91)$ \\
    $d_3$ & $-4.51(23)$ & $-4.46(22)$ & $-4.51(22)$ & $-4.56(22)$ & $-4.61(22)$ & $-4.66(22)$ \\
    $d_5$ & $-0.03(14)$ & $-0.07(14)$ & $-0.09(14)$ & $-0.15(14)$ & $-0.16(14)$ & $-0.17(14)$ \\
    $d_{14}\!-\!d_{15}$ & $-9.75(27)$ & $-9.72(27)$ & $-9.78(27)$ & $-9.88(27)$ & $-9.94(27)$ & $-10.00(27)$ \\
    $e_{14}$ & $+0.65(90)$ & $+0.48(90)$ & $+0.40(89)$ & $+0.16(90)$ & $+0.06(89)$ & $-0.03(89)$ \\
    $e_{15}$ & $-10.5(25)$ & $-11.9(25)$ & $-11.5(24)$ & $-11.8(24)$ & $-11.3(24)$ & $-10.6(24)$ \\
    $e_{16}$ & $+5.5(42)$ & $+6.4(42)$ & $+5.4(42)$ & $+5.2(41)$ & $+3.8(41)$ & $+2.1(40)$ \\
    $e_{17}$ & $+1.1(18)$ & $+0.9(18)$ & $+1.0(18)$ & $+1.1(18)$ & $+1.1(18)$ & $+1.2(18)$ \\
    $e_{18}$ & $+6.5(17)$ & $+5.6(17)$ & $+4.9(16)$ & $+2.9(16)$ & $+2.1(16)$ & $+1.4(16)$ \\
\end{tabular}\end{ruledtabular}
\end{table*}

\end{document}